\RequirePackage{ifpdf}
\documentclass[hyper,12pt,letterpaper]{JHEP3}

\usepackage{cite}

\usepackage{graphicx}
\usepackage{latexsym,amsmath,amsfonts,amssymb}
\usepackage{mathrsfs}
\usepackage{amssymb}
\usepackage{amsfonts}
\usepackage{amsmath}
\usepackage{amscd}
\usepackage[makeroom]{cancel}
\usepackage{bbm}
\usepackage{bm}
\usepackage{subfigure}
\usepackage{paralist}
\usepackage{url}
\input xy
\xyoption{all}

\numberwithin{equation}{section}

\newcommand{\nn}{\nonumber}
\newcommand{\mat}[1]{\begin{pmatrix} #1 \end{pmatrix}}

\newcommand{\be}{\begin{equation}} 
\newcommand{\ee}{\end{equation}}
\newcommand{\bea}{\begin{equation} \begin{aligned}} \newcommand{\eea}{\end{aligned} \end{equation}}

\newcommand{\bit}{\begin{itemize}} 
\newcommand{\eit}{\end{itemize}}

\newcommand{\Z}{\mathbb{Z}}
\newcommand{\C}{\mathbb{C}}
\newcommand{\R}{\mathbb{R}}

\renewcommand{\t}{\widetilde }
\renewcommand{\d}{\partial }

\renewcommand{\b}{\bar }

\newcommand{\half}{{1\over 2}}

\newcommand{\bz}{{\b z}}

\newcommand{\CE}{\mathcal{E}}
\newcommand{\CF}{\mathcal{F}}

\newcommand{\CK}{\mathcal{K}}
\newcommand{\CL}{\mathcal{L}}

\newcommand{\CN}{\mathcal{N}}
\newcommand{\CO}{\mathcal{O}}

\newcommand{\CQ}{\mathcal{Q}}

\newcommand{\CV}{\mathcal{V}}
\newcommand{\CW}{\mathcal{W}}

\newcommand{\FR}{\mathfrak{R}}
\newcommand{\Fg}{\mathfrak{g}}

\newcommand{\GG}{\mathbf{G}}

\newcommand{\bb}{{\bb b}}

\DeclareMathOperator{\Tr}{Tr}

\newcommand{\SL}{{\mathscr L}}

\newcommand{\ov}{\over}

\newcommand{\tp}{\theta^+}

\newcommand{\Ext}{{\rm Ext}}

\title{B-branes and supersymmetric quivers in 2d}

\author{Cyril~Closset,$^\flat$ Jirui~Guo$^\natural$ and Eric~Sharpe$^{\sharp}$\\

{}$^{\flat}$Theory Department, CERN\\
CH-1211, Geneva 23, Switzerland\\

{}$^{\natural}$ Department of Physics and Center for Field Theory and Particle Physics \\
Fudan University,  220 Handan Road, 200433 Shanghai, China\\

{}$^{\sharp}$ Department of Physics MC 0435, 850 West Campus Drive  \\
 Virginia Tech,  Blacksburg, VA 24061, USA\\

}

\preprint{CERN-TH-2017-244}

\abstract{We  study 2d $\mathcal{N}=(0,2)$ supersymmetric quiver gauge theories that describe the low-energy dynamics of D1-branes at Calabi-Yau fourfold (CY$_4$) singularities. On general grounds, the holomorphic sector of these theories---matter content and (classical) superpotential interactions---should be fully captured by the topological $B$-model on the CY$_4$. 
By studying a number of examples, we confirm this expectation and flesh out the dictionary between B-brane category and supersymmetric quiver:  the matter content of the supersymmetric quiver is encoded in morphisms between B-branes (that is, Ext groups of coherent sheaves), while the superpotential interactions are encoded in the $A_\infty$ algebra satisfied by the morphisms. This provides us with a derivation of the supersymmetric quiver directly from the CY$_4$ geometry.
We also suggest a relation between triality of $\mathcal{N}=(0,2)$ gauge theories and certain mutations of exceptional collections of sheaves.  
0d $\mathcal{N}=1$ supersymmetric quivers, corresponding to D-instantons probing CY$_5$ singularities, can be discussed similarly.
}

\begin{document}

\tableofcontents

\section{Introduction}
Many supersymmetric quantum field theories can  be engineered on systems of branes in string theory. The string theory embedding often provides us with an elegant geometric understanding of field theory phenomena. 
In particular,  rich classes of field theories, the {\it supersymmetric quiver gauge theories}, can be engineered by considering parallel D$p$-branes at the tip of a conical local Calabi-Yau (CY) $n$-fold ${\bf X}_n$, with $p= 9-2n$, in type IIB string theory. One obtains the following types of supersymmetric gauge theories in the open-string sector:
\begin{itemize}
\item 6d $\CN=(0,1)$ quiver theories on D5-branes at the tip of a CY$_2$ cone.
\item 4d $\CN=1$ quiver theories on D3-branes at the tip of a CY$_3$ cone.
\item 2d $\CN=(0,2)$ quiver theories on D1-branes at the tip of a CY$_4$ cone.
\item 0d $\CN=1$ quiver theories on D-instantons at the tip of a CY$_5$ cone. 
\end{itemize}
All these quiver gauge theories consist of unitary gauge groups $\prod_I U(N_I)$ with matter fields in adjoint and bifundamental representations.~\footnote{That is, fields $X_{IJ}$ in the fundamental of $U(N_I)$ and in the antifundamental of $U(N_J)$.} 
The 6d case corresponds to D5-branes at the tip of an ADE singularity $\C^2/\Gamma$, and the quiver gauge theories are the corresponding ADE quivers \cite{Douglas:1996sw}. The 4d case has been thoroughly studied from various points of views---see {\it e.g.} \cite{Douglas:1997de, Kachru:1998ys, Klebanov:1998hh, Feng:2000mi, Wijnholt:2002qz, Herzog:2003zc, Herzog:2004qw, Aspinwall:2004vm,  Aspinwall:2004bs, Hanany:2005ve, Franco:2005rj, Franco:2005sm, Feng:2005gw, Herzog:2006bu} for a very partial list. 
It is a special and important case because the D3-branes admit a smooth near-horizon limit of the form $AdS_5 \times X_5$ \cite{Kachru:1998ys, Klebanov:1998hh, Morrison:1998cs} and the 4d quiver gauge theories flow to non-trivial 4d $\CN=1$ superconformal fixed points.

The 2d and 0d cases had attracted less attention until more recently---see  \cite{Mohri:1997ef, GarciaCompean:1998kh} for some early work. A recent breakthrough was the introduction of ``brane brick models''  \cite{Franco:2015tna, Franco:2015tya}, which gave an algorithm to determine the $\CN=(0,2)$ quiver gauge theory corresponding to D1-branes probing a {\it toric} CY$_4$ singularity, similarly to brane tiling methods for D3-branes at toric CY$_3$ singularities \cite{Hanany:2005ve, Franco:2005rj, Franco:2005sm}. The brane brick models were derived using mirror symmetry in \cite{Franco:2016qxh}.  There are also hints that a similar structure exists for D$(-1)$-branes at toric CY$_5$ singularities.~\footnote{Very recently, these 2d and 0d quivers were also related to cluster algebras \cite{Franco:2017lpa}.} Note that this line of work (and the present paper) is only concerned with the {\it classical} structure of the $\CN=(0,2)$ gauge theory.
In a parallel development, there has been some important progress in our understanding of two-dimensional $\CN=(0,2)$ gauge theories  as full-fledged quantum field theories   \cite{Gadde:2013lxa, Gadde:2014ppa, Gadde:2015kda}. 
Incidentally, it was discovered that the simplest SQCD-like $\CN=(0,2)$ theories enjoy a beautiful {\it triality} \cite{Gadde:2013lxa}---an infrared ``duality'' of order three similar to Seiberg duality. Triality also seems to be a generic property of D1-branes quivers \cite{Franco:2016nwv}.  There has also been some interesting recent work on engineering $\CN=(0,2)$ models from F-theory \cite{Schafer-Nameki:2016cfr, Apruzzi:2016iac, Apruzzi:2016nfr, Lawrie:2016axq, Lawrie:2016rqe}. See also \cite{Katz:2004nn, Witten:2005px, Adams:2005tc, Tan:2006qt, McOrist:2007kp, Tan:2008mi, Donagi:2011uz, Donagi:2011va, Dedushenko:2015opz, Closset:2015ohf, Guo:2015caf, Gu:2017nye} for related works on quantum aspects of $\CN=(0,2)$ theories.

In this paper, we study 2d and 0d quivers from the point of view of {\it B-branes} on the CY $n$-fold ${\bf X}_n$. A B-brane is simply a (half-BPS) D-brane in the topological B-model on ${\bf X}_n$.   The B-model is a $g_s=0$ limit of type II string theory which (somewhat trivially) captures all $\alpha'$ corrections. It can thus be used to  accurately describe the local physics of branes at a Calabi-Yau {\it singularity.}  Since the B-model is independent of K{\"a}hler deformations, we can use any convenient limit, such as, for instance, the large volume limit of a resolved singularity, to study the quantities of interest. In this way, we loose a lot of important information---for instance, we do not keep track of the central charges of the branes, which determines their stability properties;  yet, the B-model is sufficient in order to extract all the information about the {\it holomorphic} sector of the low-energy open strings. That is, we can read off the matter spectrum and the superpotential interactions of the low-energy quiver gauge theories on D$p$-branes from the B-branes alone.~\footnote{An important caveat is that we need to be given a particular set of B-branes, the ``fractional branes'', as we will mention. The determination of whether a given set of B-branes is an allowable set of fractional branes may require information beyond the $B$-model.}

This approach was successfully carried out for D3-branes at CY$_3$ singularities \cite{Wijnholt:2002qz, Herzog:2003zc, Herzog:2004qw, Aspinwall:2004vm,  Aspinwall:2004bs, Herzog:2006bu}. What we present here is a straightforward extension of some of those earlier works. It provides a string theory derivation of some brane brick models results, without the need to rely on mirror symmetry. Our techniques are also more general, since they are valid beyond the realm of toric geometry.~\footnote{This is as a matter of principle. In this paper, all our examples will be toric geometries, somewhat by happenstance, and also so that we can compare our results to the brane brick model literature.}

Mathematically, a B-brane $\CE$ on ${\bf X}_n$ is an object in the (bounded) derived category of coherent sheaves of ${\bf X}_n$:
\be
\CE \in {\bf D}^b ({\bf X}_n)
\ee
We can think of B-branes $\CE$ as coherent sheaves; more generally they are chain complexes of coherent sheaves (up to certain equivalences).  
Given two $B$-branes $\CE$, $\CF$, we may compute their $\Ext$ groups:
\be\label{extEF}
\Ext^i_{{\bf X}_n}(\CE, \CF)~,
\ee
which are the morphisms in the derived category. Physically, they encode the low-energy modes of the {\it open strings} stretched between the D-branes $\CE$ and $\CF$ \cite{Sharpe:1999qz, Douglas:2000gi, Katz:2002gh, Katz:2002jh}. 
We refer to \cite{Sharpe:2003dr, Aspinwall:2004jr} for comprehensive  reviews of the derived category approach to $B$-branes.

\subsection*{D-branes quivers from B-branes}
Consider a D$(9-2n)$-brane transverse to the Calabi-Yau singularity ${\bf X}_n$. Away from the singularity, the brane is locally in flat space. From the point of view of ${\bf X}_n$, it is a point-like brane, which is described by a skyscraper sheaf $\CO_p$ at a point $p\in {\bf X}_n$. When at the singularity, it is expected that $\CO_p$  ``fractionates'' into marginally stable constituents. The resulting ``fractional branes'' $\{\CE_I\}$ realize a gauge group:
\be
\prod_I U(N_I)
\ee
on their worldvolume in the transverse directions. There are also massless open strings connecting the  fractional branes among themselves, which realize bifundamental (or adjoint) matter fields $X_{IJ}$. In this way, the low-energy open string sector at the singularity is described by a {\it supersymmetric quiver gauge theory}: to each fractional brane $\CE_I$, we associate a node in the quiver, denoted by $e_I$. The matter fields corresponds to various quiver {\it arrows} connecting the nodes: $e_I \longrightarrow e_J$. There are also interaction terms among the matter fields, which we will discuss  below. 

In all cases, the fractional branes are such that:
\be
\Ext^0_{{\bf X}_n}(\CE_I, \CE_J)=  \delta_{IJ}~.
\ee
These $\Ext^0\cong \text{Hom}$ groups  are identified with {\it vector multiplets} in $d=10-2n$ dimensions; a single vector multiplet is assigned to each node $e_I$, realizing the gauge group $U(N_I)$. The other $\Ext^i$ groups  (with $i=1, \cdots, n-1$)  correspond to matter fields charged under the gauge groups.

We should probably emphasize that, in this paper, we will be mostly interested in the supersymmetric quiver as an abstract algebraic object, consisting of nodes, arrows and relations. The assignment of particular gauge groups $U(N_I)$ is part of the data of a quiver {\it representation}, and the gauge group ranks can vary depending on the physical setup (that is, which D-branes are we using to probe the singularity). In other words, our concept of supersymmetric quiver can encode many different supersymmetric theories with the same structure but distinct gauge groups.~\footnote{Not all unitary gauge groups are allowed, however. Gauge anomalies provide strong constraints on the allowed quiver ranks.}

\vskip0.3cm
A crucial property of $\Ext$ groups on a Calabi-Yau variety ${\bf X}_n$ is the Serre duality relation:
\be\label{Serre dual intro}
\Ext^i_{{\bf X}_n}(\CE_I, \CE_J) \cong \Ext^{n-i}_{{\bf X}_n}(\CE_J, \CE_I)~, \qquad i= 0, \cdots, n~.
\ee
This corresponds to the CPT symmetry of the d-dimensional quiver quantum field theory.  Generalizing some relatively well-known results for D3-branes, it is  natural to propose the following identification of $\Ext$ groups with supersymmetry multiplets in various dimensions:

\paragraph{D5-brane quivers.} For D5-branes on ${\bf X}_2$, we have:
\be
\Ext^1_{{\bf X}_2}(\CE_J, \CE_I) \qquad \Leftrightarrow \qquad e_I  \; \small{\text{------} }\; e_J   \qquad \Leftrightarrow \qquad X_{IJ}~,
\ee
where $X_{IJ}$ are 6d $\CN=(0,1)$ hypermultiplets in the bifundamental representation of $U(n_I)\times U(n_J)$.
Note that the quiver link $e_I\, \text{---}\, e_J$  is unoriented since the hypermultiplet is non-chiral---this corresponds to the Serre duality $\Ext^{1}(\CE_I, \CE_J)\cong \Ext^{1}(\CE_J, \CE_I)$ on ${\bf X}_2$. In this case, ${\bf X}_2$ must be an ADE singularity while the supersymmetric quivers are extended Dynkin diagrams. 

\paragraph{D3-brane quivers.}  For D3-branes on ${\bf X}_3$, we have:
\be
\Ext^1_{{\bf X}_3}(\CE_J, \CE_I) \qquad \Leftrightarrow \qquad e_I \longrightarrow e_J   \qquad \Leftrightarrow \qquad X_{IJ}~,
\ee
where $X_{IJ}$ are 4d $\CN=1$ chiral multiplets in the bifundamental of $U(n_I) \times U(n_J)$, or in the adjoint of $U(n_I)$ if $I=J$. The arrows are oriented. Therefore,  such quiver gauge theories are generally {\it chiral} theories. More precisely, we denote by:
\be
d_{IJ}^1 \equiv \dim \Ext^1_{{\bf X}_3}(\CE_J, \CE_I)
\ee
the number of arrows from $e_I$ to $e_J$ in the 4d $\CN=1$ quiver. D3-brane quivers are ``ordinary'' quivers (with relations), consisting of nodes and arrows, of the type most studied by both physicists and mathematicians. 

\paragraph{D1-brane quivers.}   D1-branes on ${\bf X}_4$ lead to the richer structure of 2d $\CN=(0,2)$ quiver gauge theories. Those quivers have {\it two} distinct types of arrows, corresponding to $(0,2)$ chiral multiplets $X_{IJ}$ and $(0,2)$ fermi multiplets $\Lambda_{IJ}$, respectively. We propose the identification:
\bea
&\Ext^1_{{\bf X}_4}(\CE_J, \CE_I) \qquad &\Leftrightarrow& \qquad e_I \longrightarrow e_J   \qquad &\Leftrightarrow& \qquad X_{IJ}~,\cr
&\Ext^2_{{\bf X}_4}(\CE_J, \CE_I) \qquad &\Leftrightarrow& \qquad e_I \; \small{\text{- - -} }\; e_J   \qquad &\Leftrightarrow& \qquad \Lambda_{IJ}~.\cr
\eea
 Note that the $\Ext^2_{{\bf X}_4}(\CE_I, \CE_J)\cong \Ext^2_{{\bf X}_4}(\CE_J, \CE_I)$ by Serre duality. Thus the second type of arrow is unoriented. This corresponds to the self-duality of the fermi multiplet in such theories. We also define:
\be
d_{IJ}^1 \equiv \dim \Ext^1_{{\bf X}_4}(\CE_J, \CE_I)~, \qquad\quad
d_{IJ}^2 \equiv \dim \Ext^2_{{\bf X}_4}(\CE_J, \CE_I)~.
\ee
Here $d^1_{IJ}$ is the number of chiral multiplets from $e_I$ to $e_J$ (in bifundamental  representations if $I\neq J$ or adjoint representation if $I=J$). Similarly, $d^2_{IJ}= d^2_{JI}$ denotes the number of  bifundamental fermi multiplets if $I \neq J$, while  $\half d^2_{IJ}$ is the number of adjoint fermi multiplets if $I=J$.

\paragraph{D$(-1)$-brane quivers.} Finally, we may consider D-instantons on ${\bf X}_5$, which results in a 
quiver with two types of oriented arrows:
\bea
&\Ext^1_{{\bf X}_5}(\CE_J, \CE_I) \qquad &\Leftrightarrow& \qquad e_I \longrightarrow e_J   \qquad &\Leftrightarrow& \qquad X_{IJ}~,\cr
&\Ext^2_{{\bf X}_5}(\CE_J, \CE_I) \qquad &\Leftrightarrow& \qquad e_I \dashrightarrow e_J   \qquad &\Leftrightarrow& \qquad \Lambda_{IJ}~.\cr
\eea
The corresponding $\CN=1$ gauged matrix model contains two types of ``matter'' multiplets, the chiral and fermi multiplets \cite{Franco:2016tcm}. In this case, the quantities:
\be
d_{IJ}^1 \equiv \dim \Ext^1_{{\bf X}_5}(\CE_J, \CE_I)~, \qquad\quad
d_{IJ}^2 \equiv \dim \Ext^2_{{\bf X}_5}(\CE_J, \CE_I)~,
\ee
give the number of arrows of either types from $e_I$ to $e_J$. We will briefly discuss these gauged matrix models in section \ref{sec: GMM}.

\subsection*{Elusive fractional branes} The above identifications between $\Ext$ groups and supersymmetric multiplets in D-brane quivers are {\it conjectures}, that we may check in many explicit computations.   The practical usefulness of these identifications rely on identifying the fractional branes $\CE_I$ in the first place, as distinguished objects in the B-brane category on ${\bf X}_n$.  To the best of our knowledge, this remains an open problem in general. In this note, we will deal with simple examples where we can describe the fractional branes explicitly.

\subsection*{Interactions terms: Product structure in the derived category}
Importantly, the D-brane quivers have interactions terms, which are encoded in {\it superpotentials} in various dimensions. On D5-branes, the interactions are fully determined by supersymmetry, while D3-brane quivers have a non-trivial 4d $\CN=1$ superpotential $W(X)$. The 2d $\CN=(0,2)$ theories have two types of ``superpotential'' interactions, encoded in  holomorphic functions $J(X)$ and $E(X)$  \cite{Witten:1993yc}. The 0d $\CN=1$ matrix models also have two kinds of holomorphic ``superpotentials'', distinct from the 2d superpotentials, denoted by $F(X)$ and $H(X)$ \cite{Franco:2016tcm}.~\footnote{What we call $F$-term was called $J$-term in \cite{Franco:2016tcm}. We choose this notation in order to distinguish between the 2d and 0d interactions.}

These interactions terms can be recovered from the fractional branes by considering the product structure between Ext groups. Let $A$ denote the graded algebra Ext$^\bullet$ for a given set of fractional branes, where the grading is by the degree of the Ext groups. (It is also the ghost number of the $B$-model.)  There exists multi-products:
\be
m_k :  A^{\otimes k} \rightarrow A~,
\ee
of degree $2-k$, satisfying relations amongst themselves, that generate a minimal $A_{\infty}$ structure.~\footnote{A minimal $A_\infty$ structure is an $A_\infty$ structure in which $m_1=0$ \cite{1999math.....10179K}.} In particular,  if $a\in \Ext^\bullet(\CE_1, \CE_2)$ and $b\in \Ext^\bullet(\CE_2, \CE_3)$, then  $m_2(b, a) \in \Ext^\bullet(\CE_1, \CE_3)$ is the product obtained by composition.
 These multi-products correspond to disk correlators in the topological $B$-model. 

 It is known that the $A_{\infty}$ structure encodes the 4d $\CN=1$ superpotential of D3-brane quivers \cite{Aspinwall:2004bs, Aspinwall:2005ur}. Following the same methods, we will be able to derive the  2d $\CN=(0,2)$ and 0d $\CN=1$ quiver interactions.  We note that similar ideas were previously reported in
\cite{eagertalk}.

\vskip0.4cm

\noindent This paper is organized as follows. In section \ref{sec: 2d02 quivers}, we discuss the construction of 2d $\CN=(0,2)$ supersymmetric quiver gauge theories from the knowledge the B-branes on a CY fourfold. In section \ref{sec: duality}, we discuss triality of 2d $\CN=(0,2)$ quivers in this context, and relate
   triality to mutations of exceptional collections of sheaves.
In section \ref{sec: GMM}, we discuss the similar construction of 0d $\CN=1$ quiver theories from B-branes on a CY fivefold. A few complementary points are discussed in  Appendices.

\section{D1-brane quivers and 2d $\CN=(0,2)$ theories}\label{sec: 2d02 quivers}
Two-dimensional gauge theories with $\CN=(0,2)$ supersymmetry are built out of three types of supermultiplets: Vector, chiral and fermi multiplets \cite{Witten:1993yc}. 
 In Wess-Zumino gauge, the vector multiplet $(V, V_z)$ contains a gauge field $A_\mu$, left-moving gaugini and an auxiliary scalar $D$, transforming in the adjoint of the Lie algebra $\Fg = {\rm Lie}(G)$, with $\GG$ the gauge group.

The charged matter fields consist of chiral multiplets $\Phi$ and fermi multiplets $\Lambda$---and of their charge-conjugate multiplets, the anti-chiral multiplet $\t \Phi$ and the anti-fermi multiplet $\t \Lambda$, respectively. They satisfy the half-BPS conditions:
\be
\b D_+ \Phi= 0~, \qquad \qquad\qquad \b D_+ \Lambda=   E(\Phi)~.
\ee
The $E$-potential $E= E(\Phi)$ is itself a chiral superfield, given by a holomorphic function of the chiral multiplets $\Phi$.  In components, the chiral superfield reads:
\be
\Phi = \phi +  \theta^+ \psi_+ - 2 i \theta^+ \b \theta^+ \d_{\b z} \phi\,
\ee
with $\phi$ a complex scalar and $\psi_+$ a right-moving fermion. The fermi superfield is given by:
\be
\Lambda_- = \lambda_- -\theta^+ G - 2 i \theta^+ \b\theta^+ \d_{\b z} \lambda_- - \b \theta^+ E\, ,
\ee
with $\lambda_-$ a left-moving fermion an $G$ an auxiliary field. 
The chiral and fermi multiplets are valued in some representations $\FR_{\Phi}$ and $\FR_\Lambda$ of $\Fg$, respectively. Consequently, the potential $E(\Phi)$ is valued in $\FR_\Lambda$ as well.
The canonical kinetic Lagrangian for the matter fields is:
\be
\SL_{\rm kin} =  \int d\tp d\b\theta^+\, \left( i \b\Phi \mathcal{D}_z \Phi - \b \Lambda_- \Lambda_-\right)~,
\ee
with  $\mathcal{D}_z$ the gauge covariant derivative, and with the trace over $\Fg$ kept implicit. A standard super-Yang-Mills term can also be constructed for the vector multiplet. To every fermi multiplet $\Lambda$, we also associate an ``$\CN=(0,2)$ superpotential'' $J=J(\Phi)$ transforming in the conjugate representation $\b \FR_\Lambda$, such that:
\be\label{EJ constraint}
\Tr\left(E J\right)=0~,
\ee
with the trace over $\Fg$,  $\Tr : \FR_\Lambda \otimes \b\FR_\Lambda \rightarrow \C$.
The interaction Lagrangian reads:
\be
\SL_{J}= - \int d\tp\, \Lambda_- J(\Phi) -\int d\b\theta^+\,\b \Lambda_- \b J(\b\Phi)~,
\ee
with $\b J$ the conjugate potential for the anti-fermi multiplet. This Lagrangian is supersymmetric provided that  \eqref{EJ constraint} is satisfied. 
The auxiliary fields $G$, $\b G$ can be integrated out, which sets $G= \b J$ and $\b G= J$. We then obtain the following Lagrangian for the fermi multiplets:
\be
\CL_{\rm fermi}=   - 2i \b\lambda_- \d_\bz \lambda_-  + \b E E +\b J J+ \b\lambda_- {\d E\ov \d\phi}\psi_+ - \lambda_- {\d\b E\ov \d\b\phi}\b\psi_+ 
+ \lambda_- {\d J\ov \d\phi}\psi_+  -\b\lambda_- {\d \b J\ov \d\b\phi}\b\psi_+~.
\ee
Note that there is a symmetry that exchanges fermi and anti-fermi multiplets:
\be\label{fermi antiferm}
\Lambda \leftrightarrow \b \Lambda~, \qquad \qquad
E \leftrightarrow  J~, \qquad\qquad \b E \leftrightarrow \b J~.
\ee
In the presence of several fermi multiplets in distinct irreducible representations, each fermi multiplet can be ``dualized'' independently.~\footnote{
 See  \cite{Sharpe:2006qd} 
for a discussion of some subtleties in this symmetry
in $(0,2)$ NLSMs. }

\subsection{$\CN=(0,2)$ quiver gauge theory from $B$-branes at a CY$_4$ singularity}
Systems of D1-branes at CY$_4$ singularities engineer a simple yet rich class of gauge theories with  product gauge group:
\be\label{quiver G}
\GG= \prod_{I} U(N_I)~.
\ee
To each $U(N_I)$ gauge group, one associates an $\CN=(0,2)$ vector multiplet, 
denoted by a node $e_I$ in a quiver diagram. The matter fields in chiral 
multiplets are in bifundamental representations 
${\bf N}_I\otimes \b {\bf N}_J$ between unitary gauge groups. 
To each chiral multiplet $X_{IJ}$ in the fundamental of $U(N_I)$ and 
antifundamental of $U(N_J)$, we associate a solid arrow $e_I \longrightarrow e_J$ 
in the quiver diagram. The matter fields in fermi multiplets are also in 
bifundamental representations. To each bifundamental fermi multiplet 
$\Lambda_{IJ}$, we associate the dashed link 
$e_I \, \small{\text{- - -} }\, e_J$ in the quiver diagram. While $\Lambda_{IJ}$ denotes a fermi multiplet in the bifundamental ${\bf N}_I\otimes \b {\bf N}_J$ of $U(N_I)\times U(N_J)$, the associated link in the quiver is unoriented, reflecting the fermi duality~\eqref{fermi antiferm}.~\footnote{In practice, we still find it convenient to write oriented dashed arrows for fermi multiplets, reflecting a choice of representation for the fermi multiplets (that is, which is $\Lambda$ and which is $\b \Lambda$). This is because such a choice is needed to write down the off-shell supersymmetric action.} We may also have chiral and fermi multiplets in the adjoint representation of a single gauge group $U(N_I)$, corresponding to a special case of the above with $I=J$.

To each $\Lambda_{IJ}$, one associates an $E$-term and a $J$-term. Given that $\Lambda_{IJ}$ transform in the bifundamental representation ${\bf N}_I\otimes \b {\bf N}_J$, by convention, the potential $E_{\Lambda_{IJ}}$ transforms in ${\bf N}_I\otimes \b {\bf N}_J$ as well, while the potential $J_{\Lambda_{IJ}}$ transforms in the conjugate representation $\b {\bf N}_I\otimes {\bf N}_J$. In other words, $E_{\Lambda_{IJ}}$ is given by a direct sum of oriented paths $p$ (counted with complex coefficients) from $e_I$ to $e_J$ in the quiver, travelled along chiral multiplet arrows, and $J_{\Lambda_{IJ}}$ is similarly a direct sum of oriented paths $\t p$ from $e_J$ to $e_I$:
\bea\label{E and J superpot quiver}
&E_{\Lambda_{IJ}}(X) = \sum_{{\rm paths}\; p} c_{p}^{IJ}    X_{I K_1} X_{K_1K_2} \cdots X_{K_{k-1} J}~,\cr
&J_{\Lambda_{IJ}}(X) = \sum_{{\rm paths}\; \t p} \t c_{\t p}^{IJ}    X_{J L_1} X_{L_1L_2} \cdots X_{L_{\t k-1} I}~,\cr
\eea
where the sum is over all possible paths $p$ and $\t p$ of lengths $k$ and $\t k$, respectively.
The numerical coefficients $c_p^{IJ}$, $c_{\t p}^{IJ}$ are to be given (up to field redefinitions) as part of the definition of the $\CN=(0,2)$ supersymmetric quiver. They must be such that the supersymmetry constraint \eqref{EJ constraint} holds. This means that, for any {\it closed loop} $P$ for chiral multiplets in the quiver, we must have:
\be\label{susy constraint on c ct}
\sum_{\substack{p, \t p \\ p+ \t p=P}} c_p\, \t c_{\t p} =0~, \qquad \forall P~,
\ee
where the sum is over all pairs of quiver paths $p: e_I \rightarrow \cdots \rightarrow e_J$ and  $\t p : e_J \rightarrow \cdots \rightarrow e_I$ based at fermi multiplets $\Lambda_{IJ}$ such that the closed path $p+ \t p$ coincides with $P$.

\subsubsection{From B-branes to quiver}
Consider a D1-brane probing a local Calabi-Yau fourfold singularity ${\bf X}_4$. Away from the singularity, the D1-brane is described in the $B$-brane category as a skyscraper sheave $\CO_p$ at a point $p \in {\bf X}_4$. At the singularity, we expect that the D1-brane fractionates into a finite number $n$ of mutually-stable components:
\be
\CO_p \cong \CE_1 \oplus \cdots \oplus \CE_n~.
\ee
The {\it fractional branes} $\CE_I$, with $I=1, \cdots, n$, are a distinguished set in the derived category of coherent sheaves on the local CY fourfold.  If we normalize the central charge of the D1-brane to $Z(\CO_p)=1$, the fractional branes must be such that their central charge align at a special small-volume point---a ``quiver point''---in the quantum K{\"a}hler moduli space of ${\bf X}_4$, with $Z(\CE_I)\in \R_{>0}$ and $\sum_I Z(\CE_I)=1$. In the case of an orbifold of flat space, ${\bf X}_4 \cong \C^4/\Gamma$, the ``quiver point'' is the orbifold point, where perturbative string theory is valid, and the fractional branes are in one-to-one correspondence with the irreducible representations of $\Gamma$ \cite{Douglas:1996sw}.
We will not study stability issues at all in this work. We will only {\it assume} that we may identify (or guess) a suitable set of fractional branes. In general, there might be many allowable sets of fractional branes, some of which give the same quiver, and some of which give different quivers. This last possibility should correspond to field theory dualities. We will comment on this point in section \ref{sec: duality}.

Given the fractional branes:
\be
\CE_I \in  {\bf D}^b ({\bf X}_4)~,
\ee
as objects in the $B$-brane category, we may compute the morphisms between them. For $\CE_I$ and $\CE_J$ given as coherent sheaves on ${\bf X}_4$, the morphisms are elements of the  $\Ext$ groups:
\be
\Ext^{i}_{{\bf X}_4}(\CE_I, \CE_J)~, \qquad \qquad i=0, 1, 2, 3, 4~.
\ee
These groups encode massless open strings stretched between fractional branes \cite{Katz:2002gh}.  
We should have:
\be
\Ext^{0}_{{\bf X}_4}(\CE_I, \CE_J) = {\rm Hom}(\CE_I, \CE_J) = \delta_{IJ} \, \C~,
\ee
to obtain a physical quiver. This is because $\Ext^0$ is identified with the massless gauge field in the open string spectrum.  In our setup, we identify $\Ext^0(\CE_I, \CE_J)$ with the $\CN=(0,2)$ vector multiplet at the node $e_I$ of the quiver.

The degree-one $\Ext$ groups are identified with the chiral multiplets in the supersymmetric quiver:
\be
\Ext^1_{X_4}(\CE_J, \CE_I) \qquad \Leftrightarrow \qquad e_I \longrightarrow e_J   \qquad \Leftrightarrow \qquad X_{IJ}~,
\ee
By Serre duality, we have $\Ext^3_{X_4}(\CE_J, \CE_I) \cong \Ext^1_{X_4}(\CE_I, \CE_J)$, so that $\Ext^3_{X_4}(\CE_I, \CE_J) $ is identified with the anti-chiral multiplets $\b X_{IJ}$.
This identification of chiral multiplets with $\Ext^1$ is well-known in the case of four-dimensional $\CN=1$ quivers associated to D3-branes on a CY threefold \cite{Douglas:2000gi, Berenstein:2002fi, Herzog:2003zc, Aspinwall:2004jr}.   The new ingredient on a CY {\it fourfold} is that we also have independent {\it degree-two} $\Ext$ groups, with:
\be\label{ext2 self dual}
\Ext^2_{{\bf X}_4}(\CE_J, \CE_I) \cong \Ext^2_{{\bf X}_4}(\CE_I, \CE_J)
\ee
by Serre duality on ${\bf X}_4$. It is natural to identify these groups with the fermi multiplets $\Lambda_{IJ}$ in the $\CN=(0,2)$ quiver:
\be
\Ext^2_{{\bf X}_4}(\CE_J, \CE_I) \qquad \Leftrightarrow \qquad e_I \, \small{\text{- - -}}\; e_J   \qquad \Leftrightarrow \qquad \Lambda_{IJ}~.
\ee
The self-duality relation \eqref{ext2 self dual} for $\Ext^2$ correspond to the fact that fermi and anti-fermi multiplet are indistinguishable. For each pair of distinct nodes $I$, $J$, we may pick the basis of the $\Ext^2$ vector spaces:
\be\label{def alpha beta Fermi}
\{\alpha, \b \beta\} \in \Ext^2_{{\bf X}_4}(\CE_J, \CE_I)~, \qquad \qquad 
\{\beta, \b \alpha\} \in \Ext^2_{{\bf X}_4}(\CE_I, \CE_J)~,
\ee
where $\alpha$ and $\beta$ correspond to fermi multiplets $\Lambda_{IJ}$ and $\Lambda_{JI}$, while $\b\alpha$ and $\b \beta$ correspond to anti-fermi multiplets $\b \Lambda_{IJ}$ and $\b \Lambda_{JI}$, respectively, and such that Serre duality exchanges $\alpha$ with $\b \alpha$, and $\beta$ with $\b \beta$. This choice of basis is completely convention-dependent, however. This corresponds exactly to the freedom \eqref{fermi antiferm} of labelling fermi and anti-fermi multiplets in the supersymmetric field theory.

For  $I =J$,  $\Ext^2(\CE_I, \CE_I)$ is self-dual, and each pair of Serre-dual elements correspond to a pair of fermi and anti-fermi multiplets $\Lambda_{II}$, $\b \Lambda_{II}$ in the adjoint representation of $U(N_I)$.

As a simple consistency check of these identifications between $\Ext$ groups and $\CN=(0,2)$ superfields, it is interesting to look at the product variety ${\bf X_4}\cong {\bf X}_3 \times \C$, with ${\bf X}_3$ a CY threefold singularity. This non-isolated singularity preserves $\CN=(2,2)$ supersymmetry in two-dimension, and the 2d quiver should simply be the dimensional reduction of the  $\CN=1$ supersymmetric quiver for D3-branes on ${\bf X}_3$.  Each 4d $\CN=1$ vector multiplet decomposes into one $\CN=(0,2)$ vector multiplet and one adjoint fermi multiplet, and each 4d $\CN=1$ chiral multiplet decomposes into one $\CN=(0,2)$ chiral multiplet and one fermi multiplet. In terms of $\Ext$ groups, this means that we should have:
\bea\label{3 to 4 ext}
&\Ext^0_{{\bf X}_4}(\CE_I, \CE_J)  & \cong &\;  \Ext^0_{{\bf X}_3}(\CE_I, \CE_J)~, \cr
&\Ext^1_{{\bf X}_4}(\CE_I, \CE_J)  & \cong &\;  \Ext^0_{{\bf X}_3}(\CE_I, \CE_J)  \oplus   \Ext^1_{{\bf X}_3}(\CE_I, \CE_J)~,\cr
&\Ext^2_{{\bf X}_4}(\CE_I, \CE_J)  & \cong &\;  \Ext^1_{{\bf X}_3}(\CE_I, \CE_J)  \oplus   \Ext^2_{{\bf X}_3}(\CE_I, \CE_J)~,\cr
&\Ext^3_{{\bf X}_4}(\CE_I, \CE_J)  & \cong &\;  \Ext^2_{{\bf X}_3}(\CE_I, \CE_J)  \oplus   \Ext^3_{{\bf X}_3}(\CE_I, \CE_J)~,\cr
&\Ext^4_{{\bf X}_4}(\CE_I, \CE_J)  & \cong &\;  \Ext^3_{{\bf X}_3}(\CE_I, \CE_J)~.
\eea
This can be shown to be the case in general orbifolds $\C^3/\Gamma \times \C$---see Appendix \ref{app:dim red}.  Note that \eqref{3 to 4 ext} is consistent with Serre duality \eqref{Serre dual intro}. One can similarly consider the decomposition   ${\bf X_4}\cong {\bf X}_2 \times \C^2$, which preserves 2d $\CN=(4,4)$ supersymmetry.

\paragraph{A comment on conventions.} To avoid any possible confusion, let us note that we are using the physicist notation for the chiral multiplets in the $\CN=(0,2)$ superpotentials, and the mathematical notation of composition when discussing elements of $\Ext^\bullet$.  For instance, we have:
\be\label{exp xy comp}
x\in \Ext^1(\CE_2, \CE_1)~, \qquad y  \in \Ext^1(\CE_3, \CE_2)~,\qquad x\cdot y \in \Ext^1(\CE_3, \CE_1)~,
\ee
where $x\cdot y \equiv x \circ y$.
When talking about the fractional branes, we write these maps as:
\be
\CE_3 \xrightarrow[]{y} \CE_2 \xrightarrow[]{x} \CE_1~.
\ee
On the other hand, we have chosen the convention that $\Ext(\CE_J, \CE_I)$ corresponds to the chiral multiplet $X_{IJ}$, so that the direction of the arrows in the quiver are flipped: a map $\CE_J \rightarrow \CE_I$ corresponds to a quiver arrow $e_I \rightarrow e_J$. In our example \eqref{exp xy comp}, denoting by $X$ and $Y$ the chiral multiplets associated to the $\Ext$ group elements, we have:
\be
e_1  \xrightarrow[]{X} e_2 \xrightarrow[]{Y} e_3\qquad \cong\qquad U(N_1)  \xrightarrow[]{X} U(N_2) \xrightarrow[]{Y} U(N_3)~,
\ee
where on the right-hand-side we associated a gauge group $U(N_I)$ to each node $e_I$. In these conventions, we can write $x\cdot y$ as the matrix product $XY$ for the chiral multiplets.

\paragraph{Anomaly-free condition and quiver ranks.}  Consider an $\CN=(0,2)$ quiver with nodes $\{e_I\}$ and gauge group \eqref{quiver G}. For each $U(N_I)$ factor, the cancellation of the non-abelian anomaly requires:
\be
\sum_{J\neq I}\left( d_{IJ}^1 + d_{IJ}^3 - d_{IJ}^2\right) N_J + 2 N_I \left(-1 + d_{II}^1 -\half d_{II}^2\right)=0~.
\ee
Here the first sum is over the chiral and fermi multiplets in bifundamental representations, while the second term denote the contribution from the vector multiplet (with $d_{II}^0=1$) and from adjoint matter. Using Serre duality, this can be written as:
\be\label{AF cond 2d}
\sum_J \sum_{i=0}^4 (-1)^i  N_I \dim \Ext^i_{X_4}(\CE_I, \CE_J) = 0~.
\ee
This condition imposes constraint on the allowed ranks $N_I$ in the quiver. If we consider a single D1-brane, the ranks $N_I$ should be fixed from first principle; however, the explicit dictionary between brane-charge basis and quiver-rank basis is not always known. The anomaly-free condition then provides a strong constraint. The solutions to \eqref{AF cond 2d}, as a linear system for the positive integers $N_I$, correspond to all stable D-brane configurations at the singularity. In particular, the unique solution $\{N_I\}$ such that each $N_I$ is the smallest possible positive integer is expected to correspond to a single D1-brane. In the special case of toric Calabi-Yau singularities, we know from \cite{Franco:2015tna, Franco:2016nwv, Franco:2016qxh} that there exists ``toric quiver'' with equal ranks, $N_I= N$, corresponding to $N$ D1-branes. 

We should also mention that the {\it abelian} quadratic anomalies, from the $U(1)_I$ factors in $U(N_I)$, do not vanish in general. Instead, they should be cancelled by closed string contributions {\it \`a la} Green-Schwarz \cite{Douglas:1996sw, Mohri:1997ef, Ibanez:1998qp}.

\subsubsection{$A_{\infty}$ structure and $\CN=(0,2)$ superpotential}\label{subsec: Ainf and EJ}
To complete the determination of the $\CN=(0,2)$ supersymmetric quiver from the fractional branes on ${\bf X}_4$, we need to discuss the $E$- and $J$-terms  \eqref{E and J superpot quiver}.
It is convenient to package them into a gauge-invariant 
``$(0,2)$ superpotential''  $\CW$ defined as:~\footnote{
This expression is only formal. The $\CN=(0,2)$ superpotential that appears in the gauge theory Lagrangian is the usual $ \Tr\big( \Lambda_I  J^I(X))$, since superspace treats $\Lambda$ and $\b\Lambda$ asymmetrically. 
This formal $\CW$ first appeared in \cite{Franco:2017lpa}.
It elegantly  encodes the algebraic structure of the  $\CN=(0,2)$ quiver relations. This point is further discussed in Appendix~\ref{app: Ainfinity}.}
\be\label{def W 02 text}
\CW= \Tr\big(\Lambda_I J^I(X) + \b\Lambda^I E_I(X)\big)~.
\ee
Here, the index $I$ runs over all the fermi multiplets.
This $\CW$ can be computed by following the methods of \cite{Aspinwall:2004bs}, which studied 4d $\CN=1$ quiver theories on D3-branes at CY$_3$ singularities.
On general ground, the superpotential coupling constants are encoded in open string correlation functions. Those can be described in the language of $A_\infty$ algebra---see {\it e.g.} \cite{Herbst:2004jp} and references therein.

An $A_\infty$ algebra is a (graded) algebra $A$ together with a set of multiplications $m_k: A^{\otimes k}\rightarrow A$ that satisfy the $A_\infty$ relations:
\be
\sum_{r+s+t=n} (-1)^{r + st} m_{n+1-s}(a_1, \cdots, a_r,  m_s(a_{r+1}, \cdots, a_{r+s}), a_{r+s+1}, \cdots, a_n)=0~, 
\ee
for all integer $n>0$. The first relation states that $(m_1)^2 = 0$, so one can think of $m_1: A\rightarrow A$ as a
differential. The $\Ext$ group elements between $B$-branes, on the other hand, generate a {\it minimal} $A_\infty$ algebra, for which $m_1=0$.

To compute the multi-products on the $\Ext^\bullet$ algebra, we proceed as follows. 
Given an $A_{\infty}$ algebra $\t A$, one defines $H^{\bullet}(\t A)$ to be
the cohomology of $m_1$.  If $\t A$ has no multiplications beyond $m_2$,
then it has been shown \cite{kad} that one can define an $A_{\infty}$ structure
on $H^{\bullet}(\t A)$ in such a way that there is an $A_{\infty}$ map:~\footnote{That is, a family of maps satisfying certain consistency conditions \cite{kad}.}
\be
f: H^{\bullet}(\t A) \rightarrow \t A~,
\ee 
with $f_1$ equal to a particular representation $H^{\bullet}(\t A) \hookrightarrow \t A$ in which cohomology classes map to (noncanonical) representatives in $\t A$, and such that $m_1=0$ in the
$A_{\infty}$ algebra on $H^{\bullet}(\t A)$. One can then use the 
consistency conditions satisfied by elements of an $A_{\infty}$ map
to solve algebraically for $f_1 \circ m_k$.

In terms of B-branes, the algebra $\t A$ is the algebra of complexes of coherent sheaves, with chain maps between complexes. In that construction,  $m_1$ is essentially the BRST charge $Q$ of the $B$-model. The ``physical'' open string states then live in the cohomology $H^\bullet(\t A)$, which gives us the derived category ${\bf D}^b({\bf X})$---we refer to \cite{Aspinwall:2004jr} for a pedagogical discussion. We can identify the minimal $A_\infty$ algebra $A\equiv H^\bullet(\t A)$ with the $\Ext$ algebra  we are interested in.

Practically, in the examples discussed in this paper, each $B$-brane will be a single coherent sheaf, which can be represented in the derived category by a locally-free resolution. The $\Ext$ elements can then be represented by chain maps between resolutions, modulo chain homotopies. The $m_2$ products in $A$ are given by  chain map composition. The higher products can then be computed by the procedure just described.

We elaborate on this procedure in  Appendix~\ref{app: mk algo}, and we illustrate the computation of the higher products, in a specific example, in Appendix \ref{app: higher prod P1P1}. All of the other examples below will actually have $m_k=0$ for $k>2$.

\paragraph{Open string correlators and $A_\infty$ products.}
Let $A$ denote the $\Ext$ algebra associated to a local Calabi-Yau $n$-fold.  There exists a natural ``trace map'' of degree $-n$, which we denote by $\gamma: A\rightarrow\C$.
 Note that $A$ is a graded algebra, with $a$ of degree $q$ if $a\in \Ext^q$. Serre duality defines a natural pairing of degree $-n$:
\be\label{pairing def}
\langle a, b\rangle \equiv   \gamma\left(m_2(a, b)\right)~.
\ee
Consider a correlation function of $r$ boundary vertex operators $a_i \in A$ on the open-string worldsheet. In the $A_\infty$ language, this can be written as:
\be
\langle a_1 \cdots a_r \rangle \: = \:
\langle a_1, m_{r-1}(a_2, \cdots, a_r ) \rangle~,
\ee
in terms of the higher-product $m_{r-1}$ and the pairing \eqref{pairing def} \cite{Aspinwall:2004bs}.
Each $\Ext$ elements $x \in A$ is {\it dual} to a ``field'' $X$ in the supersymmetric quiver---see  Appendix~\ref{app: Ainfinity} for further details.
In the case of a 2d $\CN=(0,2)$ quiver describing B-branes on a CY$_4$ geometry, we have the $\Ext$ algebra:
\be
A\cong \Ext^0 \oplus \Ext^1 \oplus \Ext^2 \oplus \Ext^3 \oplus \Ext^4~,
\ee
where the summands denote all $\Ext$ groups between the various fractional branes, of degree $0, \cdots, 4$. Let us denote by $x\in A$ the $\Ext^1$ elements corresponding to the chiral multiplets $X$, and by  $\alpha, \t \alpha \in A$  the $\Ext^2$ elements corresponding to the fermi and anti-fermi multiplets $\Lambda, \b\Lambda$, as in \eqref{def alpha beta Fermi}.
The coupling constants $c_J$ and $c_E$ appearing as:
\be
c_J \Tr(\Lambda X_1 \cdots X_r) +c_E \Tr(\b\Lambda X'_1 \cdots X'_{r'})
\ee
in the superpotential \eqref{def W 02 text} can be computed as the open-string correlators:
\be
c_J = \langle \alpha \, x_1 \cdots x_r \rangle~, \qquad 
c_E = \langle \t\alpha \, x'_1 \cdots x'_{r'} \rangle~.
\ee

\paragraph{Explicit formula for the $E$- and $J$-terms.}
We can now spell out the precise formula for the coupling constants appearing in \eqref{E and J superpot quiver}.
Consider a fermi multiplet $\Lambda_{IJ}$ corresponding to $\alpha\in \Ext^2(\CE_J, \CE_I)$, and the charge-conjugate anti-fermi multiplet $\b \Lambda_{IJ}$ corresponding to $\b\alpha \in \Ext^2(\CE_I, \CE_J)$.
For each path $p$ as in \eqref{E and J superpot quiver}, 
we have the elements $x\in \Ext^1$ corresponding to the 
chiral multiplets $X$. We thus have:
\be\label{cp def}
c_p^{IJ} = \big\langle \b \alpha\, \, x_{I K_1} \cdots  \,  x_{K_{k-1} J}\big\rangle =  \gamma\big(m_2(\b\alpha, \, m_k(x_{I K_1},\, \cdots, \, x_{K_{k-1} K})) \big)~,
\ee
for the $E$-term coefficients, 
and
\be\label{cpt def}
c_{\t p}^{IJ} = \big\langle   \alpha\, x_{J L_1}\, \cdots  \,  x_{L_{\t k-1}I} \big\rangle =  \gamma\big(m_2(\alpha, \, m_{\t k}(x_{J L_1} ,\, \cdots , \, x_{L_{\t k-1}I} )) \big)~,
\ee
for the $J$-term coefficients. 
We can check this identification for a number of geometries  previously studied  by independent techniques, and we find perfect agreement.

Last but not least, we should note that, according to the 
dictionary \eqref{cp def}-\eqref{cpt def}, 
the $\Tr(EJ)=0$  constraint \eqref{susy constraint on c ct} translates into a 
very non-trivial relation amongst products of open string correlators. 
In Appendix~\ref{app: Ainfinity}, we give a general argument for why this
constraint will hold for $E$ and $J$ defined by the $A_{\infty}$ algebra 
as above.
In addition, we will check, in every example below, that the condition $\Tr(EJ)=0$ indeed holds, thus providing an additional consistency check on our computations. 
It would be interesting to also understand the first-principle origin of this constraint in the Calabi-Yau fourfold geometry.

\subsection{D1-brane on $\C^4$}\label{exp: C4}
To illustrate our methods, we start by considering the simplest case, ${\bf X}_4= \C^4$. In flat space, there is a single ``fractional brane'', the skyscraper sheaf $\CO_p$, which corresponds to a single transverse D1-brane. Consider $\CO_p$ at the origin of $\C^4$, without loss of generality. 
One can show that:
\be\label{ext0 C4}
\Ext^0(\CO_p, \CO_p) \cong\Ext^4(\CO_p, \CO_p) \cong \C~,
\ee
and
\be\label{ext12 C4}
\Ext^1(\CO_p, \CO_p) \cong\Ext^3(\CO_p, \CO_p) \cong \C^4~, \qquad
\Ext^2(\CO_p, \CO_p) \cong  \C^6~.
\ee
From this result, we directly read off the $\CN=(0,2)$ supermultiplet content according to the general rules. We have a single vector multiplet, $4$ chiral multiplets and $3$  fermi multiplets. If there are $N$ fractional branes at a point, all these fields are in the adjoint of  a $U(N)$ gauge group. This reproduces the field content of maximally supersymmetric  $\CN=(8,8)$ Yang-Mills theory in 2d, as expected. To compute the interaction terms, we will need to describe the $\Ext$ algebra more explicitly. 

\subsubsection{An explicit basis for $\Ext^\bullet(\CO_p, \CO_p)$}
The $\Ext$ algebra can be computed from the Koszul resolution of $\CO_p$, which reads:
\begin{equation}\label{koszul4}
0 \xrightarrow[]{} \mathcal{O} \xrightarrow[]{D} \mathcal{O}^4
\xrightarrow[]{C} \mathcal{O}^6 \xrightarrow[]{B} \mathcal{O}^4
\xrightarrow[]{A} \mathcal{O} \xrightarrow[]{} \mathcal{O}_p
\xrightarrow[]{} 0~,
\end{equation}
where:~\footnote{Here and in the following, we denote a map $M: \C^n \rightarrow \C^m$ by an $m\times n$ matrix, so that composition of maps corresponds to matrix multiplication (for instance, $A\circ B= AB$).}{\small
\[
A = \left(\begin{array}{cccc} x & y & z & w
\end{array} \right)~,
\]
\[
B = \left(\begin{array}{cccccc}
y & z & w & 0 & 0 & 0 \\
-x & 0 & 0 & z & w & 0 \\
0 & -x & 0 & -y & 0 & w \\
0 & 0 & -x & 0 & -y & -z
\end{array} \right)~, \qquad
C = \left(\begin{array}{cccc}
z & w & 0 & 0 \\
-y & 0 & w & 0 \\
0 & -y & -z & 0 \\
x & 0 & 0 & w \\
0 & x & 0 & -z \\
0 & 0 & x & y
\end{array} \right)~,\qquad
D = \left(\begin{array}{c}
w \\ -z \\ y \\ -x
\end{array} \right)~.
\]}
\vskip0.2cm
\noindent Let us present explicit expressions for the generators of $\Ext^\bullet$. 
We will use the notation:
\be\label{def X Ext}
X^i_j \in \Ext^i(\CO_p, \CO_p)~, \qquad j=1, \cdots, \dim \Ext^i(\CO_p, \CO_p)~.
\ee
Every $\Ext$ element can be represented by a chain map between two copies of the Koszul resolution; the actual $\Ext$ element is given by the corresponding element in its cohomology, by the definition of $\Ext$ as a derived functor.
First of all, $ \mathrm{Ext}^0 (\mathcal{O}_p, \mathcal{O}_p) $ is spanned by the single element:
\[
\begin{CD}
\mathcal{O} @>{D}>> \mathcal{O}^4 @>{C}>> \mathcal{O}^6 @ >{B}>> \mathcal{O}^4 @ >{A}>> \mathcal{O} \\
@V1VV @V1VV @V1VV @V1VV @V1VV\\
\mathcal{O} @>{D}>> \mathcal{O}^4 @>{C}>> \mathcal{O}^6 @ >{B}>> \mathcal{O}^4 @ >{A}>> \mathcal{O}
\end{CD}
\]
\vskip0.2cm
\noindent Secondly, $\mathrm{Ext}^1 (\mathcal{O}_p , \mathcal{O}_p) $ is spanned by maps of the form:
\[
\begin{CD}
@. \mathcal{O} @>{D}>> \mathcal{O}^4 @>{C}>> \mathcal{O}^6 @ >{B}>> \mathcal{O}^4 @ >{A}>> \mathcal{O} \\
@. @V{\alpha}VV
@V{\beta}VV
@V{\gamma}VV
@V{\sigma}VV
@.\\
\mathcal{O} @>{D}>> \mathcal{O}^4 @>{C}>> \mathcal{O}^6 @ >{B}>> \mathcal{O}^4 @ >{A}>> \mathcal{O} @.
\end{CD}
\]
\vskip0.2cm
\noindent
A basis can be obtained by taking
{\small \[
\alpha = \left(\begin{array}{c}
1 \\ 0 \\ 0 \\ 0
\end{array} \right)~,\quad
\left(\begin{array}{c}
0 \\ 1 \\ 0 \\ 0
\end{array} \right)~,\quad
\left(\begin{array}{c}
0 \\ 0 \\ 1 \\ 0
\end{array} \right)~,\quad
\left(\begin{array}{c}
0 \\ 0 \\ 0 \\ 1
\end{array} \right)
\]}
\vskip0.2cm
\noindent
and demanding the diagram be anti-commutative. 
For example, when 
$\alpha = \left(
1,0,0,0 \right)^t $, we can take
{\small \[
\beta = \left(\begin{array}{cccc}
0 & 1 & 0 & 0 \\
0 & 0 & 1 & 0 \\
0 & 0 & 0 & 0 \\
0 & 0 & 0 & 1 \\
0 & 0 & 0 & 0 \\
0 & 0 & 0 & 0
\end{array} \right)~,\quad
\gamma = \left(\begin{array}{cccccc}
0 & 0 & 1 & 0 & 0 & 0\\
0 & 0 & 0 & 0 & 1 & 0\\
0 & 0 & 0 & 0 & 0 & 1\\
0 & 0 & 0 & 0 & 0 & 0\\
\end{array} \right)~,\quad
\sigma = \left( \begin{array}{cccc}
0 & 0 & 0 & 1
\end{array} \right)
\]}\vskip0.2cm
\noindent
Similarly,  $\mathrm{Ext}^2 (\mathcal{O}_p , \mathcal{O}_p) $ is spanned by maps of the form:
 \[
\begin{CD}
@. @. \mathcal{O} @>{D}>> \mathcal{O}^4 @>{C}>> \mathcal{O}^6 @ >{B}>> \mathcal{O}^4 @ >{A}>> \mathcal{O} \\
@. @.
@V{\varphi}VV
@V{\psi}VV
@V{\omega}VV
@. @.\\
\mathcal{O} @>{D}>> \mathcal{O}^4 @>{C}>> \mathcal{O}^6 @ >{B}>> \mathcal{O}^4 @ >{A}>> \mathcal{O} @.
\end{CD}
\]\vskip0.2cm
\noindent
As before, we can choose $ \varphi $ to be one of the unit column vectors with six entries, and then make
the diagram commutative.
For example,
{\small \[
\varphi = \left(\begin{array}{c}
0 \\ 0 \\ 1 \\ 0 \\ 0 \\ 0
\end{array} \right)~,\qquad
\psi = \left(\begin{array}{cccc}
1 & 0 & 0 & 0 \\
0 & 0 & 0 & 0 \\
0 & 0 & 0 & 0 \\
0 & 0 & 0 & 1 \\
\end{array} \right)~,\qquad
\omega = \left( \begin{array}{cccccc}
0 & 0 & 0 & 1 & 0 & 0
\end{array} \right)~.
\]}\vskip0.2cm
\noindent
$ \mathrm{Ext}^3 (\mathcal{O}_p , \mathcal{O}_p) $ is spanned by maps of the form:
{\small \[
\begin{CD}
@. @. @. \mathcal{O} @>{D}>> \mathcal{O}^4 @>{C}>> \mathcal{O}^6 @ >{B}>> \mathcal{O}^4 @ >{A}>> \mathcal{O} \\
@. @. @.
@V{\rho}VV
@V{\tau}VV
@. @. @.\\
\mathcal{O} @>{D}>> \mathcal{O}^4 @>{C}>> \mathcal{O}^6 @ >{B}>> \mathcal{O}^4 @ >{A}>> \mathcal{O} @. @. @.
\end{CD}
\]}\vskip0.2cm
\noindent
Here $\rho$ is one of the unit vectors with four entries and $\tau$ is such that $A \cdot \rho = - \tau \cdot D$. For example, when $ \rho = (0,0,0,1)^t $, $\tau = (-1,0,0,0)$.
Finally, $ \mathrm{Ext}^4 (\mathcal{O}_p , \mathcal{O}_p) $ is spanned by:
{\small \[
\begin{CD}
@. @. @. @. \mathcal{O} @>{D}>> \mathcal{O}^4 @>{C}>> \mathcal{O}^6 @ >{B}>> \mathcal{O}^4 @ >{A}>> \mathcal{O} \\
@. @. @. @.
@V1VV
@. @. @. @.\\
\mathcal{O} @>{D}>> \mathcal{O}^4 @>{C}>> \mathcal{O}^6 @ >{B}>> \mathcal{O}^4 @ >{A}>> \mathcal{O} @. @. @. @.
\end{CD}
\]}
\vskip0.3cm

\subsubsection{Multiplication of maps}
 The multiplication rule can be determined by composing these maps. For example, $ X^1_3 \cdot X^1_2 $ is computed by:
\[
\begin{CD}
@. @. \mathcal{O} @>{D}>> \mathcal{O}^4 @>{C}>> \mathcal{O}^6 @ >{B}>> \mathcal{O}^4 @ >{A}>> \mathcal{O} \\
@. @.
@V{\alpha^1_2}VV
@V{\beta^1_2}VV
@V{\gamma^1_2}VV
@V{\sigma^1_2}VV @.\\
@. \mathcal{O} @>{D}>> \mathcal{O}^4 @>{C}>> \mathcal{O}^6 @ >{B}>> \mathcal{O}^4 @ >{A}>> \mathcal{O} @. \\
@.
@V{\alpha^1_3}VV
@V{\beta^1_3}VV
@V{\gamma^1_3}VV
@V{\sigma^1_3}VV @. @. \\
\mathcal{O} @>{D}>> \mathcal{O}^4 @>{C}>> \mathcal{O}^6 @ >{B}>> \mathcal{O}^4 @ >{A}>> \mathcal{O} @. @.
\end{CD}
\]
with:
\[
\alpha^1_2 = \left( \begin{array} {c}
0 \\ 1 \\ 0 \\ 0 \end{array} \right)~,\;
\beta^1_2 = \left(\begin{array}{cccc}
-1 & 0 & 0 & 0 \\
0 & 0 & 0 & 0 \\
0 & 0 & 1 & 0 \\
0 & 0 & 0 & 0 \\
0 & 0 & 0 & 1 \\
0 & 0 & 0 & 0
\end{array} \right)~,\;
\gamma^1_2 = \left(\begin{array}{cccccc}
0 & -1 & 0 & 0 & 0 & 0\\
0 & 0 & 0 & -1 & 0 & 0\\
0 & 0 & 0 & 0 & 0 & 0\\
0 & 0 & 0 & 0 & 0 & 1\\
\end{array} \right)~,\;
\sigma^1_2 = \left( \begin{array}{cccc}
0 & 0 & -1 & 0
\end{array} \right)~,
\]
\[
\alpha^1_3 = \left( \begin{array} {c}
0 \\ 0 \\ 1 \\ 0 \end{array} \right)~,\;
\beta^1_3 = \left(\begin{array}{cccc}
0 & 0 & 0 & 0 \\
-1 & 0 & 0 & 0 \\
0 & -1 & 0 & 0 \\
0 & 0 & 0 & 0 \\
0 & 0 & 0 & 0 \\
0 & 0 & 0 & 1
\end{array} \right)~,\;
\gamma^1_3 = \left(\begin{array}{cccccc}
1 & 0 & 0 & 0 & 0 & 0\\
0 & 0 & 0 & 0 & 0 & 0\\
0 & 0 & 0 & -1 & 0 & 0\\
0 & 0 & 0 & 0 & -1 & 0\\
\end{array} \right)~,\;
\sigma^1_3 = \left( \begin{array}{cccc}
0 & 1 & 0 & 0
\end{array} \right)~,
\]
from which we see that $\beta_3^1\cdot \alpha_2^1= - \varphi_3$, and so on and so forth, so that
$ X^1_3 \cdot X^1_2 = - X^2_3$.
Proceeding in this way, we find the multiplication rules:
\be\label{mij C4}
m_2(X^1_i, X^1_j)= X^1_i\cdot X^1_j = \mat{0 & X_1^2 & X_2^2 & X_4^2\\ 
                                        -X_1^2 & 0 & X_3^2 & X_5^2 \\
                                        	-X_2^2 & -X_3^2 &0 & X_6^2\\
					-X_4^2 &-X_5^2 &-X_6^2 &0}~.
\ee
The product $X^1_i\cdot X^1_j$ is given by the matrix element $ij$  in \eqref{mij C4}. 
One can also compute the products:
\[
X^2_6 \cdot X^2_1 = -X^2_5 \cdot X^2_2 = X^2_4 \cdot X^2_3 = X^4_1~,
\]
which commute. All other products between degree-two maps vanish.
This shows that the Serre dual of $X^2_1, X^2_2, X^3_3$ are $X^2_6, -X^2_5, X^2_4$ respectively.

One can also show that the higher products vanish in this case---that is, $m_k=0$ if $k>2$.
Therefore, any nonzero correlation function can be reduced to one of the following:
\bea\label{correlation4}
&\langle X^2_1~X^1_3~X^1_4 \rangle = 1~, ~\qquad && \langle  X^2_2~X^1_2~X^1_4 \rangle = -1~, \\
&\langle X^2_3~X^1_1~X^1_4 \rangle = 1~, ~\qquad&& \langle  X^2_4~X^1_2~X^1_3 \rangle = 1~, \\
&\langle X^2_5~X^1_1~X^1_3 \rangle = -1~, ~\qquad&& \langle  X^2_6~X^1_1~X^1_2 \rangle = 1~.
\eea

\subsubsection{The $\C^4$ quiver: $\CN=(8,8)$ SYM}\label{88SYM}
From \eqref{ext0 C4}-\eqref{ext12 C4}, we  see that the $\CN=(0,2)$ gauge theory corresponding to D1-branes on $\C^4$ has the field content of $\CN=(8,8)$ SYM. We can also verify that the product structure encoded in \eqref{correlation4} reproduces the correct supersymmetric interactions. In $\CN=(0,2)$ notation, this theory consists of four chiral multiplets, denoted $\Sigma$  and  $\Phi_a$ ($a=1,2,3$), and three fermi multiplets $\Lambda_a$ ($a=1,2,3$), with the $E$ and $J$ terms:
\be\label{E J C4}
E_a = [\Sigma, \Phi_a]~, \qquad \qquad J^a= \epsilon^{abc}\Phi_b \Phi_c~.
\ee
This is reproduced by our computation, with the identifications:
\be
\Sigma = -X_4^1~, \qquad \Phi_a= (X_3^1~,\, -X_2^1~,\, X_1^1)~,
\ee
for the chiral multiplets,  and
\be
\Lambda_a= (X_6^2~,\, -X_5^2~,\, X_4^2)~, \qquad \b\Lambda_a = (X_1^2~,\, X_2^2~,\, X_3^2)
\ee
for the fermi multiplets, as one can easily check using  \eqref{cp def}-\eqref{cpt def}, and
\begin{eqnarray*}
E_a & = & \sum_{ij} \langle \b\Lambda_a,  m_2(\phi_i, \phi_j) \rangle\;
\phi_i \phi_j, \\
J^a & = & \sum_{ij} \langle \Lambda_a, m_2(\phi_i, \phi_j) \rangle\;
\phi_i \phi_j,
\end{eqnarray*}
for $\{\phi_i\}$ the set of all chiral superfields---here, by abuse of notation, we identified the quiver fields with the corresponding $\Ext$ elements in the open-string correlators.
Note that the condition \eqref{EJ constraint} is satisfied, $\Tr(E_a J^a)=0$. The interaction terms \eqref{E J C4} display an $SU(3)$ flavory symmetry. On-shell, there is a larger $SU(4)$ flavor symmetry, with $(\Lambda_a, \b\Lambda_a)$ sitting in the ${\bf 6}$ of $SU(4)$. It will often be the case that the flavor symmetry displayed by the $\CN=(0,2)$ quiver is smaller than the symmetry expected from the CY$_4$ geometry. Those larger geometric symmetries can be thought to arise in the infrared of the gauge theory, as accidental symmetries \cite{Franco:2015tya}.

\subsection{Orbifolds $\C^4/\Gamma$}
The next simplest class of examples are supersymmetric orbifolds of flat space. Consider the CY$_4$ singularity $\C^4/\Gamma$, with $\Gamma$ a discrete subgroup of $SU(4)$. There exists one fractional brane $\CE_I$ for each irreducible representation $\rho_I$ of $\Gamma$ \cite{Katz:2002jh}.  We also denote by $\rho_I$ the trivial line bundle $\CO$ with the corresponding $\Gamma$-equivariant structure. The fractional branes are given by:
\be
\CE_I = \rho_I \otimes \CO_p~,
\ee
with $\CO_p$ the skyscraper sheaf supported at the origin.
In the following, we consider a few examples with $\Gamma$ abelian, for simplicity.

\subsubsection{${\mathbb C}^4/{\mathbb Z}_2(1,1,1,1)$ }
\begin{figure}[t]
\begin{center}
\includegraphics[width=7cm]{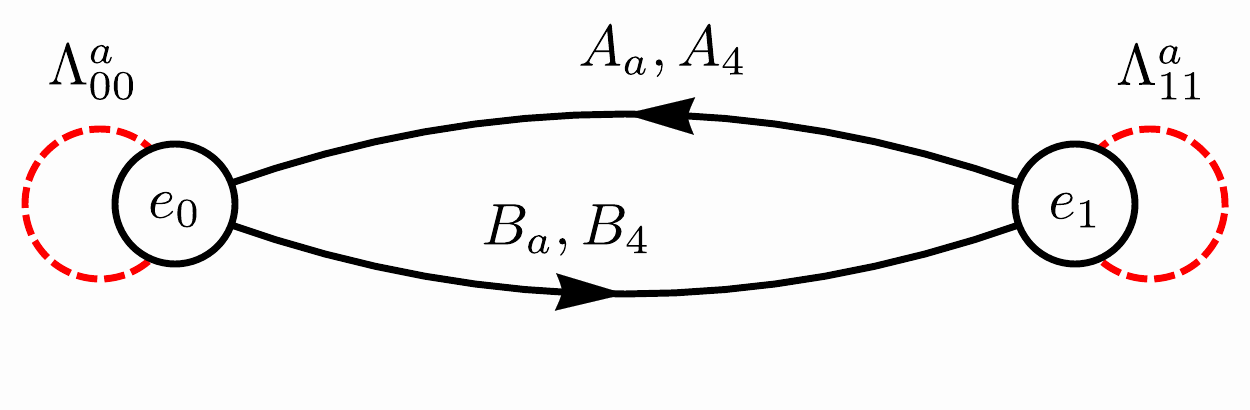}
\caption{The $\C^2/\Z_{2}(1,1,1,1)$ quiver. The index $a$ takes the values $a=1,2,3$. There are three adjoint fermi multiplets $\Lambda^a$ at each nodes, and two sets of four chiral multiplets, $A_i=(A_a, A_4)$ and $B_i=(B_a, B_4)$, in bifundamental representations. The quiver arrows have multiplicities equal to the number of distinct chiral or fermi multiplets.}
\label{fig: C4Z2}
\end{center}
\end{figure}

Consider ${\mathbb C}^4/{\mathbb Z}_2$, where the
generator of ${\mathbb Z}_2$ acts on the  $\C^4$ coordinates $(x,y,z,w)$  as:
\be
(x,y,z,w) \: \mapsto \: (-x,-y,-z,-w)~.
\ee
We have two fractional branes:
\be
\CE_0 = \rho_0 \otimes \CO_p~, \qquad  \CE_1 = \rho_1 \otimes \CO_p~.
\ee
for the trivial and non-trivial representation of $\Z_2$, respectively.
The dimensions of the $\Ext$ groups can be computed following the methods of \cite{Katz:2002jh}. 
We have:
\bea\label{Ext groups Z2}
&\Ext^0(\CE_0, \CE_0)\cong \C~, \quad  
&&\Ext^1(\CE_0, \CE_0)\cong 0~, \quad
&&\Ext^2(\CE_0, \CE_0)\cong \C^6~, \cr
&\Ext^0(\CE_1, \CE_1)\cong \C~, \quad  
&&\Ext^1(\CE_1, \CE_1)\cong 0~, \quad
&&\Ext^2(\CE_1, \CE_1)\cong \C^6~, \cr
&\Ext^0(\CE_0, \CE_1)\cong 0~, \quad  
&&\Ext^1(\CE_0, \CE_1)\cong \C^4~, \quad
&&\Ext^2(\CE_0, \CE_1)\cong 0~, \cr
&\Ext^0(\CE_1, \CE_0)\cong 0~, \quad  
&&\Ext^1(\CE_1, \CE_0)\cong \C^4~, \quad
&&\Ext^2(\CE_1, \CE_0)\cong 0~, \cr
\eea
with the higher $\Ext$ groups determined by Serre duality. 
We can also recover this spectrum from the results of section \ref{exp: C4}. Let us replace $X$ in \eqref{def X Ext} by  $a$, $b$, $c$, $d$ according to the following diagram:
\[
\xymatrix{
\mathcal{E}_0 \ar@(ul,dl)[]_c \ar@/^/[rrr]^a
&&& \mathcal{E}_1 \ar@/^/[lll]^b \ar@(dr,ur)[]_d}~, \]
which encodes all possible $\Ext$ groups. 
From the Koszul resolution \eqref{koszul4} and the fact that the maps $A, B, C, D$ are all odd under $\mathbb{Z}_2$, we see that the superscript of $a$ and $b$ can only take values $1,3$, while the superscript of $c$ and $d$ can only take values $0,2,4$, in agreement with \eqref{Ext groups Z2}.
This gives us the $\CN=(0,2)$ quiver indicated in Figure \ref{fig: C4Z2}.

The $B$-model correlation functions can be read off from \eqref{correlation4}. The $\CN=(0,2)$ superpotential immediately follows. 
Let $\Lambda^1_{00}, \Lambda^2_{00}, \Lambda^3_{00}, \Lambda^1_{11}, \Lambda^2_{11}, \Lambda^3_{11}$ denote the fermi superfields corresponding to
$c^2_4, c^2_5, c^2_6, d^2_4, d^2_5, d^2_6$, respectively. Note that they are Serre dual to
$c^2_3, -c^2_2, c^2_1, d^2_3, -d^2_2, d^2_1$.
Let us also denote the chiral superfields corresponding to $a^1_j, b^1_j$ by $A_j, B_j$. 
We then have, for instance:
\bea
& J_{\Lambda^1_{00}} =\sum_{i,j} \left\langle c_4^2 b^1_i a^1_j \right\rangle\,  B_i A_j= B_2 A_3-B_3 A_2~, \cr
& E_{\Lambda^1_{00}} =\sum_{i,j} \left\langle c_3^2 b^1_i a^1_j \right\rangle\,  B_i A_j= B_1 A_4-B_4 A_1~,
\eea
and so on and so forth. 
It is convenient to introduce the notation:
\be
\Lambda_{00}^a~, \qquad \Lambda_{11}^a~, \qquad A_i =(A_a, A_4)~, \qquad B_i =(B_a, B_4)~,
\ee
with the index $a=1,2,3$,
to emphasize an $SU(3)$ flavor symmetry. The interaction terms are given by:
\bea
&J_{\Lambda^a_{00}} = \epsilon^{abc} B_b A_c~, \qquad && J_{\Lambda^a_{11}}=  A_a B_4-A_4 B_a~,\cr
&E_{\Lambda^a_{00}} =  B_a A_4-B_4 A_a~, \qquad && E_{\Lambda^a_{11}}= \epsilon^{abc}A_b B_c~.
\eea
This satisfies $\Tr(EJ)=0$, and it is in perfect agreement with the results of \cite{Franco:2015tna}.
Note that, while the Lagrangian of the theory only has an $SU(3) \times U(1)$ global symmetry, the $E$ and $J$ terms of either node, taken together, fit into the ${\bf 6}$ of $SU(4)$, while the fields $A_i$ and $B_i$ each sit in the ${\bf 4}$ of $SU(4)$. This is the sign of an enhanced global symmetry in the infrared of the gauge theory, which can also be seen in the geometry.

\subsubsection{${\mathbb C}^4 / {\mathbb Z}_3(1,1,2,2)$}
Consider $\C^4/\Z_3$ with the orbifold action:
\be\label{orbZ311222}
(x,y,z,w) \: \mapsto \: (\omega x, \omega y, \omega^2 z, \omega^2 w)~, \qquad \omega=e^{2\pi i \ov 3}~.
\ee
As before, we denote by $\rho_i$ $(i=1,2,3)$ the trivial line bundle with
equivariant structure $i$, in conventions in which $\rho_0$ has the trivial equivariant structure, and $\rho_1^* = \rho_2$. The possible $\Ext$ groups can be organized in the following diagram:
\begin{figure}[t]
\begin{center}
\includegraphics[height=5.5cm]{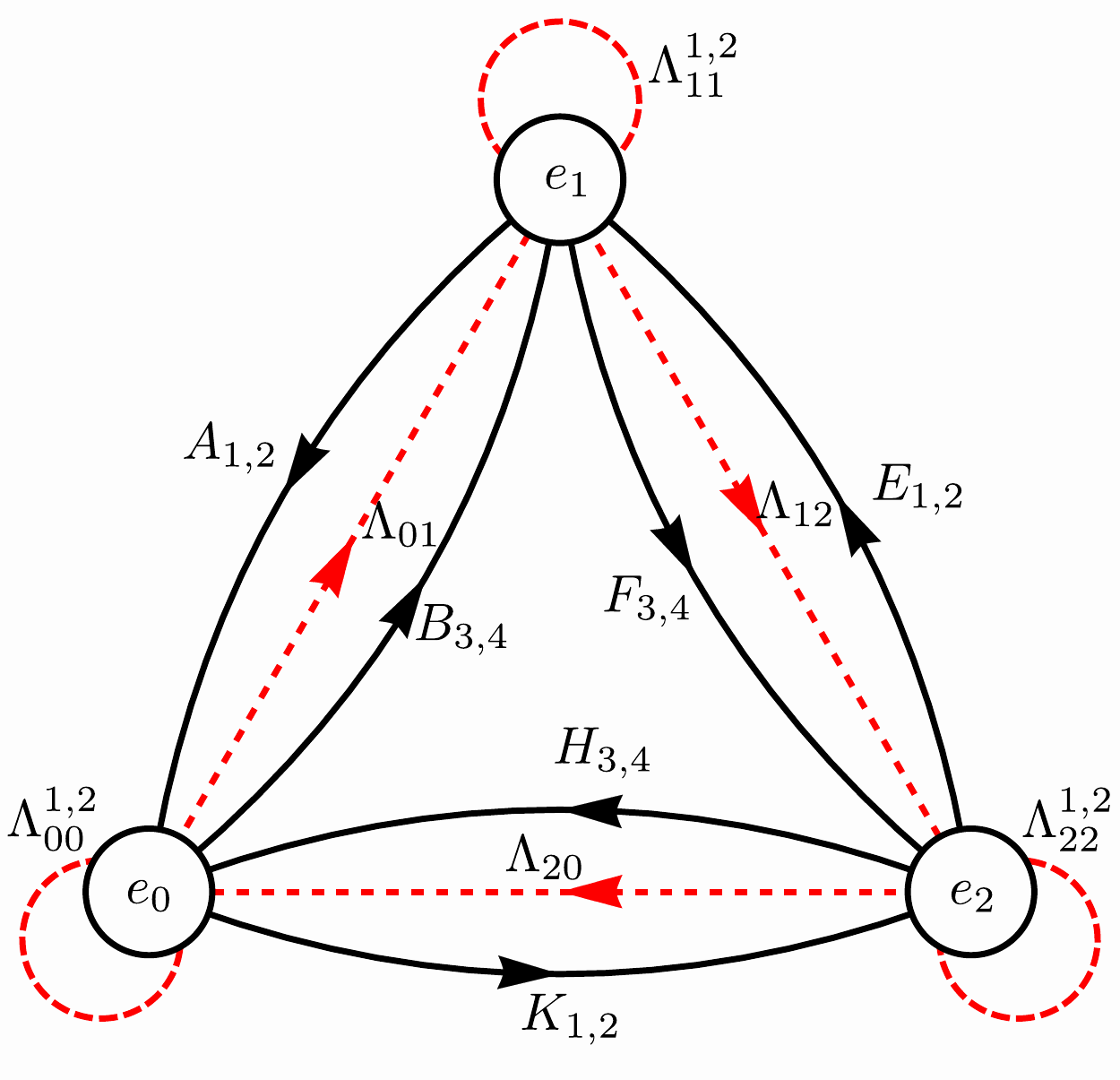}
\caption{The $\C^3/\Z_{3}(1,1,2,2)$ supersymmetric quiver, with all the chiral and fermi multiplets indicated explicitly.}
\label{fig: C4Z3}
\end{center}
\end{figure}
\[
\xymatrix{
 & & \mathcal{E}_1 \ar@(ul,ur)[]^d \ar[dldl]^b \ar@<1ex>[drdr]^e & & \\
 \\
 \mathcal{E}_0 \ar@(ul,dl)[]_c \ar@<1ex>[urur]^a \ar[rrrr]^h & & & &
 \mathcal{E}_2 \ar@(ur,dr)[]^g \ar[ulul]^f \ar@<1ex>[llll]^k
}
\]
They are of the form:
\bea
& \Ext^0(\CE_i, \CE_j)\cong \C \delta_{ij}~,\cr
& \Ext^1(\CE_i, \CE_j) \cong  \C^2 \delta_{i, j+1}\oplus  \C^2 \delta_{i, j+2}\cr
& \Ext^2(\CE_i, \CE_j) \cong   \C^4 \delta_{ij} \oplus \C \delta_{i, j+1}\oplus  \C \delta_{i, j+2}
\eea
with $i$ defined mod $3$, and the higher $\Ext$ groups determined by Serre duality. 
From the orbifold  weights \eqref{orbZ311222} on the coordinates, we can determine the weights for
the sheaves in the Koszul resolution of $\mathcal{E}_i$. The result is:
{\small\[
\left( \begin{array} {c}
i \end{array} \right) \xrightarrow[]{D} \left( \begin{array} {c}
i+2 \\ i+2 \\ i+1 \\ i+1 \end{array} \right)
\xrightarrow[]{C} \left( \begin{array} {c}
i+1 \\ i \\ i \\ i \\ i \\ i+2 \end{array} \right) \xrightarrow[]{B} \left( \begin{array} {c}
i+2 \\ i+2 \\ i+1 \\ i+1 \end{array} \right)
\xrightarrow[]{A} \left( \begin{array} {c}
i \end{array} \right)
\]}
Thus the spectrum is given explicitly by:
\bea\nn
 &X^1_1, X^1_2, X^2_6, X^3_1, X^3_2~, \qquad&  \text{for}\quad  X=a,e,k~, \cr
 & X^1_3, X^1_4, X^2_1, X^3_3, X^3_4 ~, \qquad&  \text{for}\quad X=b,h,f~, \cr
 & X^0_1, X^2_2, X^2_3, X^2_4, X^2_5, X^4_1 ~, &\qquad  \text{for}\quad X=c,d,g~.
 \eea
Let us denote by $\Lambda_{01}, \Lambda_{12}, \Lambda_{20}$ the fermi multiplets corresponding
to $b^2_1, f^2_1, h^2_1$, respectively, with the charge-conjugate fermi multiplets $\t \Lambda_{01}, \t\Lambda_{12},\t \Lambda_{20}$ corresponding to the Serre dual elements $a^2_6, e^2_6, k^2_6$;
 let us denote by $\Lambda^1_{00}, \Lambda^2_{00}, \Lambda^1_{11}, \Lambda^2_{11}, \Lambda^1_{22}, \Lambda^2_{22}$ the fermi multiplets  corresponding
to $c^2_4, c^2_5, d^2_4, d^2_5, g^2_4, g^2_5$ (dual to $c^2_3, -c^2_2, d^2_3, -d^2_2, g^2_3, -g^2_2$), respectively. We also denote by $A_i, B_j, \cdots$ the chiral multiplets associated to $a^1_i, b^1_j, \cdots$, so that we have the 12 chiral multiplets:
\be
A_1, A_2, B_3, B_4, \qquad E_1, E_2, F_3, F_4, \qquad K_1, K_2, H_3, H_4~,
\ee
in the spectrum.
The corresponding $\CN=(0,2)$ supersymmetric quiver is shown in Figure \ref{fig: C4Z3}. 
We can  directly compute the superpotential terms from \eqref{correlation4}. One finds:
\bea
& J_{\Lambda^1_{00}}= K_2 H_3- B_3 A_2~,  \;
&& J_{\Lambda^1_{11}}=A_2 B_3 - F_3 E_2~, \;
&& J_{\Lambda^1_{22}}=E_2 F_3-H_3 K_2~, 
 \cr
&J_{\Lambda^2_{00}}=  B_3 A_1 -K_1 H_3~,  \;
&&J_{\Lambda^2_{11}}=    F_3 E_1 -A_1 B_3~,\;
&& J_{\Lambda^2_{22}}=   H_3 K_1 -E_1 F_3~,\cr
&J_{\Lambda_{01}}=   F_3 H_4 - F_4 H_3~,\;
&& J_{\Lambda_{12}}=          H_3 B_4 -H_4 B_3~,\; 
&& J_{\Lambda_{20}}=  B_3 F_4 -B_4 F_3~.
\eea
and
\bea
&E_{\Lambda^1_{00}} =K_1 H_4 - B_4 A_1~,\;
&& E_{\Lambda^1_{11} }= A_1 B_4 - F_4 E_1~, \;
&& E_{ \Lambda^1_{22} }= E_1 F_4 -H_4 K_1~, \\
&E_{\Lambda^2_{00} }= K_2 H_4 - B_4 A_2~, \;
&& E_{ \Lambda^2_{11} }= A_2 B_4 - F_4 E_2~,\;
&&E_{ \Lambda^2_{22} }= E_2 F_4 -H_4 K_2~,\\
&E_{ \Lambda_{01} }= K_1 E_2 -K_2 E_1~,\;
&&E_{ \Lambda_{12} }= A_1 K_2 - A_2 K_1~,\; 
&&E_{ \Lambda_{20} }= E_1 A_2 - E_2 A_1~.
\eea
One can check that $\Tr(EJ)=0$. This again agrees with the results of \cite{Franco:2015tna}. Note that this quiver theory has only a $U(1)^3$ (toric) flavor symmetry, though there is an expected enhancement to $SU(2)^2\times U(1)$ in the infrared.

\subsubsection{$\C^4/\Z_4(1,1,1,1)$ }\label{subsec: C4Z4orb}
Consider $\C^/\Z_4$, where ${\mathbb Z}_4$ acts on the coordinates $\C^4$ coordinates as
\be\label{defC4Z4}
(x,y,z,w) \: \mapsto \: (\omega x, \omega y,\omega z, \omega w),\qquad \omega=e^{2\pi i \ov 4}~.
\ee
We have four fractional branes, $\CE_i$, $i=0,1,2,3$.
In this case, the weights for
the sheaves in the Koszul resolution of $\mathcal{E}_i$ are given as follows:
{\small\[
\left( \begin{array} {c}
i \end{array} \right) \xrightarrow[]{D} \left( \begin{array} {c}
i+1 \\ i+1 \\ i+1 \\ i+1 \end{array} \right)
\xrightarrow[]{C} \left( \begin{array} {c}
i+2 \\ i+2 \\ i+2 \\ i+2 \\ i+2 \\ i+2 \end{array} \right) \xrightarrow[]{B} \left( \begin{array} {c}
i+3 \\ i+3 \\ i+3 \\ i+3 \end{array} \right)
\xrightarrow[]{A} \left( \begin{array} {c}
i \end{array} \right)
\]}
The  $\Ext$ groups can be  summarized by the diagram:
\[
\xymatrix{
 & & \mathcal{E}_1 \ar[dldl]^3 \ar@<0.5ex>[drdr]^2 \ar@<0.5ex>[d]^1 & & \\
 & & \mathcal{E}_0 \ar@<0.5ex>[u]^{3} \ar@<0.5ex>[ldl]^2 \ar@<0.5ex>[rdr]^1 & & \\
 \mathcal{E}_2 \ar@<0.5ex>[urur]^1 \ar[rur]^2 \ar[rrrr]^3 & & & &
 \mathcal{E}_3 \ar[lul]^3 \ar[ulul]^2 \ar@<0.5ex>[llll]^1}
\]
Here, the number attached to each arrow is the degree, and
 we omitted the degree-$0$ and degree-$4$ operators from one sheaf to itself,
which also survive the orbifold projection and correspond to $\CN=(0,2)$ vector multiplets.  
We can similarly compute the interaction terms. They will be presented in section \ref{sec local P3} below, after we reconsider the same quiver in a different guise. Any other supersymmetry-preserving orbifold of $\C^4$ can be   worked out similarly.

\subsection{Fractional branes on a local $\mathbb{P}^3$}\label{sec local P3}
Another interesting class of examples are given by  Calabi-Yau fourfold singularities ${\bf X}_4$ that admit a crepant resolution:
\be
\pi: \t {\bf X}_4\rightarrow  {\bf X}_4~.
\ee
One of the simplest such singularity is the $\C^4/\Z_4$ orbifold \eqref{defC4Z4}, which admits a crepant resolution as the total space of the canonical line bundle over $\mathbb{P}^3$:
\be\label{local P3 def}
 \t {\bf X}_4 = {\rm Tot}\left(\CO(-4)\rightarrow \mathbb{P}^3\right)~. 
\ee
For a Calabi-Yau threefold the total space of the canonical line bundle over a Fano surface, nice bases of fractional branes can be found in terms of strongly exceptional collections \cite{Cachazo:2001sg, Herzog:2003zc, Aspinwall:2004vm, Hanany:2006nm, Herzog:2006bu}.  We can similarly construct a well-behaved set of sheaves on $ \t {\bf X}_4$ starting from what is known as a {\it strongly exceptional collection of sheaves} on $\mathbb{P}^3$. We will discuss this procedure in section \ref{subsec: mutations}. In the rest of this section, we will just postulate the sets of fractional branes, without further explanation.

We will discuss two distinct sets of fractional branes on \eqref{local P3 def}, which give rise to two distinct supersymmetric quivers. In section \ref{sec: duality}, we will show that those two quivers are related by a field theory infrared duality, and by a mutation of the corresponding exceptional collections. We should emphasize that these two quiver gauge theories are only two relatively simple examples among an infinite number of dual theories for D1-branes probing the same CY$_4$ geometry. 
We refer to Appendix~\ref{sect:warmup:local-p2} for a review of the  simpler case of a CY threefold.

\subsubsection{Fractional branes and $\Ext$ algebra (I)}
\label{sect:fourfold:p2:first}

Consider the following strongly exceptional collection on $\mathbb{P}^3$:
\be
\{\Omega^3(3)[3], \Omega^2(2)[2], \Omega(1)[1], \mathcal {O}\} \cong
\{\wedge^3Q^*[3],\wedge^2Q^*[2],Q^*[1],\mathcal {O} \}~.
\ee
Let $i$ denote the embedding
$i:\mathbb{P}^3\hookrightarrow  \t {\bf X}_4 $.
The four fractional branes are identified with $\CE_j= i_\ast \Omega^j(j)[j]$  ($j=0,1,2,3$), namely:
\be\label{fractional branes TheoryI}
\CE_0 =i_*\mathcal{O}~, \quad
\CE_1= i_*\Omega(1)[1]~,\quad
\CE_2=  i_*\Omega^2(2)[2]~,\quad
\CE_3=  i_*\Omega^3(3)[3]~.
\ee
One can compute the $\Ext$ groups explicitly. One finds:~\footnote{One needs to use the fact that $\mathrm{Ext}^i(\mathcal {F}[m],\mathcal {G}[n])=\mathrm{Ext}^{i-m+n}(\mathcal {F},\mathcal {G})$.}
\be
\dim \Ext^1(\CE_I, \CE_J) ={\small \mat{0& 0 &0 & 4 \\  4 & 0 & 0 & 0 \\ 0 & 4 & 0 &0 \\ 0&0&4&0}}~, \qquad \quad
\dim \Ext^2(\CE_I, \CE_J) ={\small \mat{0&0&6&0\\ 0&0&0&6\\ 6&0&0&0\\ 0&6&0&0}}~,
\ee
where $I, J=0, 1, 2, 3$.
The corresponding quiver diagram for the $\Ext$ groups reads:
\[
\xymatrix{ \Omega^3(3)[3] \ar[rr]^4_c \ar@{--}[dr]_6 & &
\Omega^2(2)[2] \ar[dd]^4_b \\
& \ar@{--}[dr]_{\lambda} \ar@{--}[ur]^{\psi} & \\
\mathcal{O} \ar[uu]^4_d \ar@{--}[ur]^6 & & \Omega(1)[1] \ar[ll]_4^a
}
\]
Here the arrows stand for elements of $\Ext^1$ and the dashed lines stand for elements of $\Ext^2$, with the multiplicities indicated. 
This coincides with the $[\C^4/\Z_4]$ orbifold quiver in section \ref{subsec: C4Z4orb}.

Interestingly, the $\Ext$ groups fill out irreducible representations of $GL(4)$, which are induced from the underlying $GL(4)$ symmetry of $\mathbb{P}^3$. The precise representations can be worked out from the Bott-Borel-Weil theorem \cite{10.2307/1969996, 10.2307/1970237}. The $\Ext^1$ elements $a$, $b$, $c$, $d$ naturally span the ${\bf 4}$ or ${\bf 4}'$ (fundamental or fully anti-symmetric representations), while the $\Ext^2$ elements $\psi$ and $\lambda$ fall into ${\bf 6}$'s (anti-symmetric representations)  of $GL(4)$.~\footnote{We refer to Appendix A of \cite{Guo:2015gha} for a short introduction to the Bott-Borell-Weyl theorem in a related context. For our purposes in this paper, we only wish to point out the geometric origin of the $SU(4)$ global symmetry in the $\CN=(0,2)$ supersymmetric quivers to be discussed below.}

\paragraph{An explicit basis for $\Ext^\bullet$.}
Let us  compute the $\Ext$ generators explicitly.~\footnote{In the following, we use the notation $X= \t {\bf X}_4$ to avoid clutter.} We take $(x_i, y_i, z_i)$ to be the coordinates on the patch $U_i$ such that the $i$-th homogeneous coordinate of $\mathbb{P}^3$ is nonzero, $i =
0,1,2,3$. We also take $w_i$ to be the coordinate of the fiber of
$\mathcal{O}(-4)$ over $U_i$. A sheaf of the form $i_* \mathcal{E}$
has a Koszul resolution:
\[
0 \xrightarrow[]{} \mathcal{E}(4) \xrightarrow[]{w_0} \mathcal{E}
\xrightarrow[]{} i_* \mathcal{E} \xrightarrow[]{} 0~.
\]
Every state in the $\Ext$ quiver can be represented by a chain map between the corresponding locally-free resolutions of sheaves, as follows:

\vskip0.2cm
\noindent $\mathrm{Ext}^1(i_*\Omega(1)[1], i_* \mathcal{O})$ is generated by
$a_i \in \check{C}^0(X, \mathit{Hom}^{1}(i_*\Omega(1)[1], i_*
\mathcal{O}))$:
{\small\[
a_1:~~~~~\;
\begin{CD}
\Omega(5) @>{}>> \Omega(1) \\
@V{(-1,0,0)}VV @V{(1,0,0)}VV \\
\mathcal{O}(4) @>{}>> \mathcal{O}
\end{CD}~~~~~~~~~~~~~
a_2:~~~~~\;
\begin{CD}
\Omega(5) @>{}>> \Omega(1) \\
@V{(0,-1,0)}VV @V{(0,1,0)}VV \\
\mathcal{O}(4) @>{}>> \mathcal{O}
\end{CD}~~~
\]
\[
a_3:~~~~~\;
\begin{CD}
\Omega(5) @>{}>> \Omega(1) \\
@V{(0,0,-1)}VV @V{(0,0,1)}VV \\
\mathcal{O}(4) @>{}>> \mathcal{O}
\end{CD}~~~~~~~~~~~~~
a_4:~~~~~\;
\begin{CD}
\Omega(5) @>{}>> \Omega(1) \\
@VV{(-x_0,-y_0,-z_0)}V @VV{(x_0,y_0,z_0)}V \\
\mathcal{O}(4) @>{}>> \mathcal{O}
\end{CD}~~~
\]}
\vskip0.2cm
\noindent
$\mathrm{Ext}^1(i_*\Omega^2(2)[2], i_*\Omega(1)[1])$ is generated by
$b_i \in \check{C}^0(X, \mathit{Hom}^{1}(i_*\Omega^2(2)[2],
i_*\Omega(1)[1]))$:
{\small\[
b_1:\quad\qquad
\begin{CD}
\Omega^2(6) @>{}>> \Omega^2(2) \\
@V{\tiny \left(\begin{array}{ccc} 0 & -1 & 0 \\ 1 & 0 & 0 \\ 0 & 0 &
0 \end{array}\right)}VV
@VV{\tiny \left(\begin{array}{ccc} 0 & 1 & 0 \\ -1 & 0 & 0 \\ 0 & 0 & 0 \end{array}\right)}V \\
\Omega(5) @>{}>> \Omega(1)
\end{CD}\qquad\qquad\quad
b_2:\qquad\quad
\begin{CD}
\Omega^2(6) @>{}>> \Omega^2(2) \\
@V{\tiny \left(\begin{array}{ccc} 0 & 0 & -1 \\ 0 & 0 & 0 \\ 1 & 0 &
0
\end{array}\right)}VV
@VV{\tiny \left(\begin{array}{ccc} 0 & 0 & 1 \\ 0 & 0 & 0 \\ -1 & 0 & 0 \end{array}\right)}V \\
\Omega(5) @>{}>> \Omega(1)
\end{CD}\qquad
\]
\[
b_3:\qquad\quad
\begin{CD}
\Omega^2(6) @>{}>> \Omega^2(2) \\
@V{\tiny \left(\begin{array}{ccc} 0 & 0 & 0 \\ 0 & 0 & -1 \\ 0 & 1 &
0
\end{array}\right)}VV
@VV{\tiny \left(\begin{array}{ccc} 0 & 0 & 0 \\ 0 & 0 & 1 \\ 0 & -1 & 0 \end{array}\right)}V \\
\Omega(5) @>{}>> \Omega(1)
\end{CD}\qquad\qquad\quad
b_4:\qquad\qquad\quad
\begin{CD}
\Omega^2(6) @>{}>> \Omega^2(2) \\
@V{\tiny \left(\begin{array}{ccc} 0 & -z_0 & y_0 \\ z_0 & 0 & -x_0 \\
-y_0 & x_0 & 0
\end{array}\right)}VV
@VV{\tiny \left(\begin{array}{ccc} 0 & z_0 & -y_0 \\ -z_0 & 0 & x_0 \\ y_0 & -x_0 & 0 \end{array}\right)}V \\
\Omega(5) @>{}>> \Omega(1)
\end{CD}\qquad
\]}
$\mathrm{Ext}^1(i_*\Omega^3(3)[3], i_*\Omega^2(2)[2])$ is generated
by $c_i \in \check{C}^0(X, \mathit{Hom}^{1}(i_*\Omega^3(3)[3],
i_*\Omega^2(2)[2]))$:
\vskip0.2cm
\noindent
{\small\[
c_1:\qquad
\begin{CD}
\mathcal{O}(3) @>{}>> \mathcal{O}(-1) \\
@V{\tiny \left(\begin{array}{c} -1 \\ 0 \\ 0
\end{array}\right)}VV
@VV{\tiny \left(\begin{array}{c} 1 \\ 0 \\ 0 \end{array}\right)}V \\
Q(3) @>{}>> Q(-1)
\end{CD}\qquad\qquad
c_2:\qquad
\begin{CD}
\mathcal{O}(3) @>{}>> \mathcal{O}(-1) \\
@V{\tiny \left(\begin{array}{c} 0 \\ -1 \\ 0
\end{array}\right)}VV
@VV{\tiny \left(\begin{array}{c} 0 \\ 1 \\ 0 \end{array}\right)}V \\
Q(3) @>{}>> Q(-1)
\end{CD}\qquad\]
\[
c_3:\qquad
\begin{CD}
\mathcal{O}(3) @>{}>> \mathcal{O}(-1) \\
@V{\tiny \left(\begin{array}{c} 0 \\ 0 \\ -1
\end{array}\right)}VV
@VV{\tiny \left(\begin{array}{c} 0 \\ 0 \\ 1 \end{array}\right)}V \\
Q(3) @>{}>> Q(-1)
\end{CD}\qquad\qquad
c_4:\qquad
\begin{CD}
\mathcal{O}(3) @>{}>> \mathcal{O}(-1) \\
@V{\tiny \left(\begin{array}{c} -x_0 \\ -y_0 \\ -z_0
\end{array}\right)}VV
@VV{\tiny \left(\begin{array}{c} x_0 \\ y_0 \\ z_0 \end{array}\right)}V \\
Q(3) @>{}>> Q(-1)
\end{CD}\qquad
\]}
\vskip0.2cm
\noindent
$\mathrm{Ext}^1(i_*\mathcal{O}, i_*\Omega^3(3)[3])$ is generated by
$d_i \in \check{C}^3(X, \mathit{Hom}^{-2}(i_*\mathcal{O},
i_*\Omega^3(3)[3]))$:
{\small
\[
d_1:\qquad
\begin{CD}
@. \mathcal{O}(4) @>{}>> \mathcal{O} \\
@. @V{x_0^{-1}y_0^{-1}z_0^{-1}}VV @.\\
\mathcal{O}(3) @>{}>> \mathcal{O}(-1) @.
\end{CD}\qquad
d_2:\qquad
\begin{CD}
@. \mathcal{O}(4) @>{}>> \mathcal{O} \\
@. @V{x_0^{-2}y_0^{-1}z_0^{-1}}VV @.\\
\mathcal{O}(3) @>{}>> \mathcal{O}(-1) @.
\end{CD}\qquad
\]
\[
d_3:\qquad
\begin{CD}
@. \mathcal{O}(4) @>{}>> \mathcal{O} \\
@. @V{x_0^{-1}y_0^{-2}z_0^{-1}}VV @.\\
\mathcal{O}(3) @>{}>> \mathcal{O}(-1) @.
\end{CD}\qquad
d_4:\qquad
\begin{CD}
@. \mathcal{O}(4) @>{}>> \mathcal{O} \\
@. @V{x_0^{-1}y_0^{-1}z_0^{-2}}VV @.\\
\mathcal{O}(3) @>{}>> \mathcal{O}(-1) @.
\end{CD}\qquad
\]}
\vskip0.2cm 
\noindent $\mathrm{Ext}^2(i_*\Omega^2(2)[2], i_*\mathcal{O})$ is generated by
$\psi_n \in \check{C}^0(X, \mathit{Hom}^{2}(i_*\Omega^2(2)[2],
i_*\mathcal{O}))$:
{\small
\[
\psi_1:~~~~~
\begin{CD}
Q(3) @>{}>> Q(-1) \\
@V{(1,0,0)}VV @V{(1,0,0)}VV \\
\mathcal{O}(4) @>{}>> \mathcal{O}
\end{CD}~~~~~~~~~~~~~
\psi_2:~~~~~
\begin{CD}
Q(3) @>{}>> Q(-1) \\
@V{(0,1,0)}VV @V{(0,1,0)}VV \\
\mathcal{O}(4) @>{}>> \mathcal{O}
\end{CD}~~~
\]
\[
\psi_3:~~~~~
\begin{CD}
Q(3) @>{}>> Q(-1) \\
@V{(0,0,1)}VV @V{(0,0,1)}VV \\
\mathcal{O}(4) @>{}>> \mathcal{O}
\end{CD}~~~~~~~~~~~~~
\psi_4:~~~~~\quad
\begin{CD}
Q(3) @>{}>> Q(-1) \\
@V{(y_0,-x_0,0)}VV @V{(y_0,-x_0,0)}VV \\
\mathcal{O}(4) @>{}>> \mathcal{O}
\end{CD}~~~
\]
\[
\psi_5:~~~~~\quad
\begin{CD}
Q(3) @>{}>> Q(-1) \\
@V{(z_0,0,-x_0)}VV @V{(z_0,0,-x_0)}VV \\
\mathcal{O}(4) @>{}>> \mathcal{O}
\end{CD}~~~~~~~~~~~~~
\psi_6:~~~~~\quad
\begin{CD}
Q(3) @>{}>> Q(-1) \\
@V{(0,z_0,-y_0)}VV @V{(0,z_0,-y_0)}VV \\
\mathcal{O}(4) @>{}>> \mathcal{O}
\end{CD}~~~
\]}
\vskip0.2cm
\noindent
We denote the Serre dual of $\psi$ by $\psi'$, with
$\psi'_n\in \check{C}^3(X, \mathit{Hom}^{-1}(i_*\mathcal{O}, i_*\Omega^2(2)[2]
))$:
{\small
\[
\psi_1':\qquad
\begin{CD}
@. \mathcal{O}(4) @>{}>> \mathcal{O} \\
@. @V{\tiny \left(\begin{array}{c} x_0^{-1}y_0^{-1}z_0^{-1} \\ 0 \\
0 \end{array}\right)}VV @.\\
Q(3) @>{}>> Q(-1) @.
\end{CD}\qquad
\psi_2':\qquad
\begin{CD}
@. \mathcal{O}(4) @>{}>> \mathcal{O} \\
@. @V{\tiny \left(\begin{array}{c} 0 \\ x_0^{-1}y_0^{-1}z_0^{-1} \\ 0 \end{array}\right)}VV @.\\
Q(3) @>{}>> Q(-1) @.
\end{CD}\qquad
\]
\[
\psi_3':\qquad
\begin{CD}
@. \mathcal{O}(4) @>{}>> \mathcal{O} \\
@. @V{\tiny \left(\begin{array}{c} 0 \\ 0 \\ x_0^{-1}y_0^{-1}z_0^{-1} \end{array}\right)}VV @.\\
Q(3) @>{}>> Q(-1) @.
\end{CD}\qquad
\psi_4':\qquad
\begin{CD}
@. \mathcal{O}(4) @>{}>> \mathcal{O} \\
@. @V{\tiny \left(\begin{array}{c}  x_0^{-1}y_0^{-2}z_0^{-1} \\ 0 \\ 0 \end{array}\right)}VV @.\\
Q(3) @>{}>> Q(-1) @.
\end{CD}\qquad
\]
\[
\psi_5':\qquad
\begin{CD}
@. \mathcal{O}(4) @>{}>> \mathcal{O} \\
@. @V{\tiny \left(\begin{array}{c} x_0^{-1}y_0^{-1}z_0^{-2} \\ 0 \\ 0 \end{array}\right)}VV @.\\
Q(3) @>{}>> Q(-1) @.
\end{CD}\qquad
\psi_6':\qquad
\begin{CD}
@. \mathcal{O}(4) @>{}>> \mathcal{O} \\
@. @V{\tiny \left(\begin{array}{c} 0 \\  x_0^{-1}y_0^{-1}z_0^{-2} \\ 0 \end{array}\right)}VV @.\\
Q(3) @>{}>> Q(-1) @.
\end{CD}\qquad
\]}
\vskip0.1cm
\noindent 
Note that, since the maps given by:
{\small\[
\left(\begin{array}{c} x_0^{-1}y_0^{-2}z_0^{-1} \\
x_0^{-2}y_0^{-1}z_0^{-1} \\ 0
\end{array}\right),
\left(\begin{array}{c} x_0^{-1}y_0^{-1}z_0^{-2} \\ 0 \\
x_0^{-2}y_0^{-1}z_0^{-1}
\end{array}\right),
\left(\begin{array}{c} 0 \\ x_0^{-1}y_0^{-1}z_0^{-2} \\
x_0^{-1}y_0^{-2}z_0^{-1}
\end{array}\right)
\]}
\vskip0.1cm
\noindent are exact, $\psi'_4, \psi'_5, \psi'_6$ can be equivalently
represented by:
{\small\[
\left(\begin{array}{c} 0 \\
-x_0^{-2}y_0^{-1}z_0^{-1} \\ 0
\end{array}\right),
\left(\begin{array}{c} 0 \\ 0 \\
-x_0^{-2}y_0^{-1}z_0^{-1}
\end{array}\right),
\left(\begin{array}{c} 0 \\ 0 \\
-x_0^{-1}y_0^{-2}z_0^{-1}
\end{array}\right)~,
\]}
\vskip-0.1cm
\noindent  respectively. 

\vskip0.2cm
\noindent $\mathrm{Ext}^2(i_*\Omega^3(3)[3],
i_*\Omega(1)[1])$ is generated by $\lambda_n \in \check{C}^0(X,
\mathit{Hom}^{2}(i_*\Omega^3(3)[3], i_*\Omega(1)[1]))$:
{\small\[
\lambda_1:\qquad
\begin{CD}
\mathcal{O}(3) @>{}>> \mathcal{O}(-1) \\
@V{\tiny \left(\begin{array}{c} 1 \\ 0 \\ 0
\end{array}\right)}VV
@VV{\tiny \left(\begin{array}{c} 1 \\ 0 \\ 0 \end{array}\right)}V \\
Q^*(4) @>{}>> Q^*
\end{CD}\qquad\qquad
\lambda_2:\qquad
\begin{CD}
\mathcal{O}(3) @>{}>> \mathcal{O}(-1) \\
@V{\tiny \left(\begin{array}{c} 0 \\ 1 \\ 0
\end{array}\right)}VV
@VV{\tiny \left(\begin{array}{c} 0 \\ 1 \\ 0 \end{array}\right)}V \\
Q^*(4) @>{}>> Q^*
\end{CD}\qquad
\]
\[
\lambda_3:\qquad
\begin{CD}
\mathcal{O}(3) @>{}>> \mathcal{O}(-1) \\
@V{\tiny \left(\begin{array}{c} 0 \\ 0 \\ 1
\end{array}\right)}VV
@VV{\tiny \left(\begin{array}{c} 0 \\ 0 \\ 1 \end{array}\right)}V \\
Q^*(4) @>{}>> Q^*
\end{CD}\qquad\qquad
\lambda_4:\qquad
\begin{CD}
\mathcal{O}(3) @>{}>> \mathcal{O}(-1) \\
@V{\tiny \left(\begin{array}{c} y_0 \\ -x_0 \\ 0
\end{array}\right)}VV
@VV{\tiny \left(\begin{array}{c} y_0 \\ -x_0 \\ 0 \end{array}\right)}V \\
Q^*(4) @>{}>> Q^*
\end{CD}\qquad
\]
\[
\lambda_5:\qquad
\begin{CD}
\mathcal{O}(3) @>{}>> \mathcal{O}(-1) \\
@V{\tiny \left(\begin{array}{c} z_0 \\ 0 \\ -x_0
\end{array}\right)}VV
@VV{\tiny \left(\begin{array}{c} z_0 \\ 0 \\ -x_0 \end{array}\right)}V \\
Q^*(4) @>{}>> Q^*
\end{CD}\qquad\qquad
\lambda_6:\qquad
\begin{CD}
\mathcal{O}(3) @>{}>> \mathcal{O}(-1) \\
@V{\tiny \left(\begin{array}{c} 0 \\ z_0 \\ -y_0
\end{array}\right)}VV
@VV{\tiny \left(\begin{array}{c} 0 \\ z_0 \\ -y_0 \end{array}\right)}V \\
Q^*(4) @>{}>> Q^*
\end{CD}\qquad
\]}
\vskip0.2cm
\noindent
We denote the Serre dual of $\lambda$ by $\lambda'$, with
$\lambda'_n\in \check{C}^3(X, \mathit{Hom}^{-1}(i_*\Omega(1)[1], i_*\Omega^3(3)[3]
))$:
{\small\[
\lambda'_1:\qquad
\begin{CD}
@. Q^*(4) @>{}>> Q^* \\
@. @V{(x_0^{-1}y_0^{-1}z_0^{-1},0,0)}VV @.\\
\mathcal{O}(3) @>{}>> \mathcal{O}(-1) @.
\end{CD}\qquad
\lambda'_2:\qquad
\begin{CD}
@. Q^*(4) @>{}>> Q^* \\
@. @V{(0,x_0^{-1}y_0^{-1}z_0^{-1},0)}VV @.\\
\mathcal{O}(3) @>{}>> \mathcal{O}(-1) @.
\end{CD}\qquad
\]
\[
\lambda'_3:\qquad
\begin{CD}
@. Q^*(4) @>{}>> Q^* \\
@. @V{(0,0,x_0^{-1}y_0^{-1}z_0^{-1})}VV @.\\
\mathcal{O}(3) @>{}>> \mathcal{O}(-1) @.
\end{CD}\qquad
\lambda'_4:\qquad
\begin{CD}
@. Q^*(4) @>{}>> Q^* \\
@. @V{(x_0^{-1}y_0^{-2}z_0^{-1},0,0)}VV @.\\
\mathcal{O}(3) @>{}>> \mathcal{O}(-1) @.
\end{CD}\qquad
\]
\[
\lambda'_5:\qquad
\begin{CD}
@. Q^*(4) @>{}>> Q^* \\
@. @V{(x_0^{-1}y_0^{-1}z_0^{-2},0,0)}VV @.\\
\mathcal{O}(3) @>{}>> \mathcal{O}(-1) @.
\end{CD}\qquad
\lambda'_6:\qquad
\begin{CD}
@. Q^*(4) @>{}>> Q^* \\
@. @V{(0,x_0^{-1}y_0^{-1}z_0^{-2},0)}VV @.\\
\mathcal{O}(3) @>{}>> \mathcal{O}(-1) @.
\end{CD}\qquad
\]}
\vskip0.1cm 
\noindent 
Note that, since the maps given by 
$(x_0^{-1}y_0^{-2}z_0^{-1},x_0^{-2}y_0^{-1}z_0^{-1},0)$, 
 $(x_0^{-1}y_0^{-1}z_0^{-2},0,x_0^{-2}y_0^{-1}z_0^{-1})$, 
 and $(0,x_0^{-1}y_0^{-1}z_0^{-2},x_0^{-1}y_0^{-2}z_0^{-1})$,
are exact, $\lambda'_4, \lambda'_5, \lambda'_6$ can be equivalently
represented by $ (0,-x_0^{-2}y_0^{-1}z_0^{-1},0)$,
 $ (0,0,-x_0^{-2}y_0^{-1}z_0^{-1})$,
 $(0,0,-x_0^{-1}y_0^{-2}z_0^{-1})$,   respectively.
\vskip0.3cm 
\noindent 
Finally, $\mathrm{Ext}^4(i_*\mathcal{O}, i_*\mathcal{O})$ is generated by $t
\in \check{C}^3(X, \mathit{Hom}^{1}(i_*\mathcal{O},
i_*\mathcal{O}))$:
{\small\[
t:\qquad
\begin{CD}
@. \mathcal{O}(4) @>{}>> \mathcal{O} \\
@. @V{x_0^{-1}y_0^{-1}z_0^{-1}}VV @.\\
\mathcal{O}(4) @>{}>> \mathcal{O} @.
\end{CD}\qquad
\]}
\vskip0.1cm 
\noindent  $\mathrm{Ext}^4(i_*\Omega^3(3)[3], i_*\Omega^3(3)[3])$ is generated
by $t' \in \check{C}^3(X, \mathit{Hom}^{1}(i_*\Omega^3(3)[3],
i_*\Omega^3(3)[3]))$:
{\small\[
t':\qquad
\begin{CD}
@. \mathcal{O}(3) @>{}>> \mathcal{O}(-1) \\
@. @V{x_0^{-1}y_0^{-1}z_0^{-1}}VV @.\\
\mathcal{O}(3) @>{}>> \mathcal{O}(-1) @.
\end{CD}\qquad
\]}
\vskip0.1cm 
\noindent 
And similarly for the other two $\Ext^4$ generators. 
\vskip0.3cm 
\noindent 
From this data, we determine the multiplication rules $m_2(x,y)$ by composition.  One finds:
\be\label{ab bc prod}
a_i \cdot b_j = \mat{\psi_2 & \psi_3 &0 & \psi_6 \cr
                                  - \psi_1 & 0 & \psi_3 & -\psi_5 \cr
      				 0 &- \psi_1& -\psi_2& \psi_4\cr
	                         -\psi_4  & -\psi_5&  -\psi_6&  0 \cr}~, \qquad\quad
b_i \cdot c_j = \mat{ -\lambda_2 & \lambda_1 &0 & \lambda_4\cr
					-\lambda_3 & 0& \lambda_1 &\lambda_5\cr
					0 &-\lambda_3 &\lambda_2 & \lambda_6\cr
					-\lambda_6 & \lambda_5 & -\lambda_4 & 0}~,	                         
\ee
and
\be\label{cd da prod}
c_i \cdot d_j = \mat{\psi_1' & 0 &\psi'_4 & \psi'_5 \cr
                                   \psi'_2 & -\psi'_4 & 0 & \psi'_6 \cr
      				 \psi_3' &- \psi_5& -\psi'_6&0 \cr
	                         0  & \psi'_1&  \psi'_2&  \psi'_3 \cr}~, \qquad\quad
d_i \cdot a_j = \mat{ -\lambda_1' & -\lambda'_2 & -\lambda'_3 &0\cr
					0 & \lambda'_4& \lambda'_5 &-\lambda'_1\cr
					-\lambda_4' &0 &\lambda'_6 & -\lambda'_2\cr
					-\lambda'_5 & -\lambda'_6 &0 & -\lambda_3'}~,	                         
\ee
with all other products $\Ext^1\cdot \Ext^1$ vanishing. (All higher products also vanish.) It is convenient to define the basis:
\bea\label{redef abcd}
&({\bf a}_1, {\bf a}_2, {\bf a}_3, {\bf a}_4) \equiv(a_3, -a_2, a_1, a_4)~, \quad 
&& ({\bf b}_1, {\bf b}_2, {\bf b}_3, {\bf b}_4)  \equiv (b_1, b_2, b_3, b_4)~, \cr
&({\bf c}_1, {\bf c}_2, {\bf c}_3, {\bf c}_4)  \equiv (c_3, -c_2, c_1, c_4)~, \quad
&& ({\bf d}_1, {\bf d}_2, {\bf d}_3, {\bf d}_4)  \equiv (d_4, -d_3, d_2, -d_1)~. 
\eea
such that the matrices $({\bf a}_i \cdot {\bf b}_j)$, $({\bf b}_i \cdot {\bf c}_j)$, $({\bf c}_i \cdot {\bf d}_j)$ and $({\bf d}_i \cdot {\bf a}_j)$ are all antisymmetric. This is simply a manifestation of the $GL(4)$ symmetry mentioned above.

\begin{figure}[t]
\begin{center}
\includegraphics[height=5cm]{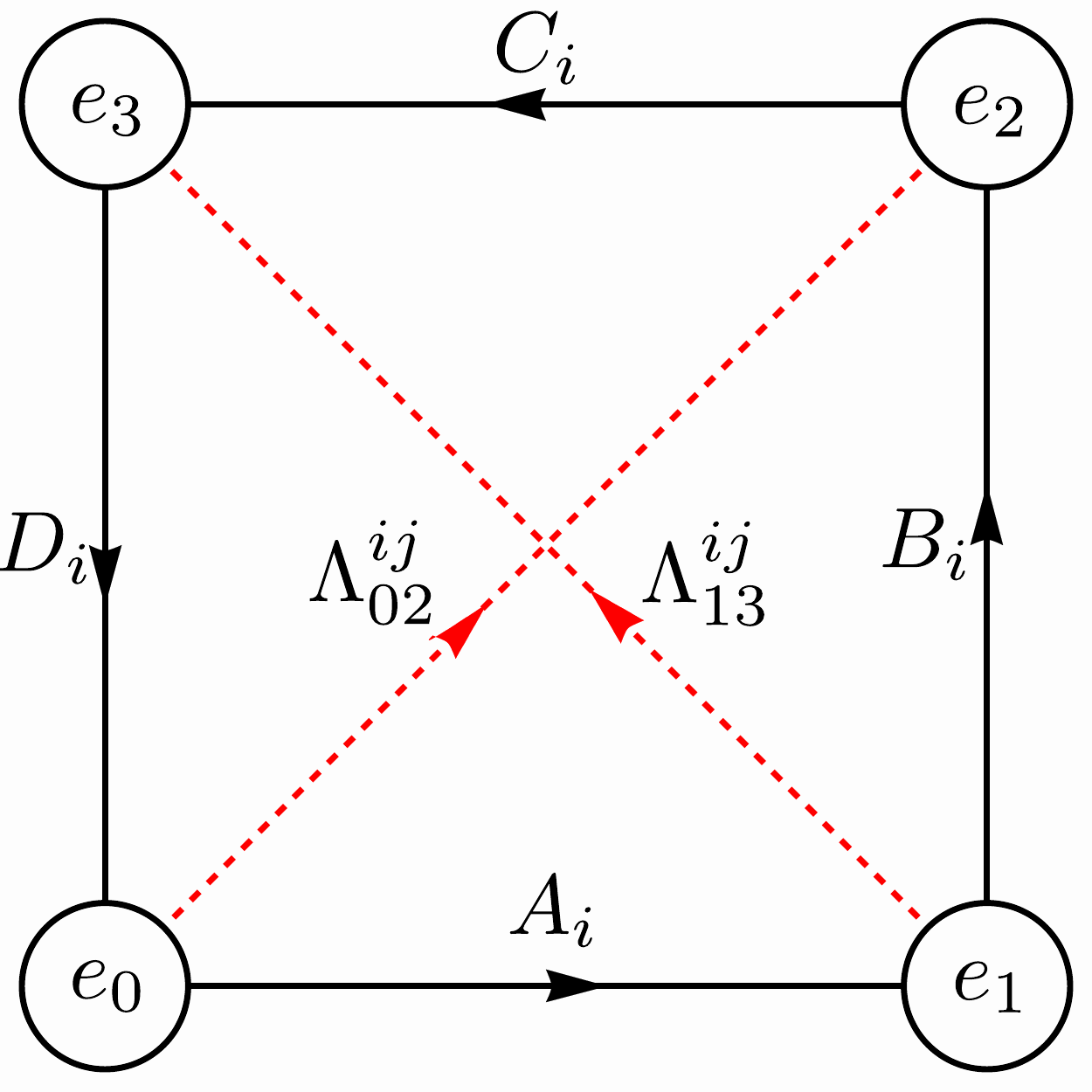}
\caption{The $\C^4/\Z_4(1,1,1,1)$ supersymmetric quiver---``theory (I)''.}
\label{figC4Z4 01}
\end{center}
\end{figure}
\subsubsection{Supersymmetric quiver (I)}
From the above results, we have a complete description of the 2d $\CN=(0,2)$ supersymmetric quiver for D1-branes on the $\C^4/\Z_4$ singularity. The chiral multiplets are identified with the $\Ext^1$ group elements according to:
\be\label{def ABCD}
A_i = {\bf a}_i~, \qquad B_j = {\bf b}_j~, \qquad C_k = {\bf c}_k~, \qquad D_l = {\bf d}_l~, 
\ee
with $i,j,k,l\in 1, \cdots, 4$, and the elements ${\bf a}, {\bf b}, {\bf c}, {\bf d}$ defined in \eqref{redef abcd}. As expected, the quiver theory has an $SU(4)$ global symmetry, with the fields \eqref{def ABCD} in the $\bf 4$ of $SU(4)$. The fermi multiplets $\Lambda_{02}^{n} \sim \psi_n$ and $\Lambda_{13}^m\sim \lambda_m$  naturally fit in the ${\bf 6}$ of $SU(4)$, which we denote by $\Lambda^{n}= \Lambda^{ij}= - \Lambda^{ji}$. We define the fermi multiplets in terms of the elements of $\Ext^2$ according to:
\bea
(\Lambda_{02}^{12}, \Lambda_{02}^{13}, \Lambda_{02}^{14}, \Lambda_{02}^{23}, \Lambda_{02}^{24}, \Lambda_{02}^{34}) =
(-\psi_1, -\psi_2, \psi_4, -\psi_3, \psi_5, \psi_6)~, \cr
(\Lambda_{2}^{13}, \Lambda_{13}^{13}, \Lambda_{13}^{14}, \Lambda_{13}^{23}, \Lambda_{13}^{24}, \Lambda_{13}^{34}) =
(-\lambda_1,-\lambda_2, \lambda_4, -\lambda_3, \lambda_5, \lambda_6)~,
\eea
and similarly for the (Serre dual) anti-fermi multiplets. The supersymmetric quiver is displayed  in Figure \ref{figC4Z4 01}. The $SU(4)$-preserving interactions terms encoded in \eqref{ab bc prod}-\eqref{cd da prod} take the simple  form:
\bea
&J_{\Lambda_{02}^{ij}} =\epsilon^{ijkl} C_k D_l~, \qquad \qquad 
&&J_{\Lambda_{13}^{ij}} =\epsilon^{ijkl} D_k A_l~,\cr
&E_{\Lambda_{02}^{ij}} = A_i B_j - A_j B_i~, \qquad \qquad 
&&E_{\Lambda_{13}^{ij}} = B_i C_j - B_j C_i~.
\eea
This satisfies $\Tr(EJ)=0$. It again agrees with the results of \cite{Franco:2015tna} for the $\C^4/\Z_4$ orbifold.

\subsubsection{Fractional branes and $\Ext$ algebra (III)}
\label{sect:fourfold:p2:second}

Another interesting set of fractional branes on $\t{\bf X}_4={\rm Tot}(\CO(-4)\rightarrow \mathbb{P}^3)$ is given by:
\be\label{P3 frac b 2}
\CE_0=  i_*\mathcal{O}[1]~, \qquad 
\CE_1= i_*\mathcal{O}(1)~, \qquad
\CE_2= \Omega^2(2)[2]~, \qquad
\CE_3=i_*\mathcal{O}(-1)[3]~.
\ee
Here we simply postulate this set of fractional branes, which we will further discuss in section \ref{sec: duality}. The $\Ext$ groups between the sheaves \eqref{P3 frac b 2} have a slightly more complicated structure than in the previous example. One finds:
\be
\dim \Ext^1(\CE_I, \CE_J) ={\small \mat{ 0& 4 &0 & 0 \\  0 & 0 & 0 & 10 \\ 6 & 0 & 0 &0 \\ 0&0&4&0}}~, \qquad \quad
\dim \Ext^2(\CE_I, \CE_J) ={\small \mat{0&0&0&4 \\ 0&0&20&0\\ 0&20&0&0\\ 4&0&0&0}}~,
\ee
This is conveniently summarized in the following $\Ext$ quiver:
\[
\xymatrix{ \mathcal{O}(-1)[3]  \ar[rr]_4^{c'}   & &
 \Omega^2(2)[2] \ar[ldld]^{{\hskip-0.8cm} d' \; \; \;\; 6}\\
& \ar[lu]_{b'}    &  \\
 \mathcal{O}[1]  \ar[rr]_{a'}^4 \ar@{--}[uu]^{\theta}_4  & 
 & \mathcal{O}(1)   \ar@{--}[uu]_\gamma^{20} \ar@{-}[lu]^{\hskip0.5cm 10} }
\]
Comparing to the $\Ext$ quiver of section \ref{sect:fourfold:p2:first}, we see that the upper row is the same (since $\Omega^3(3) \cong \CO(-1)$ on $\mathbb{P}^3$). We can thus identity $c'$ with the elements $c$ there. 
All $\Ext$ groups again form irreducible representations of $GL(4)$. In particular, the elements $b'$ span the ${\bf 10}$ (symmetric representation) and the elements $\gamma$ span the ${\bf 20}'$ (the mixed-symmetry three-tensor).

\paragraph{An explicit basis for $\Ext^\bullet$.}
We now turn to the explicit description of the $\Ext$ algebra.
$\mathrm{Ext}^1(i_*\Omega^2(2)[2], i_* \mathcal{O}[1])$ is generated
by $d'_n \in \check{C}^0(X, \mathit{Hom}^{1}(i_*\Omega^2(2)[2], i_*
\mathcal{O}[1])), n=1,...,6$:
{\small\[
d'_1:~~~~~\;
\begin{CD}
Q(3) @>{}>> Q(-1) \\
@V{(-1,0,0)}VV @V{(1,0,0)}VV \\
\mathcal{O}(4) @>{}>> \mathcal{O}
\end{CD}~~~~~~~~~~~~~
d'_2:~~~~~\;
\begin{CD}
Q(3) @>{}>> Q(-1) \\
@V{(0,-1,0)}VV @V{(0,1,0)}VV \\
\mathcal{O}(4) @>{}>> \mathcal{O}
\end{CD}~~~
\]
\[
d'_3:~~~~~\;
\begin{CD}
Q(3) @>{}>> Q(-1) \\
@V{(0,0,-1)}VV @V{(0,0,1)}VV \\
\mathcal{O}(4) @>{}>> \mathcal{O}
\end{CD}~~~~~~~~~~~~~
d'_4:~~~~~\;\quad
\begin{CD}
Q(3) @>{}>> Q(-1) \\
@V{(-y_0,x_0,0)}VV @V{(y_0,-x_0,0)}VV \\
\mathcal{O}(4) @>{}>> \mathcal{O}
\end{CD}~~~
\]
\[
d'_5:~~~~~\;\quad
\begin{CD}
Q(3) @>{}>> Q(-1) \\
@V{(-z_0,0,x_0)}VV @V{(z_0,0,-x_0)}VV \\
\mathcal{O}(4) @>{}>> \mathcal{O}
\end{CD}~~~~~~~~~~~~~
d'_6:~~~~~\;\quad
\begin{CD}
Q(3) @>{}>> Q(-1) \\
@V{(0,-z_0,y_0)}VV @V{(0,z_0,-y_0)}VV \\
\mathcal{O}(4) @>{}>> \mathcal{O}
\end{CD}~~~
\]}
\vskip0.2cm\noindent
$\mathrm{Ext}^1(i_*\mathcal{O}[1], i_*\mathcal(1))$ is generated by
$a'_i \in \check{C}^0(X, \mathit{Hom}^{1}(i_*\mathcal{O}[1],
i_*\mathcal(1))), i=1,2,3,4$:
{\small \[
a'_1:\qquad
\begin{CD}
\mathcal{O}(4) @>{}>> \mathcal{O} \\
@V{1}VV
@VV{-1}V \\
\mathcal{O}(5) @>{}>> \mathcal{O}(1)
\end{CD}\quad
a'_2:\qquad
\begin{CD}
\mathcal{O}(4) @>{}>> \mathcal{O} \\
@V{x_0}VV
@VV{-x_0}V \\
\mathcal{O}(5) @>{}>> \mathcal{O}(1)
\end{CD}\quad
\]
\[
a'_3:\qquad
\begin{CD}
\mathcal{O}(4) @>{}>> \mathcal{O} \\
@V{y_0}VV
@VV{-y_0}V \\
\mathcal{O}(5) @>{}>> \mathcal{O}(1)
\end{CD}\quad
a'_4:\qquad
\begin{CD}
\mathcal{O}(4) @>{}>> \mathcal{O} \\
@V{z_0}VV
@VV{-z_0}V \\
\mathcal{O}(5) @>{}>> \mathcal{O}(1)
\end{CD}\quad
\]}
\vskip0.2cm\noindent
$\mathrm{Ext}^1(i_*\mathcal{O}(-1)[3], i_*\Omega^2(2)[2])$ is
generated by $c'_i \in \check{C}^0(X,
\mathit{Hom}^{1}(i_*\mathcal{O}(-1)[3], i_*\Omega^2(2)[2])),
i=1,2,3,4$:
{\small\[
c'_1:\qquad
\begin{CD}
\mathcal{O}(3) @>{}>> \mathcal{O}(-1) \\
@V{\tiny \left(\begin{array}{c} -1 \\ 0 \\ 0
\end{array}\right)}VV
@VV{\tiny \left(\begin{array}{c} 1 \\ 0 \\ 0 \end{array}\right)}V \\
Q(3) @>{}>> Q(-1)
\end{CD}\qquad
c'_2:\qquad
\begin{CD}
\mathcal{O}(3) @>{}>> \mathcal{O}(-1) \\
@V{\tiny \left(\begin{array}{c} 0 \\ -1 \\ 0
\end{array}\right)}VV
@VV{\tiny \left(\begin{array}{c} 0 \\ 1 \\ 0 \end{array}\right)}V \\
Q(3) @>{}>> Q(-1)
\end{CD}\qquad\]
\[
c'_3:\qquad
\begin{CD}
\mathcal{O}(3) @>{}>> \mathcal{O}(-1) \\
@V{\tiny \left(\begin{array}{c} 0 \\ 0 \\ -1
\end{array}\right)}VV
@VV{\tiny \left(\begin{array}{c} 0 \\ 0 \\ 1 \end{array}\right)}V \\
Q(3) @>{}>> Q(-1)
\end{CD}\qquad
c'_4:\qquad
\begin{CD}
\mathcal{O}(3) @>{}>> \mathcal{O}(-1) \\
@V{\tiny \left(\begin{array}{c} -x_0 \\ -y_0 \\ -z_0
\end{array}\right)}VV
@VV{\tiny \left(\begin{array}{c} x_0 \\ y_0 \\ z_0 \end{array}\right)}V \\
Q(3) @>{}>> Q(-1)
\end{CD}\qquad
\]}
\vskip0.2cm\noindent
$\mathrm{Ext}^1(i_*\mathcal{O}(1), i_*\mathcal{O}(-1)[3])$ is
generated by $b'_l \in \check{C}^3(X,
\mathit{Hom}^{-2}(i_*\mathcal{O}(1), i_*\mathcal{O}(-1)[3])),
l=1,...,10$:
{\small\[
b'_1:\qquad
\begin{CD}
@. \mathcal{O}(5) @>{}>> \mathcal{O}(1) \\
@. @V{x_0^{-1}y_0^{-1}z_0^{-1}}VV @.\\
\mathcal{O}(3) @>{}>> \mathcal{O}(-1) @.
\end{CD}\qquad
b'_2:\qquad
\begin{CD}
@. \mathcal{O}(5) @>{}>> \mathcal{O}(1) \\
@. @V{x_0^{-2}y_0^{-1}z_0^{-1}}VV @.\\
\mathcal{O}(3) @>{}>> \mathcal{O}(-1) @.
\end{CD}\qquad
\]
\[
b'_3:\qquad
\begin{CD}
@. \mathcal{O}(5) @>{}>> \mathcal{O}(1) \\
@. @V{x_0^{-1}y_0^{-2}z_0^{-1}}VV @.\\
\mathcal{O}(3) @>{}>> \mathcal{O}(-1) @.
\end{CD}\qquad
b'_4:\qquad
\begin{CD}
@. \mathcal{O}(5) @>{}>> \mathcal{O}(1) \\
@. @V{x_0^{-1}y_0^{-1}z_0^{-2}}VV @.\\
\mathcal{O}(3) @>{}>> \mathcal{O}(-1) @.
\end{CD}\qquad
\]
\[
b'_5:\qquad
\begin{CD}
@. \mathcal{O}(5) @>{}>> \mathcal{O}(1) \\
@. @V{x_0^{-3}y_0^{-1}z_0^{-1}}VV @.\\
\mathcal{O}(3) @>{}>> \mathcal{O}(-1) @.
\end{CD}\qquad
b'_6:\qquad
\begin{CD}
@. \mathcal{O}(5) @>{}>> \mathcal{O}(1) \\
@. @V{x_0^{-1}y_0^{-3}z_0^{-1}}VV @.\\
\mathcal{O}(3) @>{}>> \mathcal{O}(-1) @.
\end{CD}\qquad
\]
\[
b'_7:\qquad
\begin{CD}
@. \mathcal{O}(5) @>{}>> \mathcal{O}(1) \\
@. @V{x_0^{-1}y_0^{-1}z_0^{-3}}VV @.\\
\mathcal{O}(3) @>{}>> \mathcal{O}(-1) @.
\end{CD}\qquad
b'_8:\qquad
\begin{CD}
@. \mathcal{O}(5) @>{}>> \mathcal{O}(1) \\
@. @V{x_0^{-2}y_0^{-2}z_0^{-1}}VV @.\\
\mathcal{O}(3) @>{}>> \mathcal{O}(-1) @.
\end{CD}\qquad
\]
\[
b'_9:\qquad
\begin{CD}
@. \mathcal{O}(5) @>{}>> \mathcal{O}(1) \\
@. @V{x_0^{-2}y_0^{-1}z_0^{-2}}VV @.\\
\mathcal{O}(3) @>{}>> \mathcal{O}(-1) @.
\end{CD}\qquad
b'_{10}:\qquad
\begin{CD}
@. \mathcal{O}(5) @>{}>> \mathcal{O}(1) \\
@. @V{x_0^{-1}y_0^{-2}z_0^{-2}}VV @.\\
\mathcal{O}(3) @>{}>> \mathcal{O}(-1) @.
\end{CD}\qquad
\]}
\vskip0.2cm\noindent
$\mathrm{Ext}^2(i_*\Omega^2(2)[2], i_*\mathcal{O}(1))$ is generated
by $\gamma_s \in \check{C}^0(X, \mathit{Hom}^{2}(i_*\Omega^2(2)[2],
i_*\mathcal{O}(1))), s=1,...,20$:
{\small\[
\gamma_1:~~~~~
\begin{CD}
Q(3) @>{}>> Q(-1) \\
@V{(1,0,0)}VV @V{(1,0,0)}VV \\
\mathcal{O}(5) @>{}>> \mathcal{O}(1)
\end{CD}~~~~~~~~~~~~~
\gamma_2:~~~~~
\begin{CD}
Q(3) @>{}>> Q(-1) \\
@V{(0,1,0)}VV @V{(0,1,0)}VV \\
\mathcal{O}(5) @>{}>> \mathcal{O}(1)
\end{CD}~~~
\]
\[
\gamma_3:~~~~~
\begin{CD}
Q(3) @>{}>> Q(-1) \\
@V{(0,0,1)}VV @V{(0,0,1)}VV \\
\mathcal{O}(5) @>{}>> \mathcal{O}(1)
\end{CD}~~~~~~~~~~~~~
\gamma_4:~~~~~
\begin{CD}
Q(3) @>{}>> Q(-1) \\
@V{(x_0,0,0)}VV @V{(x_0,0,0)}VV \\
\mathcal{O}(5) @>{}>> \mathcal{O}(1)
\end{CD}~~~
\]
\[
\gamma_5:~~~~~
\begin{CD}
Q(3) @>{}>> Q(-1) \\
@V{(0,x_0,0)}VV @V{(0,x_0,0)}VV \\
\mathcal{O}(5) @>{}>> \mathcal{O}(1)
\end{CD}~~~~~~~~~~~~~
\gamma_6:~~~~~
\begin{CD}
Q(3) @>{}>> Q(-1) \\
@V{(0,0,x_0)}VV @V{(0,0,x_0)}VV \\
\mathcal{O}(5) @>{}>> \mathcal{O}(1)
\end{CD}~~~
\]
\[
\gamma_7:~~~~~
\begin{CD}
Q(3) @>{}>> Q(-1) \\
@V{(y_0,0,0)}VV @V{(y_0,0,0)}VV \\
\mathcal{O}(5) @>{}>> \mathcal{O}(1)
\end{CD}~~~~~~~~~~~~~
\gamma_8:~~~~~
\begin{CD}
Q(3) @>{}>> Q(-1) \\
@V{(0,y_0,0)}VV @V{(0,y_0,0)}VV \\
\mathcal{O}(5) @>{}>> \mathcal{O}(1)
\end{CD}~~~
\]
\[
\gamma_9:~~~~~
\begin{CD}
Q(3) @>{}>> Q(-1) \\
@V{(0,0,y_0)}VV @V{(0,0,y_0)}VV \\
\mathcal{O}(5) @>{}>> \mathcal{O}(1)
\end{CD}~~~~~~~~~~~~~
\gamma_{10}:~~~~~
\begin{CD}
Q(3) @>{}>> Q(-1) \\
@V{(z_0,0,0)}VV @V{(z_0,0,0)}VV \\
\mathcal{O}(5) @>{}>> \mathcal{O}(1)
\end{CD}~~~
\]
\[
\gamma_{11}:~~~~~
\begin{CD}
Q(3) @>{}>> Q(-1) \\
@V{(0,z_0,0)}VV @V{(0,z_0,0)}VV \\
\mathcal{O}(5) @>{}>> \mathcal{O}(1)
\end{CD}~~~~~~~~~~~~~
\gamma_{12}:~~~~~
\begin{CD}
Q(3) @>{}>> Q(-1) \\
@V{(0,0,z_0)}VV @V{(0,0,z_0)}VV \\
\mathcal{O}(5) @>{}>> \mathcal{O}(1)
\end{CD}~~~
\]
\[
\gamma_{13}:~~~~~
\begin{CD}
Q(3) @>{}>> Q(-1) \\
@VV{(0,-z_0^2,y_0 z_0)}V @VV{(0,-z_0^2,y_0 z_0)}V \\
\mathcal{O}(5) @>{}>> \mathcal{O}(1)
\end{CD}~~~~~~~~~~~~~
\gamma_{14}:~~~~~
\begin{CD}
Q(3) @>{}>> Q(-1) \\
@VV{(0,y_0 z_0,-y_0^2)}V @VV{(0,y_0 z_0,-y_0^2)}V \\
\mathcal{O}(5) @>{}>> \mathcal{O}(1)
\end{CD}~~~
\]
\[
\gamma_{15}:~~~~~
\begin{CD}
Q(3) @>{}>> Q(-1) \\
@VV{(0,x_0 z_0,-x_0 y_0)}V @VV{(0,x_0 z_0,-x_0 y_0)}V \\
\mathcal{O}(5) @>{}>> \mathcal{O}(1)
\end{CD}~~~~~~~~~~~~~
\gamma_{16}:~~~~~
\begin{CD}
Q(3) @>{}>> Q(-1) \\
@VV{(-z_0^2,0,x_0 z_0)}V @VV{(-z_0^2,0,x_0 z_0)}V \\
\mathcal{O}(5) @>{}>> \mathcal{O}(1)
\end{CD}~~~
\]
\[
\gamma_{17}:~~~~~
\begin{CD}
Q(3) @>{}>> Q(-1) \\
@VV{(x_0 z_0,0,-x_0^2)}V @VV{(x_0 z_0,0,-x_0^2)}V \\
\mathcal{O}(5) @>{}>> \mathcal{O}(1)
\end{CD}~~~~~~~~~~~~~
\gamma_{18}:~~~~~
\begin{CD}
Q(3) @>{}>> Q(-1) \\
@VV{(y_0 z_0,0,-x_0 y_0)}V @VV{(y_0 z_0,0,-x_0 y_0)}V \\
\mathcal{O}(5) @>{}>> \mathcal{O}(1)
\end{CD}~~~
\]
\[
\gamma_{19}:~~~~~
\begin{CD}
Q(3) @>{}>> Q(-1) \\
@VV{(x_0 y_0,-x_0^2,0)}V @VV{(x_0 y_0,-x_0^2,0)}V \\
\mathcal{O}(5) @>{}>> \mathcal{O}(1)
\end{CD}~~~~~~~~~~~~~
\gamma_{20}:~~~~~
\begin{CD}
Q(3) @>{}>> Q(-1) \\
@VV{(-y_0^2,x_0 y_0,0)}V @VV{(-y_0^2,x_0 y_0,0)}V \\
\mathcal{O}(5) @>{}>> \mathcal{O}(1)
\end{CD}~~~
\]}
\vskip0.2cm\noindent
We denote the Serre dual of $\gamma$ by $\gamma'$,  with
$\gamma_s'\in \check{C}^3(X, \mathit{Hom}^{-1}(i_*\mathcal{O}(1),
i_*\Omega^2(2)[2]))$:
{\small\[
\gamma_1':\qquad
\begin{CD}
@. \mathcal{O}(5) @>{}>> \mathcal{O}(1) \\
@. @V{\tiny \left(\begin{array}{c} x_0^{-1}y_0^{-1}z_0^{-1} \\ 0 \\ 0 \end{array}\right)}VV @.\\
Q(3) @>{}>> Q(-1) @.
\end{CD}\qquad
\gamma_2':\qquad
\begin{CD}
@. \mathcal{O}(5) @>{}>> \mathcal{O}(1) \\
@. @V{\tiny \left(\begin{array}{c} 0 \\ x_0^{-1}y_0^{-1}z_0^{-1} \\ 0 \end{array}\right)}VV @.\\
Q(3) @>{}>> Q(-1) @.
\end{CD}\qquad
\]
\[
\gamma_3':\qquad
\begin{CD}
@. \mathcal{O}(5) @>{}>> \mathcal{O}(1) \\
@. @V{\tiny \left(\begin{array}{c} 0 \\ 0 \\ x_0^{-1}y_0^{-1}z_0^{-1} \end{array}\right)}VV @.\\
Q(3) @>{}>> Q(-1) @.
\end{CD}\qquad
\gamma_4':\qquad
\begin{CD}
@. \mathcal{O}(5) @>{}>> \mathcal{O}(1) \\
@. @V{\tiny \left(\begin{array}{c} x_0^{-2}y_0^{-1}z_0^{-1} \\ 0 \\ 0 \end{array}\right)}VV @.\\
Q(3) @>{}>> Q(-1) @.
\end{CD}\qquad
\]
\[
\gamma_5':\qquad
\begin{CD}
@. \mathcal{O}(5) @>{}>> \mathcal{O}(1) \\
@. @V{\tiny \left(\begin{array}{c} 0 \\ x_0^{-2}y_0^{-1}z_0^{-1} \\ 0 \end{array}\right)}VV @.\\
Q(3) @>{}>> Q(-1) @.
\end{CD}\qquad
\gamma_6':\qquad
\begin{CD}
@. \mathcal{O}(5) @>{}>> \mathcal{O}(1) \\
@. @V{\tiny \left(\begin{array}{c} 0 \\ 0 \\ x_0^{-2}y_0^{-1}z_0^{-1} \end{array}\right)}VV @.\\
Q(3) @>{}>> Q(-1) @.
\end{CD}\qquad
\]
\[
\gamma_7':\qquad
\begin{CD}
@. \mathcal{O}(5) @>{}>> \mathcal{O}(1) \\
@. @V{\tiny \left(\begin{array}{c} x_0^{-1}y_0^{-2}z_0^{-1} \\ 0 \\ 0 \end{array}\right)}VV @.\\
Q(3) @>{}>> Q(-1) @.
\end{CD}\qquad
\gamma_8':\qquad
\begin{CD}
@. \mathcal{O}(5) @>{}>> \mathcal{O}(1) \\
@. @V{\tiny \left(\begin{array}{c} 0 \\ x_0^{-1}y_0^{-2}z_0^{-1} \\ 0 \end{array}\right)}VV @.\\
Q(3) @>{}>> Q(-1) @.
\end{CD}\qquad
\]
\[
\gamma_9':\qquad
\begin{CD}
@. \mathcal{O}(5) @>{}>> \mathcal{O}(1) \\
@. @V{\tiny \left(\begin{array}{c} 0 \\ 0 \\ x_0^{-1}y_0^{-2}z_0^{-1} \end{array}\right)}VV @.\\
Q(3) @>{}>> Q(-1) @.
\end{CD}\qquad
\gamma_{10}':\qquad
\begin{CD}
@. \mathcal{O}(5) @>{}>> \mathcal{O}(1) \\
@. @V{\tiny \left(\begin{array}{c} x_0^{-1}y_0^{-1}z_0^{-2} \\ 0 \\ 0 \end{array}\right)}VV @.\\
Q(3) @>{}>> Q(-1) @.
\end{CD}\qquad
\]
\[
\gamma_{11}':\qquad
\begin{CD}
@. \mathcal{O}(5) @>{}>> \mathcal{O}(1) \\
@. @V{\tiny \left(\begin{array}{c} 0 \\ x_0^{-1}y_0^{-1}z_0^{-2} \\ 0 \end{array}\right)}VV @.\\
Q(3) @>{}>> Q(-1) @.
\end{CD}\qquad
\gamma_{12}':\qquad
\begin{CD}
@. \mathcal{O}(5) @>{}>> \mathcal{O}(1) \\
@. @V{\tiny \left(\begin{array}{c} 0 \\ 0 \\ x_0^{-1}y_0^{-1}z_0^{-2} \end{array}\right)}VV @.\\
Q(3) @>{}>> Q(-1) @.
\end{CD}\qquad
\]
\[
\gamma_{13}':\qquad
\begin{CD}
@. \mathcal{O}(5) @>{}>> \mathcal{O}(1) \\
@. @V{\tiny \left(\begin{array}{c} 0 \\ 0 \\ x_0^{-1}y_0^{-2}z_0^{-2} \end{array}\right)}VV @.\\
Q(3) @>{}>> Q(-1) @.
\end{CD}\qquad
\gamma_{14}':\qquad
\begin{CD}
@. \mathcal{O}(5) @>{}>> \mathcal{O}(1) \\
@. @V{\tiny \left(\begin{array}{c} 0 \\ x_0^{-1}y_0^{-2}z_0^{-2} \\ 0 \end{array}\right)}VV @.\\
Q(3) @>{}>> Q(-1) @.
\end{CD}\qquad
\]
\[
\gamma_{15}':\qquad
\begin{CD}
@. \mathcal{O}(5) @>{}>> \mathcal{O}(1) \\
@. @V{\tiny \left(\begin{array}{c} 0 \\ x_0^{-2}y_0^{-1}z_0^{-2} \\ 0 \end{array}\right)}VV @.\\
Q(3) @>{}>> Q(-1) @.
\end{CD}\qquad
\gamma_{16}':\qquad
\begin{CD}
@. \mathcal{O}(5) @>{}>> \mathcal{O}(1) \\
@. @V{\tiny \left(\begin{array}{c} 0 \\ 0 \\ x_0^{-2}y_0^{-1}z_0^{-2} \end{array}\right)}VV @.\\
Q(3) @>{}>> Q(-1) @.
\end{CD}\qquad
\]
\[
\gamma_{17}':\qquad
\begin{CD}
@. \mathcal{O}(5) @>{}>> \mathcal{O}(1) \\
@. @V{\tiny \left(\begin{array}{c} x_0^{-2}y_0^{-1}z_0^{-2} \\ 0 \\ 0 \end{array}\right)}VV @.\\
Q(3) @>{}>> Q(-1) @.
\end{CD}\qquad
\gamma_{18}':\qquad
\begin{CD}
@. \mathcal{O}(5) @>{}>> \mathcal{O}(1) \\
@. @V{\tiny \left(\begin{array}{c} x_0^{-1}y_0^{-2}z_0^{-2} \\ 0 \\ 0 \end{array}\right)}VV @.\\
Q(3) @>{}>> Q(-1) @.
\end{CD}\qquad
\]
\[
\gamma_{19}':\qquad
\begin{CD}
@. \mathcal{O}(5) @>{}>> \mathcal{O}(1) \\
@. @V{\tiny \left(\begin{array}{c} x_0^{-2}y_0^{-2}z_0^{-1} \\ 0 \\ 0 \end{array}\right)}VV @.\\
Q(3) @>{}>> Q(-1) @.
\end{CD}\qquad
\gamma_{20}':\qquad
\begin{CD}
@. \mathcal{O}(5) @>{}>> \mathcal{O}(1) \\
@. @V{\tiny \left(\begin{array}{c} 0 \\ x_0^{-2}y_0^{-2}z_0^{-1} \\ 0 \end{array}\right)}VV @.\\
Q(3) @>{}>> Q(-1) @.
\end{CD}\qquad
\]}
\vskip0.2cm\noindent
$\mathrm{Ext}^2(i_*\mathcal{O}(-1)[3], i_*\mathcal{O}[1])$ is
generated by $\theta_i \in \check{C}^0(X,
\mathit{Hom}^{2}(i_*\mathcal{O}(-1)[3], i_*\mathcal{O}[1])),
i=1,...,4$:
{\small\[
\theta_1:\qquad
\begin{CD}
\mathcal{O}(3) @>{}>> \mathcal{O}(-1) \\
@V{1}VV
@VV{1}V \\
\mathcal{O}(4) @>{}>> \mathcal{O}
\end{CD}\qquad
\theta_2:\qquad
\begin{CD}
\mathcal{O}(3) @>{}>> \mathcal{O}(-1) \\
@V{x_0}VV
@VV{x_0}V \\
\mathcal{O}(4) @>{}>> \mathcal{O}
\end{CD}\qquad
\]
\[
\theta_3:\qquad
\begin{CD}
\mathcal{O}(3) @>{}>> \mathcal{O}(-1) \\
@V{y_0}VV
@VV{y_0}V \\
\mathcal{O}(4) @>{}>> \mathcal{O}
\end{CD}\qquad
\theta_4:\qquad
\begin{CD}
\mathcal{O}(3) @>{}>> \mathcal{O}(-1) \\
@V{z_0}VV
@VV{z_0}V \\
\mathcal{O}(4) @>{}>> \mathcal{O}
\end{CD}\qquad
\]}
\vskip0.2cm\noindent
We denote the Serre dual of $\theta$ by $\theta'$, with 
$\theta_i'\in \check{C}^3(X, \mathit{Hom}^{-1}(i_*\mathcal{O}[1],
i_*\mathcal{O}(-1)[3]))$:
{\small\[
\theta'_1:\qquad
\begin{CD}
@. \mathcal{O}(4) @>{}>> \mathcal{O} \\
@. @V{x_0^{-1}y_0^{-1}z_0^{-1}}VV @.\\
\mathcal{O}(3) @>{}>> \mathcal{O}(-1) @.
\end{CD}\qquad
\theta'_2:\qquad
\begin{CD}
@. \mathcal{O}(4) @>{}>> \mathcal{O} \\
@. @V{x_0^{-2}y_0^{-1}z_0^{-1}}VV @.\\
\mathcal{O}(3) @>{}>> \mathcal{O}(-1) @.
\end{CD}\qquad
\]
\[
\theta'_3:\qquad
\begin{CD}
@. \mathcal{O}(4) @>{}>> \mathcal{O} \\
@. @V{x_0^{-1}y_0^{-2}z_0^{-1}}VV @.\\
\mathcal{O}(3) @>{}>> \mathcal{O}(-1) @.
\end{CD}\qquad
\theta'_4:\qquad
\begin{CD}
@. \mathcal{O}(4) @>{}>> \mathcal{O} \\
@. @V{x_0^{-1}y_0^{-1}z_0^{-2}}VV @.\\
\mathcal{O}(3) @>{}>> \mathcal{O}(-1) @.
\end{CD}\qquad
\]}
\vskip0.2cm\noindent
$\mathrm{Ext}^4(i_*\mathcal{O}[1], i_*\mathcal{O}[1])$ is generated
by $t \in \check{C}^3(X, \mathit{Hom}^{1}(i_*\mathcal{O}[1],
i_*\mathcal{O}[1]))$:
{\small\[
t:\qquad
\begin{CD}
@. \mathcal{O}(4) @>{}>> \mathcal{O} \\
@. @V{x_0^{-1}y_0^{-1}z_0^{-1}}VV @.\\
\mathcal{O}(4) @>{}>> \mathcal{O} @.
\end{CD}\qquad
\]}
\vskip0.2cm\noindent
$\mathrm{Ext}^4(i_*\mathcal{O}(1), i_*\mathcal{O}(1))$ is generated
by $t' \in \check{C}^3(X, \mathit{Hom}^{1}(i_*\mathcal{O}(1),
i_*\mathcal{O}(1)))$:
{\small\[
t':\qquad
\begin{CD}
@. \mathcal{O}(5) @>{}>> \mathcal{O}(1) \\
@. @V{x_0^{-1}y_0^{-1}z_0^{-1}}VV @.\\
\mathcal{O}(5) @>{}>> \mathcal{O}(1) @.
\end{CD}\qquad
\]}
\vskip0cm\noindent
From these data, we can again compute the product rules, which are all of the form $m_2(x,y)= x\cdot y$ and can be obtained by composition. One finds:

{\small\[
b'_i\cdot a'_j = \left(\begin{array}{cccc} \theta'_1 & 0 & 0 & 0 \\
\theta'_2 & \theta'_1 & 0 & 0 \\ \theta'_3 & 0 & \theta'_1 & 0 \\
\theta'_4 & 0 & 0 & \theta'_1 \\ 0 & \theta'_2 & 0 & 0 \\
0 & 0 & \theta'_3 & 0 \\ 0 & 0 & 0 & \theta'_4 \\
0 & \theta'_3 & \theta'_2 & 0 \\ 0 & \theta'_4 & 0 & \theta'_2 \\
0 & 0 & \theta'_4 & \theta'_3
\end{array}\right)~,\qquad
d'_i\cdot c'_j = \left(\begin{array}{cccc} \theta_1 & 0 & 0 &
\theta_2 \\ 0 & \theta_1 & 0 & \theta_3 \\ 0 & 0 & \theta_1 &
\theta_4 \\ 0 & \theta_4 & -\theta_3 & 0 \\ \theta_4 & 0 & -\theta_2
& 0 \\ \theta_3 & -\theta_2 & 0 & 0
\end{array}\right)~,
\]
\[
a'_i\cdot d'_j = \left(\begin{array}{cccccc} -\gamma_1 & -\gamma_2 &
-\gamma_3 & \gamma_9-\gamma_{11} & \gamma_6-\gamma_{10} & \gamma_5 -
\gamma_7 \\ -\gamma_4 & -\gamma_5 & -\gamma_6 & -\gamma_{15} &
-\gamma_{17} & -\gamma_{19} \\ -\gamma_7 & -\gamma_8 & -\gamma_9 &
-\gamma_{14} & -\gamma_{18} & \gamma_{20} \\ -\gamma_{10} &
-\gamma_{11} & -\gamma_{12} & \gamma_{13} & \gamma_{16} &
\gamma_{15}-\gamma_{18}
\end{array}\right)~,
\]
\[
c'_i\cdot b'_j = \left(\begin{array}{cccccccccc} -\gamma'_1 &
-\gamma'_4 & -\gamma'_7 & -\gamma'_{10} & 0 & \gamma'_{20} &
\gamma'_{16} & -\gamma'_{19} & -\gamma'_{17} & -\gamma'_{18} \\
-\gamma'_2 & -\gamma'_5 & -\gamma'_8 & -\gamma'_{11} & \gamma'_{19}
& 0 & \gamma'_{13} & -\gamma'_{20} & -\gamma'_{15} & -\gamma'_{14} \\
-\gamma'_3 & -\gamma'_6 & -\gamma'_9 & -\gamma'_{12} & \gamma'_{17}
& \gamma'_{14} & 0 & \gamma'_{15}+\gamma'_{18} & -\gamma'_{16} & -\gamma'_{13} \\
0 & -\gamma'_1 & -\gamma'_2 & -\gamma'_3 & -\gamma'_4 & -\gamma'_8 &
-\gamma'_{12} & -\gamma'_5-\gamma'_7 & -\gamma'_6-\gamma'_{10} & -\gamma'_9-\gamma'_{11} \\
\end{array}\right)~.
\]}

\subsubsection{Supersymmetric quiver (III)}\label{subsec: quiver theory III derived}
\begin{figure}[t]
\begin{center}
\includegraphics[height=5cm]{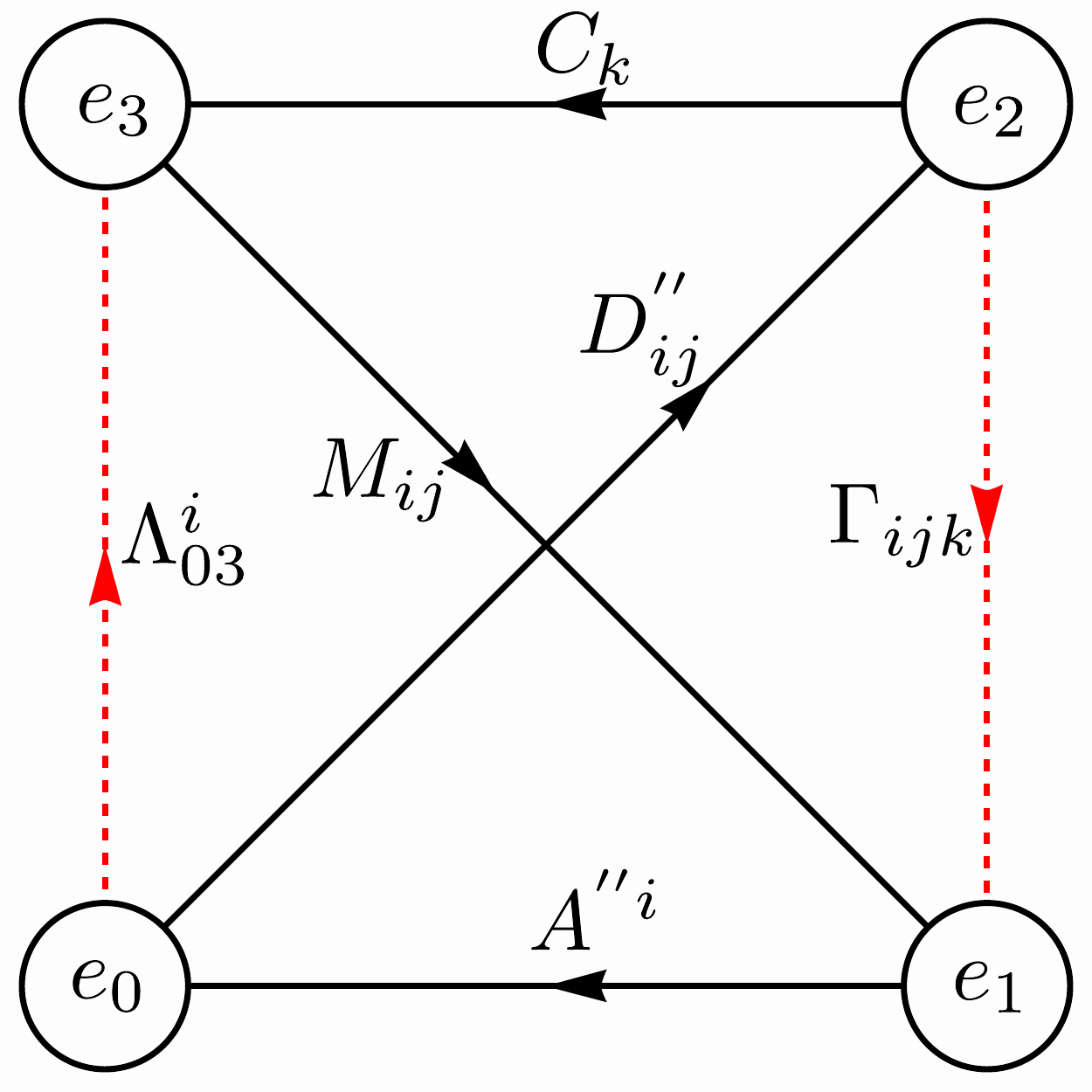}
\caption{Another $\C^4/\Z_4(1,1,1,1)$ supersymmetric quiver, which we dub ``Theory (III)''.}
\label{figC4Z4BIS 01}
\end{center}
\end{figure}

To present the final result for the $\CN=(0,2)$ quiver theory associated to the fractional branes \eqref{P3 frac b 2}, it is convenient to take advantage of the $SU(4)$ global symmetry. Let us introduce the chiral multiplets:
\bea
&A^{''i} = (A^{''1}, A^{''2}, A^{''3}, A^{''4} ) &=&\; (a'_a, -a_3', a_2', -a_1')~, \cr
&C_{i} = (C_1, C_2, C_3, C_4) &=&\; (c'_3, -c'_2, c'_1, c_4')~, 
\eea
which are identified with the $\Ext^1$ elements $a', c'$ as indicated.
We also introduce the fields $M_{ij}=M_{ji}\sim b'$ and $D^{''}_{ij}= - D^{''}_{ji}\sim d'$ in the ${\bf 10}$ and ${\bf 6}$ of $SU(4)$, respectively:
\be
M_{ij}= \mat{b'_7 &-b'_{10} & b'_9& -b'_4 \\
				-b'_{10}& b'_6& -b'_8& b'_3\\
				b'_9& -b'_8 & b'_5 &-b'_2 \\
				-b'_4 & b'_3 &-b'_2 &b'_1}~, \qquad\quad
D^{''}_{ij}=\mat{0 & -d_1'& -d_2'& d_4'\\
		d_1'& 0& -d_3' & d_5'\\
		d_2'& d_3' &0 & d_6\\
		-d_4' & -d_5' &-d_6' &0 }~.
\ee
We define the fermi multiplets $\Lambda_{03}^{i}\sim \theta'$ and $\Gamma_{ijk}\sim \gamma$ in the ${\bf 4}$ and ${\bf 20}'$ of $SU(4)$, respectively. We have:
\be
\Lambda_{03}^i = (\Lambda_{03}^1, \Lambda_{03}^2, \Lambda_{03}^3, \Lambda_{03}^4)= (\theta'_4, -\theta'_3, \theta_2', -\theta_1')~.
\ee
in terms of the $\Ext^2$ elements $\theta'$.
The  fields $\Gamma_{ijk}$ are such that:
\be
\Gamma_{ij k}= -\Gamma_{ji k}~, \qquad \qquad \epsilon^{ijkl} \Gamma_{jkl}=0~.
\ee
We choose the explicit set of 20 fields:
\bea\label{def Gammaijk}
&\Gamma_{ij 1}= \mat{0 & - \gamma_{13}'  &\gamma_{16}'& -\gamma'_{12}\\
				 & 0 & \Gamma_{231} & \gamma'_{11}\\
				 && 0 & -\gamma'_{10}\\
				 &&&0}~, \qquad  
&& \Gamma_{ij 2}= \mat{0 & - \gamma_{14}'  &\gamma_{18}'& \gamma'_{9}\\
				 & 0 & \gamma'_{20} & -\gamma'_{8}\\
				 && 0 & \gamma'_{7}\\
				 &&&0}~, \cr
& \Gamma_{ij 3}=\mat{0 & \gamma_{15}'  &-\gamma_{17}'& -\gamma'_{6}\\
				 & 0 & \gamma'_{19} & \gamma'_{5}\\
				 && 0 & -\gamma'_{4}\\
				 &&&0}~, \qquad  
&& \Gamma_{ij 4}= \mat{0 & \Gamma_{124}  & \Gamma_{134} & \gamma'_{3}\\
				 & 0 & \Gamma_{234}&-\gamma'_{2}\\
				 && 0 & \gamma'_{1}\\
				 &&&0}~,			 				 
\eea
which are identified with the $\Ext^2$ elements as indicated (the fields $\Gamma_{231}, \Gamma_{124},  \Gamma_{134},  \Gamma_{234}$ are redundant).  The supersymmetric quiver is shown in Figure \ref{figC4Z4BIS 01}.
It is  convenient to introduce the notation:
\be
\t D^{ij}= \half \epsilon^{ijkl}D^{''}_{kl}~.
\ee
The interaction terms for $\Lambda_{03}$ are given by:
\be
E_{\Lambda_{03}^i} = - \t D^{il} C_l~, \qquad J_{\Lambda_{03}^i}  = M_{ij}A^{''j}~.
\ee
The $E$-terms for $\Gamma$ read:
\be
E_{\Gamma_{ijk}} = C_i M_{jk} - C_j M_{ik}~.
\ee
To write down the $J$-terms, it is more convenient to use the explicit choice of $20$ components as in \eqref{def Gammaijk}. We find:
\be
J_{\Gamma_{ijk}}= - A^{''k} \t D^{ij}\quad \quad \text{if}\;\; k=i \;\; \text{or}\;\; k=j~.
\ee
The $J$-terms of the remaining 8 fields are given explicitly by:
\bea
& J_{\Gamma_{123}}= - A^{''3} \t D^{12}+  A^{''1} \t D^{23}~, \qquad 
&&  J_{\Gamma_{132}}= - A^{''2} \t D^{13}-  A^{''1} \t D^{23}~, \cr
& J_{\Gamma_{142}}= - A^{''2} \t D^{14}-  A^{''4} \t D^{12}~, \qquad 
&&  J_{\Gamma_{143}}= - A^{''3} \t D^{14}-  A^{''4} \t D^{13}~, \cr
& J_{\Gamma_{241}}= - A^{''1} \t D^{24}+  A^{''4} \t D^{21}~, \qquad 
&&  J_{\Gamma_{243}}= - A^{''3} \t D^{24}-  A^{''4} \t D^{21}~, \cr
& J_{\Gamma_{341}}= - A^{''1} \t D^{34}+  A^{''4} \t D^{13}~, \qquad 
&&  J_{\Gamma_{342}}= - A^{''2} \t D^{34}+  A^{''4} \t D^{23}~. \cr
\eea
One can check that:
\be
\sum_{i=1}^4 \Tr(E_{\Lambda_{03}^i}J_{\Lambda_{03}^i}) = - \sum_{s=1}^{20} \Tr(E_{\Gamma_{(s)}}J_{\Gamma_{(s)}})  = \Tr\left( C_i M_{j k} A^{'' k} \t D^{ij}\right)~,
\ee
and therefore $\Tr(EJ)=0$, as required.
Interestingly, this supersymmetric quiver cannot be realized as a brane brick model \cite{Franco:2015tna}. This is an example of a ``non-toric quiver'' (even though the CY$_4$ geometry is itself toric, in this case). 

Since both quiver theories (I) and (III) appear to describe the low-energy dynamics of D1-branes at the $[\C^4/\Z_4]$ singularity, we expect that  these two  gauge theories are related by an infrared duality. It is indeed the case, as we will discuss in section \ref{sec: duality}.


\subsection{Fractional branes on a local $\mathbb{P}^1\times \mathbb{P}^1$}
\label{sec local P1P1}
As our last example, we consider a toric singularity which is not an orbifold. 
Let ${\bf X}_4$ be the real cone over the seven-manifold known as $Q^{1,1,1}$:
\be\label{Q111 def}
 {\bf X}_4 =C(Q^{1,1,1})~, \qquad \quad Q^{1,1,1}\cong {SU(2) \times SU(2) \times SU(2)\ov U(1) \times U(1)}~.
\ee
This singularity was also discussed in \cite{Franco:2015tna}.
In order to describe fractional D1-branes on ${\bf X}_4$, we will consider the following crepant resolution, the local $\mathbb{P}^1\times \mathbb{P}^1$ geometry:
 \be\label{local p1p1 Tot}
\t {\bf X}_4 = \mathrm{Tot}(\mathcal{O}(-1,-1)\oplus\mathcal{O}(-1,-1)\rightarrow
\mathbb{P}^1\times \mathbb{P}^1)~.
\ee
We can construct a set of fractional branes on \eqref{local p1p1 Tot} in terms of a strongly exceptional collection on $\mathbb{P}^1\times \mathbb{P}^1$, similarly to the local $\mathbb{P}^3$ example. We choose the collection $\{\CO(-1,-1), \CO(0,-1), \CO(-1,0), \CO\}$ on $\mathbb{P}^1\times \mathbb{P}^1$.
The corresponding fractional branes on the resolved singularity \eqref{local p1p1 Tot}  are then  given by:
\be
\CE_0= i_\ast \CO~, \quad
\CE_1= i_\ast \CO(-1,0)[1]~, \quad
\CE_2= i_\ast \CO(0,-1)[1]~, \quad
\CE_3= i_\ast \CO(-1,-1)[2]~.
\ee
As before, $i$ denotes the embedding $i:\mathbb{P}^1\times
\mathbb{P}^1\hookrightarrow \t {\bf X}_4$. 
The normal bundle of $\mathbb{P}^1\times \mathbb{P}^1$ in $\t {\bf X}_4$ is $\mathcal
{N}=\mathcal{O}(-1,-1)\oplus\mathcal{O}(-1,-1)$, thus
$\wedge^2\mathcal {N}=\mathcal{O}(-2,-2)$, from which we can compute:
\bea
&\mathrm{Ext}^i(i_*\mathcal{O}(-1,-1)[2],i_*\mathcal{O}(0,-1)[1])=\mathbb{C}^2 \delta^{i1}~, \\
&\mathrm{Ext}^i(i_*\mathcal{O}(-1,-1)[2],i_*\mathcal{O}(-1,0)[1])=\mathbb{C}^2 \delta^{i1}~, \\
&\mathrm{Ext}^i(i_*\mathcal{O}(-1,-1)[2],i_*\mathcal{O})=\mathbb{C}^2 \delta^{i3}\oplus \mathbb{C}^4 \delta^{i2}~, \\
&\mathrm{Ext}^i(i_*\mathcal{O}(0,-1)[1],i_*\mathcal{O}(-1,0)[1])=\mathbb{C}^2 \delta^{i2}~,\\
&\mathrm{Ext}^i(i_*\mathcal{O}(0,-1)[1],i_*\mathcal{O})=\mathbb{C}^2 \delta^{i1}~, \\
&\mathrm{Ext}^i(i_*\mathcal{O}(-1,0)[1],i_*\mathcal{O})=\mathbb{C}^2 \delta^{i1}~, \\
&\mathrm{Ext}^i(i_*\mathcal{O}(-1,-1)[2],i_*\mathcal{O}(-1,-1)[2])=\mathrm{Ext}^i_X(i_*\mathcal{O}(0,-1)[1],i_*\mathcal{O}(0,-1)[1])\\
&\qquad=\mathrm{Ext}^i(i_*\mathcal{O}(-1,0)[1],i_*\mathcal{O}(-1,0)[1])=\mathrm{Ext}^i_X(i_*\mathcal{O},i_*\mathcal{O})=
\mathbb{C}\delta^{i0} \oplus \mathbb{C}\delta^{i4}~.
\eea
The corresponding $\Ext^{1,2}$ quiver diagram reads:
\[
\xymatrix{
 & & \mathcal{O}(-1,0)[1] \ar[dldl]^2 \ar@{--}[d]^2 & & \\
 & & \mathcal{O}(0,-1)[1] \ar[ldl]^{2} & & \\
 \mathcal{O} \ar[rrrr]^2 \ar@/_1pc/@{--}[rrrr]_4 & & & &
 \mathcal{O}(-1,-1)[2] \ar[lul]^2 \ar[lulu]^2}
\]
where the solid lines represent $\Ext^1$ elements, the dashed lines
represent $\Ext^2$ elements, and the number labeling each line is the
corresponding degeneracy.

\subsubsection{The $\Ext$ algebra on a local $\mathbb{P}^1\times \mathbb{P}^1$}\label{subsec: Ext P1P1}
Let us compute the $A_\infty$ structure satisfied by the $\Ext$ group elements. If we
denote by $x_0, x_1$ the homogeneous coordinates on the first
$\mathbb{P}^1$ and $y_0, y_1$ the homogeneous coordinates on the
second $\mathbb{P}^1$, then $\t {\bf X}_4$ can be covered by four open sets
$U_{ij}, i,j = 0,1$, defined by
\[
U_{ij}=\{x_i \neq 0, y_j \neq 0\}
\]
We also define local coordinates $x = x_1/x_0, w = x_0/x_1,
u=y_1/y_0, v=y_0/y_1$ in the corresponding open sets, and define
$y_{ij},z_{ij}$ to be the coordinates of the fibers in $U_{ij}$.
Thus, we have the transition functions $y_{01}=u y_{00}, z_{11}=xu z_{00}$, and so forth. 
We have the following Koszul resolutions of the fractional branes:
\[
\begin{split}
&0 \xrightarrow[]{} \mathcal{O}(2,2) \xrightarrow[]{\tiny
\left(\begin{array}{c}-z_{00}\\y_{00}\end{array}\right)}
\mathcal{O}(1,1)\oplus\mathcal{O}(1,1) \xrightarrow[]{\tiny
\left(\begin{array}{cc}y_{00}&z_{00}\end{array}\right)} \mathcal{O}
\xrightarrow[]{} i_* \mathcal{O}_{\mathbb{P}^1\times \mathbb{P}^1}
\xrightarrow[]{} 0\\
&0 \xrightarrow[]{} \mathcal{O}(1,2)
\xrightarrow[]{\tiny\left(\begin{array}{c}-z_{00}\\y_{00}\end{array}\right)}
\mathcal{O}(0,1)\oplus\mathcal{O}(0,1) \xrightarrow[]{\tiny
\left(\begin{array}{cc}y_{00}&z_{00}\end{array}\right)}
\mathcal{O}(-1,0) \xrightarrow[]{} i_*
\mathcal{O}_{\mathbb{P}^1\times \mathbb{P}^1}(-1,0)
\xrightarrow[]{} 0\\
&0 \xrightarrow[]{} \mathcal{O}(2,1) \xrightarrow[]{\tiny
\left(\begin{array}{c}-z_{00}\\y_{00}\end{array}\right)}
\mathcal{O}(1,0)\oplus\mathcal{O}(1,0) \xrightarrow[]{\tiny
\left(\begin{array}{cc}y_{00}&z_{00}\end{array}\right)}
\mathcal{O}(0,-1) \xrightarrow[]{} i_*
\mathcal{O}_{\mathbb{P}^1\times \mathbb{P}^1}(0,-1)
\xrightarrow[]{} 0\\
&0 \xrightarrow[]{} \mathcal{O}(1,1) \xrightarrow[]{\tiny
\left(\begin{array}{c}-z_{00}\\y_{00}\end{array}\right)}
\mathcal{O}\oplus\mathcal{O} \xrightarrow[]{\tiny
\left(\begin{array}{cc}y_{00}&z_{00}\end{array}\right)}
\mathcal{O}(-1,-1) \xrightarrow[]{} i_*
\mathcal{O}_{\mathbb{P}^1\times \mathbb{P}^1}(-1,-1) \xrightarrow[]{}
0
\end{split}
\]
where all the bundle maps are written on coordinate patch
$U_{00}$. Every state in the $\Ext$ quiver diagram can be represented by a
chain map between two of the above complexes.  Let us introduce the notation:
\[
\xymatrix{
 & & \mathcal{O}(-1,0)[1] \ar[dldl]^a \ar@{--}[d]^{\alpha} & & \\
 & & \mathcal{O}(0,-1)[1] \ar[ldl]^b & & \\
 \mathcal{O} \ar[rrrr]^c \ar@/_1pc/@{--}[rrrr]_{\beta} & & & &
 \mathcal{O}(-1,-1)[2] \ar[lul]^d \ar[lulu]^e}
\]

\vskip0.2cm\noindent
$\mathrm{Ext}^1(i_*\mathcal{O}(-1,0)[1],i_*\mathcal{O})$ is
generated by $a_1, a_2 \in \check{C}^0({U_{ij}},\mathit{Hom}^1(i_*\mathcal{O}(-1,0)[1],i_*\mathcal{O}))$:
{\small \[
a_1\; : \qquad \begin{CD}
 \mathcal{O}(1,2) @>{}>> \mathcal{O}(0,1)\oplus\mathcal{O}(0,1) @>{}>> \mathcal{O}(-1,0)\\
@V1VV @V{\tiny \left(\begin{array}{cc} -1 & 0 \\ 0 & -1\end{array}\right)}VV @V1VV\\
\mathcal{O}(2,2) @>{}>> \mathcal{O}(1,1)\oplus\mathcal{O}(1,1)
@>{}>> \mathcal{O}
\end{CD}
\]
\[
a_2\; : \qquad \begin{CD}
\mathcal{O}(1,2) @>{}>> \mathcal{O}(0,1)\oplus\mathcal{O}(0,1) @>{}>> \mathcal{O}(-1,0)\\
@VxVV @V{\tiny \left(\begin{array}{cc} -x & 0 \\ 0 & -x\end{array}\right)}VV @VxVV\\
\mathcal{O}(2,2) @>{}>> \mathcal{O}(1,1)\oplus\mathcal{O}(1,1)
@>{}>> \mathcal{O}
\end{CD}
\]}
\vskip0.2cm\noindent
$\mathrm{Ext}^1(i_*\mathcal{O}(0,-1)[1],i_*\mathcal{O})$ is
generated by $b_1, b_2 \in \check{C}^0({U_{ij}},\mathit{Hom}^1(i_*\mathcal{O}(0,-1)[1],i_*\mathcal{O}))$:
{\small 
\[
b_1\; : \qquad\begin{CD}
\mathcal{O}(2,1) @>{}>> \mathcal{O}(1,0)\oplus\mathcal{O}(1,0) @>{}>> \mathcal{O}(0,-1)\\
@V1VV @V{\tiny \left(\begin{array}{cc} -1 & 0 \\ 0 & -1\end{array}\right)}VV @V1VV\\
\mathcal{O}(2,2) @>{}>> \mathcal{O}(1,1)\oplus\mathcal{O}(1,1)
@>{}>> \mathcal{O}
\end{CD}
\]
and
\[
b_2\; : \qquad \begin{CD}
\mathcal{O}(2,1) @>{}>> \mathcal{O}(1,0)\oplus\mathcal{O}(1,0) @>{}>> \mathcal{O}(0,-1)\\
@VuVV @V{\tiny \left(\begin{array}{cc} -u & 0 \\ 0 & -u\end{array}\right)}VV @VuVV\\
\mathcal{O}(2,2) @>{}>> \mathcal{O}(1,1)\oplus\mathcal{O}(1,1)
@>{}>> \mathcal{O}
\end{CD}
\]}
\vskip0.2cm\noindent
The representatives for the $\Ext^1$ elements $d_{1, 2}$ and $e_{1, 2}$ are given by same maps as in  $a_{1, 2}$ and $b_{1, 2}$, respectively.
\vskip0.2cm\noindent
$\mathrm{Ext}^2(i_*\mathcal{O}(-1,0)[1],i_*\mathcal{O}(0,-1)[1])$ is generated by
$\alpha_i \in \check{C}^1({U_{ij}},\mathit{Hom}^1(i_*\mathcal{O}(-1,0),i_*\mathcal{O}(0,-1)))$, $i=1,2$:
{\small \[
\alpha_1\; : \qquad \begin{CD}
@. \mathcal{O}(1,2) @>{}>> \mathcal{O}(0,1)\oplus\mathcal{O}(0,1) @>{}>> \mathcal{O}(-1,0) \\
@. @V{\tiny \left(\begin{array}{c} \alpha \\ 0 \end{array}\right)}VV
@V{\tiny \left(\begin{array}{cc} 0 & -\alpha \end{array}\right)}VV @.\\
\mathcal{O}(2,1) @>{}>> \mathcal{O}(1,0)\oplus\mathcal{O}(1,0)
@>{}>> \mathcal{O}(0,-1) @.
\end{CD}
\]
\[
\alpha_2\; : \qquad  \begin{CD}
@. \mathcal{O}(1,2) @>{}>> \mathcal{O}(0,1)\oplus\mathcal{O}(0,1) @>{}>> \mathcal{O}(-1,0) \\
@. @V{\tiny \left(\begin{array}{c} 0 \\ \alpha \end{array}\right)}VV
@V{\tiny \left(\begin{array}{cc} \alpha & 0 \end{array}\right)}VV @.\\
\mathcal{O}(2,1) @>{}>> \mathcal{O}(1,0)\oplus\mathcal{O}(1,0)
@>{}>> \mathcal{O}(0,-1) @.
\end{CD}
\]}
with:
\[
(\alpha)_{01}=(\alpha)_{02}=u^{-1}~,\qquad 
(\alpha)_{13}=(\alpha)_{23}=-u^{-1}~,\qquad  (\alpha)_{03}=(\alpha)_{12}=0~.
\]
The Serre dual elements to $\alpha_{1, 2}$, denoted by $\alpha'_{1,2}$, can be defined in  the following way:
{\small \[
\alpha'_1\; : \qquad  \begin{CD}
@. \mathcal{O}(2,1) @>{}>> \mathcal{O}(1,0)\oplus\mathcal{O}(1,0) @>{}>> \mathcal{O}(0,-1) \\
@. @V{\tiny \left(\begin{array}{c} \alpha' \\ 0
\end{array}\right)}VV
@V{\tiny \left(\begin{array}{cc} 0 & -\alpha' \end{array}\right)}VV @.\\
\mathcal{O}(1,2) @>{}>> \mathcal{O}(0,1)\oplus\mathcal{O}(0,1)
@>{}>> \mathcal{O}(-1,0) @.
\end{CD}
\]
\[
\alpha'_2\; : \qquad  \begin{CD}
@. \mathcal{O}(2,1) @>{}>> \mathcal{O}(1,0)\oplus\mathcal{O}(1,0) @>{}>> \mathcal{O}(0,-1) \\
@. @V{\tiny \left(\begin{array}{c} 0 \\ \alpha'
\end{array}\right)}VV
@V{\tiny \left(\begin{array}{cc} \alpha' & 0 \end{array}\right)}VV @.\\
\mathcal{O}(1,2) @>{}>> \mathcal{O}(0,1)\oplus\mathcal{O}(0,1)
@>{}>> \mathcal{O}(-1,0) @.
\end{CD}
\]}
with:
\[
(\alpha')_{02}=(\alpha)_{03}=(\alpha')_{12}=(\alpha')_{13}=x^{-1}~,\quad 
(\alpha')_{01}=(\alpha')_{23}=0~.
\]
\vskip0.2cm\noindent
$\mathrm{Ext}^1(i_*\mathcal{O}, i_*\mathcal{O}(-1,-1)[2])$ is
generated by:
{\small\[
c_1\; : \qquad  \begin{CD}
@. \mathcal{O}(2,2) @>{}>> \mathcal{O}(1,1)\oplus\mathcal{O}(1,1) @>{}>> \mathcal{O} \\
@. @V{\tiny \left(\begin{array}{c} c \\ 0 \end{array}\right)}VV
@V{\tiny \left(\begin{array}{cc} 0 & -c \end{array}\right)}VV @.\\
\mathcal{O}(1,1) @>{}>> \mathcal{O}\oplus\mathcal{O} @>{}>>
\mathcal{O}(-1,-1) @.
\end{CD}
\]
\[
c_2\; : \qquad \begin{CD}
@. \mathcal{O}(2,2) @>{}>> \mathcal{O}(1,1)\oplus\mathcal{O}(1,1) @>{}>> \mathcal{O} \\
@. @V{\tiny \left(\begin{array}{c} 0 \\ c \end{array}\right)}VV
@V{\tiny \left(\begin{array}{cc} c & 0 \end{array}\right)}VV @.\\
\mathcal{O}(1,1) @>{}>> \mathcal{O}\oplus\mathcal{O} @>{}>>
\mathcal{O}(-1,-1) @.
\end{CD}
\]}
\vskip0.2cm\noindent
Closedness requires:
\[
(c)_{123}+(c)_{013}=(c)_{023}+(c)_{012}.
\]
If the two sides of the above identity were both zero, $c$ would be
exact. We deduce that one of $(c)_{013}$ and $(c)_{123}$ is $\pm
x^{-1} u^{-1}$ and the other is zero, and similarly for $(c)_{023}$ and
$(c)_{012}$. Different choices only differ by exact terms and sign
convention. In the following, we will fix:
\[
(c)_{012}=(c)_{013}=0~,\qquad  (c)_{023}=(c)_{123}=x^{-1}u^{-1}.
\]
\vskip0.2cm\noindent
$\mathrm{Ext}^2(i_*\mathcal{O}, i_*\mathcal{O}(-1,-1)[2])$ is generated by the elements $\beta_1, \cdots, \beta_4$:
{\small \[
\begin{CD}
@. @. \mathcal{O}(2,2) @>{}>> \mathcal{O}(1,1)\oplus\mathcal{O}(1,1) @>{}>> \mathcal{O} \\
@. @. @V{\beta}VV @. @.\\
\mathcal{O}(1,1) @>{}>> \mathcal{O}\oplus\mathcal{O} @>{}>>
\mathcal{O}(-1,-1) @. @.
\end{CD}
\]}
$\beta_1$ is defined by
\[
\begin{split}
(\beta_1)_{123}+(\beta_1)_{013}=x^{-1}u^{-1}~,\qquad &(\beta_1)_{123}
(\beta_1)_{013} = 0~, \\
(\beta_1)_{012}+(\beta_1)_{023}=x^{-1}u^{-1}~,\qquad &(\beta_1)_{012}
(\beta_1)_{023} = 0~.
\end{split}
\]
Again, different choices do not affect the cohomology class they
represent. $\beta_2$, $\beta_3$ and $\beta_4$ are defined similarly
with $x^{-1}u^{-1}$ replaced by $x^{-2}u^{-1}$ for $\beta_2$,
$x^{-1}u^{-2}$ for $\beta_3$ and $x^{-2}u^{-2}$ for $\beta_4$.
The Serre dual elements are given by $\beta'_1, \cdots, \beta'_4$ defined by
{\small \[
\beta'_1\; : \qquad \begin{CD}
\mathcal{O}(1,1) @>{}>> \mathcal{O}\oplus\mathcal{O} @>{}>> \mathcal{O}(-1,-1)\\
@V1VV @V{\tiny \left(\begin{array}{cc} 1 & 0 \\ 0 & 1\end{array}\right)}VV @V1VV\\
\mathcal{O}(2,2) @>{}>> \mathcal{O}{1,1}\oplus\mathcal{O}(1,1)
@>{}>> \mathcal{O}
\end{CD}
\]
\[
\beta'_2\; : \qquad  \begin{CD}
\mathcal{O}(1,1) @>{}>> \mathcal{O}\oplus\mathcal{O} @>{}>> \mathcal{O}(-1,-1)\\
@VxVV @V{\tiny \left(\begin{array}{cc} x & 0 \\ 0 & x\end{array}\right)}VV @VxVV\\
\mathcal{O}(2,2) @>{}>> \mathcal{O}{1,1}\oplus\mathcal{O}(1,1)
@>{}>> \mathcal{O}
\end{CD}
\]
\[
\beta'_3\; : \qquad  \begin{CD}
\mathcal{O}(1,1) @>{}>> \mathcal{O}\oplus\mathcal{O} @>{}>> \mathcal{O}(-1,-1)\\
@VuVV @V{\tiny \left(\begin{array}{cc} u & 0 \\ 0 & u \end{array}\right)}VV @VuVV\\
\mathcal{O}(2,2) @>{}>> \mathcal{O}{1,1}\oplus\mathcal{O}(1,1)
@>{}>> \mathcal{O}
\end{CD}
\]
\[
\beta'_4\; : \qquad  \begin{CD}
\mathcal{O}(1,1) @>{}>> \mathcal{O}\oplus\mathcal{O} @>{}>> \mathcal{O}(-1,-1)\\
@VxuVV @V{\tiny \left(\begin{array}{cc} xu & 0 \\ 0 & xu \end{array}\right)}VV @VxuVV\\
\mathcal{O}(2,2) @>{}>> \mathcal{O}{1,1}\oplus\mathcal{O}(1,1)
@>{}>> \mathcal{O}
\end{CD}
\]}
\vskip0.2cm\noindent
The generator of $\Ext^4$ at each node has the following form:
{\small \[
\xymatrix { \mathcal{O}(m+2,n+2) \ar[drr]^{t} \ar[r] &
\mathcal{O}(m+1,n+1)\oplus\mathcal{O}(m+1,n+1) \ar[r] &
\mathcal{O}(m,n) \\
\mathcal{O}(m+2,n+2) \ar[r] &
\mathcal{O}(m+1,n+1)\oplus\mathcal{O}(m+1,n+1) \ar[r] &
\mathcal{O}(m,n) }
\] }
with
\[
\begin{split}
(t)_{123}+(t)_{013}=x^{-1}u^{-1}~,\qquad &(t)_{123}
(t)_{013} = 0~, \\
(t)_{012}+(t)_{023}=x^{-1}u^{-1}~,\qquad &(t)_{012} (t)_{023} = 0~.
\end{split}
\]
\vskip0cm\noindent
It can be shown that $\alpha'_1$ is Serre dual to $\alpha_2$,
$\alpha'_2$ is Serre dual to $-\alpha_1$ and $\beta'_i$ is Serre
dual to $\beta_i$:
\be\nn
m_2(\alpha_1', \alpha_2)= t~, \quad 
m_2(\alpha_2', \alpha_1)= -t~, \quad 
m_2(\beta_i, \beta_j)= \delta_{ij}\, t~, \quad 
\ee

\vskip0.2cm\noindent From the composition of the chain maps, one
can  compute the products:
\begin{equation}\label{m2}
\begin{split}
&m_2(b_1, d_1) = \beta'_1~,\quad m_2(b_1, d_2) = \beta'_2~,\quad m_2(b_2, d_1) =
\beta'_3~,\quad m_2(b_2, d_2) = \beta'_4~, \\
&m_2(a_1, e_1) = \beta'_1~,\quad m_2(a_2, e_1) = \beta'_2~,\quad m_2(a_1, e_2) =
\beta'_3~,\quad m_2(a_2, e_2) = \beta'_4~,
\end{split}
\end{equation}
amongst the $\Ext^1$ elements.
 In addition, this model also has non-zero higher products, whose computation is rather more technical \cite{Aspinwall:2004bs, kad}. 
 We discuss it in Appendix \ref{app: higher prod P1P1}. One finds the non-zero products:
\bea\nn
&m_3(d_1, c_1, a_2) =- m_3(d_2, c_1, a_1)= \alpha_1~, 
&m_3(d_1, c_2, a_2) =- m_3(d_2, c_2, a_1)= \alpha_2~, \cr
&m_3(e_1, c_2, b_2)= -m_3(e_2, c_2, b_1)= -\alpha_2'~, 
&m_3(e_2, c_1, b_1)=- m_3(e_1, c_1, b_2)= \alpha_1'~, 
\eea
and:
\bea\label{m4 P1P1}
&m_4(c_1, b_2, d_2, c_2)= m_4(c_2, a_2, e_2, c_1)=-\beta_1~, \cr
& m_4(c_2, a_1, e_2, c_1)=m_4(c_1, b_2, d_1, c_2)= \beta_2~,\cr
&m_4(c_1, a_2, e_1, c_2)=m_4(c_2, b_1, d_2, c_1)= - \beta_3~, \cr
&m_4(c_2, b_1, d_1, c_1)=m_4(c_1, a_1, e_1, c_2)=\beta_4~,
\eea
 with all other products amongst the $\Ext^1$ elements vanishing.

\subsubsection{The local $\mathbb{P}^1\times \mathbb{P}^1$ quiver}
\begin{figure}[t]
\begin{center}
\includegraphics[height=5cm]{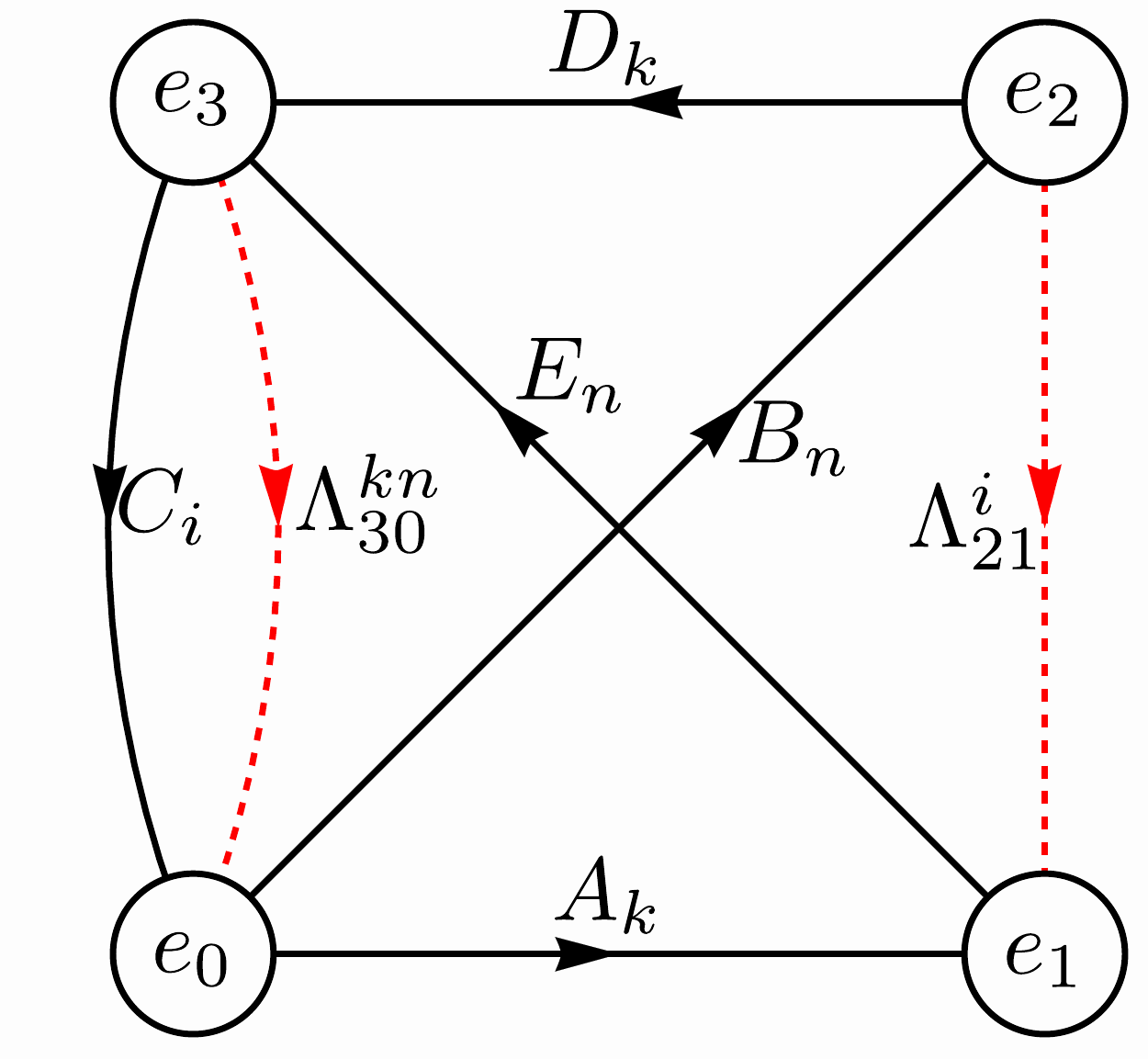}
\caption{The $C(Q^{1,1,1})$ supersymmetric quiver.}
\label{figQ111}
\end{center}
\end{figure}

Given the above result, it is straightforward to write down the corresponding quiver gauge theory, shown in Figure~\ref{figQ111}. 
From the geometric structure \eqref{Q111 def}, one would expect that the corresponding supersymmetric quiver theory has an $SU(2)^3$ global symmetry. However, the $A_\infty$ structure only preserves the minimal ``toric'' flavor symmetry $U(1)^3$, which is the apparent symmetry of the quiver gauge theory.

The  $\CN=(0,2)$ quiver has four pairs of chiral multiplet, which are identified with the above $\Ext^1$ elements according to:
\be
A_k = a_k~, \qquad B_n = -b_n~, \qquad C_i = c_i~, \qquad D_k = d_k~, \qquad E_n= e_n~,
\ee
with $k,n, i \in 1,2$. The $k$ and $n$ index are related to the $SU(2)\times SU(2)$ induced from the $\mathbb{P}^1\times \mathbb{P}^1$ geometry; however, the interaction terms break this symmetry to its maximal torus. The quiver has the fermi multiplets:
\be
\Lambda_{21}^i = \left(\Lambda_{21}^1, \Lambda_{21}^2)= (\alpha_1, \alpha_2\right)~, \qquad
\Lambda_{03}^{kn}= \left(\Lambda_{03}^{11},\Lambda_{03}^{12},\Lambda_{03}^{21}, \Lambda_{03}^{22}\right)= (\beta_1, \beta_3, \beta_2, \beta_4)~,
\ee
which are identified with the $\Ext^2$ elements as indicated.
From the $A_\infty$ product structure discussed above, we find the interaction terms:
\bea
& J_{\Lambda_{21}^1} =E_2 C_2 B_1- E_1 C_2 B_2~, \qquad 
&& E_{\Lambda_{21}^1} = D_1 C_1 A_2-D_2 C_1 A_1~,\cr
& J_{\Lambda_{21}^2} =E_1 C_1 B_2- E_2 C_1 B_1~, \qquad 
&& E_{\Lambda_{21}^2} = D_1 C_2 A_2-D_2 C_2 A_1~,
\eea
and
\bea
& J_{\Lambda_{30}^{11}} = A_1 E_1 - B_1 D_1 ~, \qquad
&& E_{\Lambda_{30}^{11}} = C_1 B_2 D_2 C_2-C_2 A_2 E_2 C_1~, \cr
& J_{\Lambda_{30}^{12}} = A_1 E_2 -B_2 D_1~, \qquad
&& E_{\Lambda_{30}^{12}} = C_2 B_1 D_2 C_1-C_1 A_2 E_1 C_2 ~, \cr
& J_{\Lambda_{30}^{21}} = A_2 E_1 -B_1 D_2~, \qquad
&& E_{\Lambda_{30}^{21}} = C_2 A_1 E_2 C_1-C_1 B_2 D_1 C_2~, \cr
& J_{\Lambda_{30}^{22}} = A_2 E_2 -B_2 D_2~, \qquad
&& E_{\Lambda_{30}^{22}} = C_1 A_1 E_1 C_2-C_2 B_1 D_1 C_1~.
\eea
This satisfies $\Tr(EJ)=0$ and agrees with  \cite{Franco:2015tna}.


\section{Triality and mutations of exceptional collections}\label{sec: duality}

For some D3-brane quiver theories, it was proposed long ago that Seiberg duality in the gauge theory can be understood in terms of {\it mutations} of the underlying branes \cite{Cachazo:2001sg}. More precisely, for a singularity  ${\bf X}_3$ whose crepant resolution $\t {\bf X}_3$ is the total space of the canonical line bundle over a del Pezzo surface $B_2$, we can construct the fractional branes on $\t {\bf X}_3$ in terms of an exceptional collection of B-branes on $B_2$ \cite{Cachazo:2001sg, Wijnholt:2002qz, Herzog:2003zc}, and Seiberg dualities can be realized as mutations of the exceptional collection \cite{Herzog:2004qw}. (See Appendix~\ref{sect:warmup:local-p2} for an explicit example.)

We may consider the Calabi-Yau fourfold analogue of this setup, which involves the singularity ${\bf X}_4$ whose crepant resolution is 
$\t {\bf X}_4  = {\rm Tot}(\CK \rightarrow B_3)$, 
with $B_3$ a Fano threefold and $\CK$ its canonical line bundle. The fractional branes on ${\bf X}_4$  can be similarly constructed from the data of an exceptional collection $\{\CE\}$ of sheaves on $B_3$, in principle. 
 In the previous section, we considered the simplest possible example, $B_3 = {\mathbb P}^3$.
   A mutation of the exceptional collection gives another exceptional collection $\{\CE'\}$, and we can again consider the corresponding $\CN=(0,2)$ quiver gauge theory. It is natural to suspect that the geometric operation amounts to a field theory {\it duality } between the different $\CN=(0,2)$ quiver gauge theories. 
A well-studied example~\footnote{
Other two-dimensional dualities are also known amongst $\CN=(2,2)$ and $\CN=(0,2)$ gauge
theories, see for example \cite{Hori:2011pd,Jia:2014ffa}.
} of an $\CN=(0,2)$ gauge theory duality is the {\it triality} of Gadde, Gukov and Putrov (GGP) \cite{Gadde:2013lxa}. We will show, in the simplest example of local ${\mathbb P}^3$, that  indeed {\it   mutation is triality}. This obviously deserves further study, which we leave for future work.

\subsection{Triality acting on $\CN=(0,2)$ supersymmetric quivers}\label{subsec: triality review}
\begin{figure}[t]
\begin{center}
\subfigure[\small Theory $(i)$.]{
\includegraphics[height=5cm]{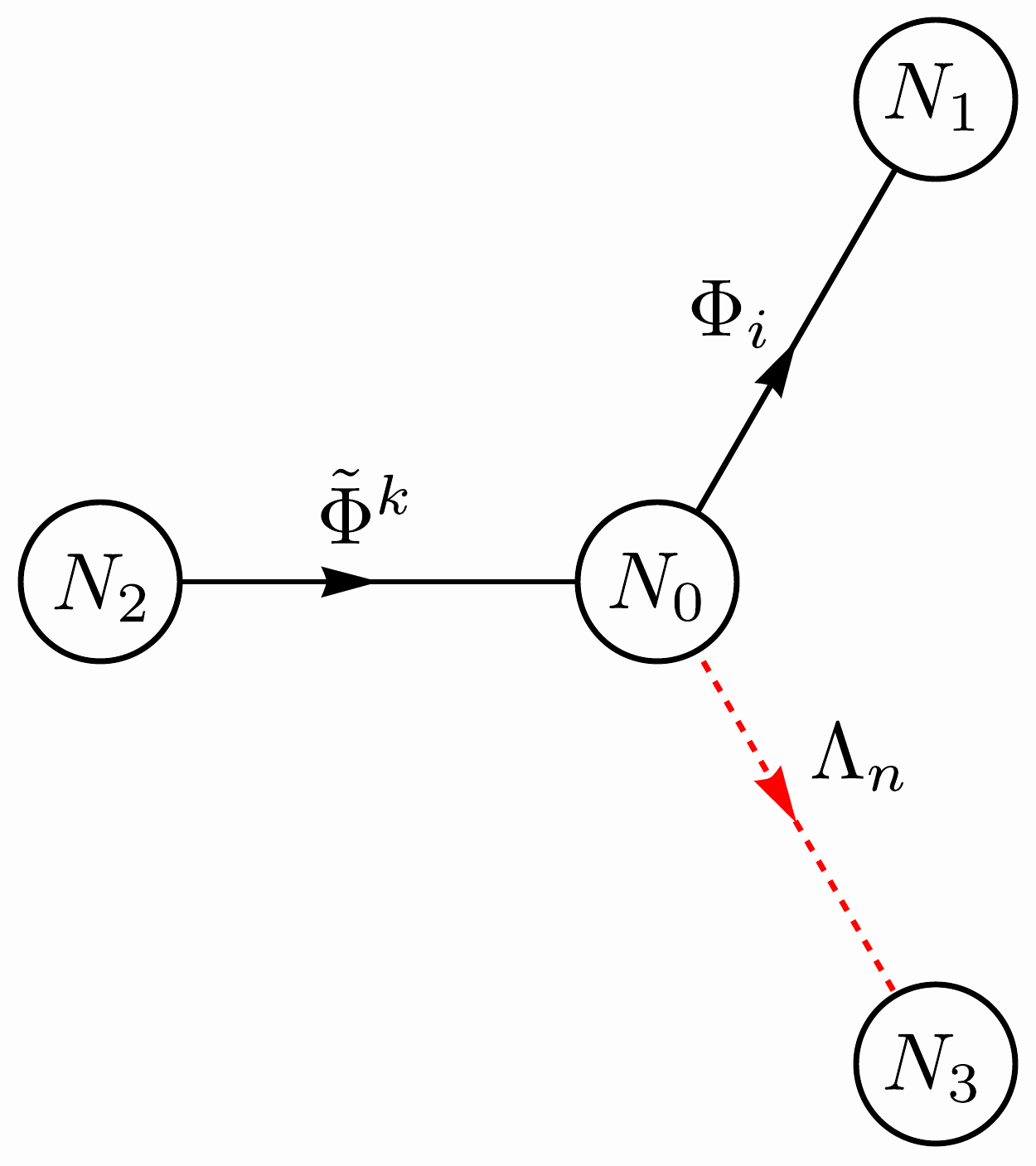}\label{fig:triality i}}\quad\;
\subfigure[\small Theory $(ii)$.]{
\includegraphics[height=5cm]{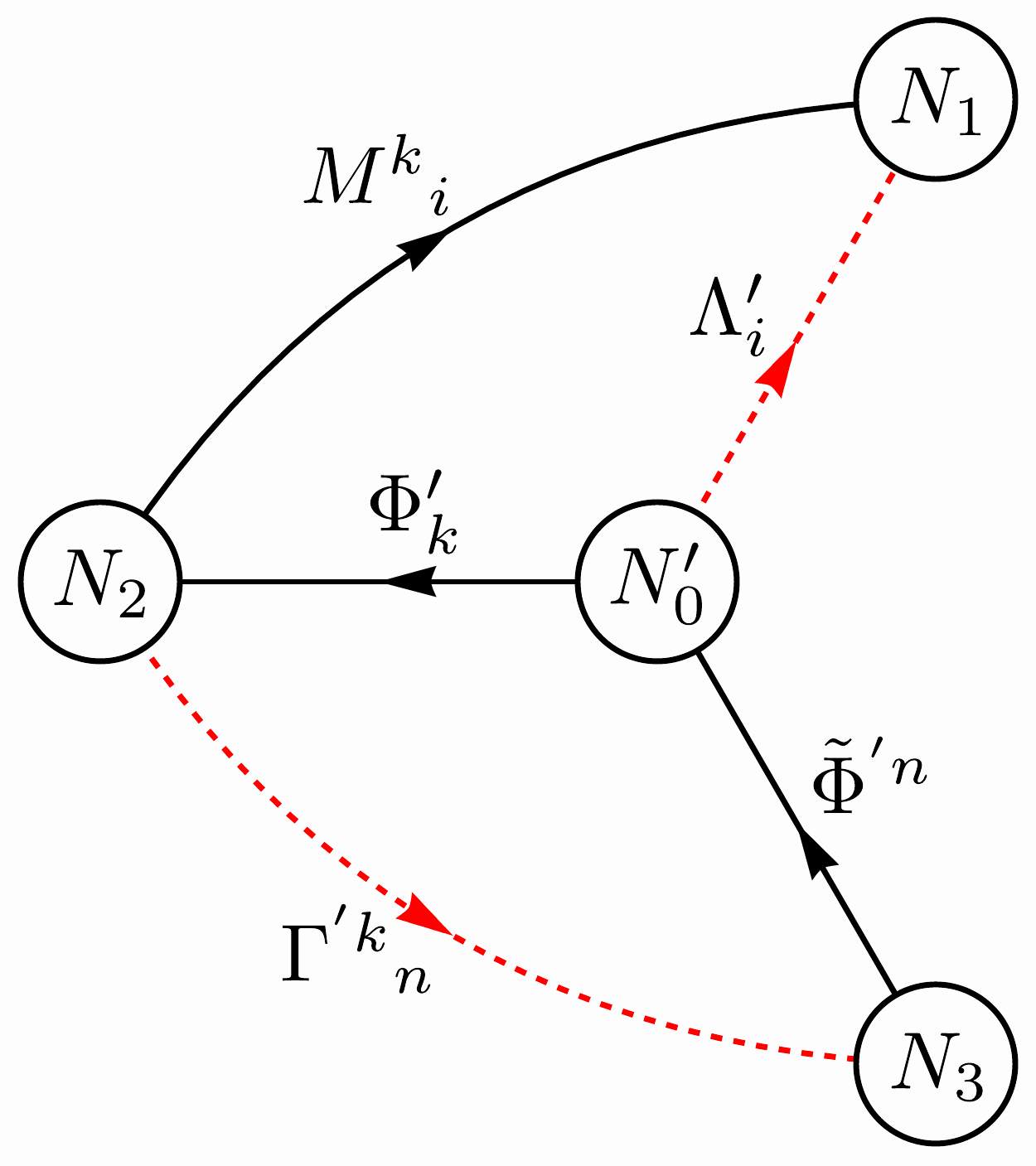}\label{fig:triality ii}}\quad\;
\subfigure[\small Theory $(iii)$.]{
\includegraphics[height=5cm]{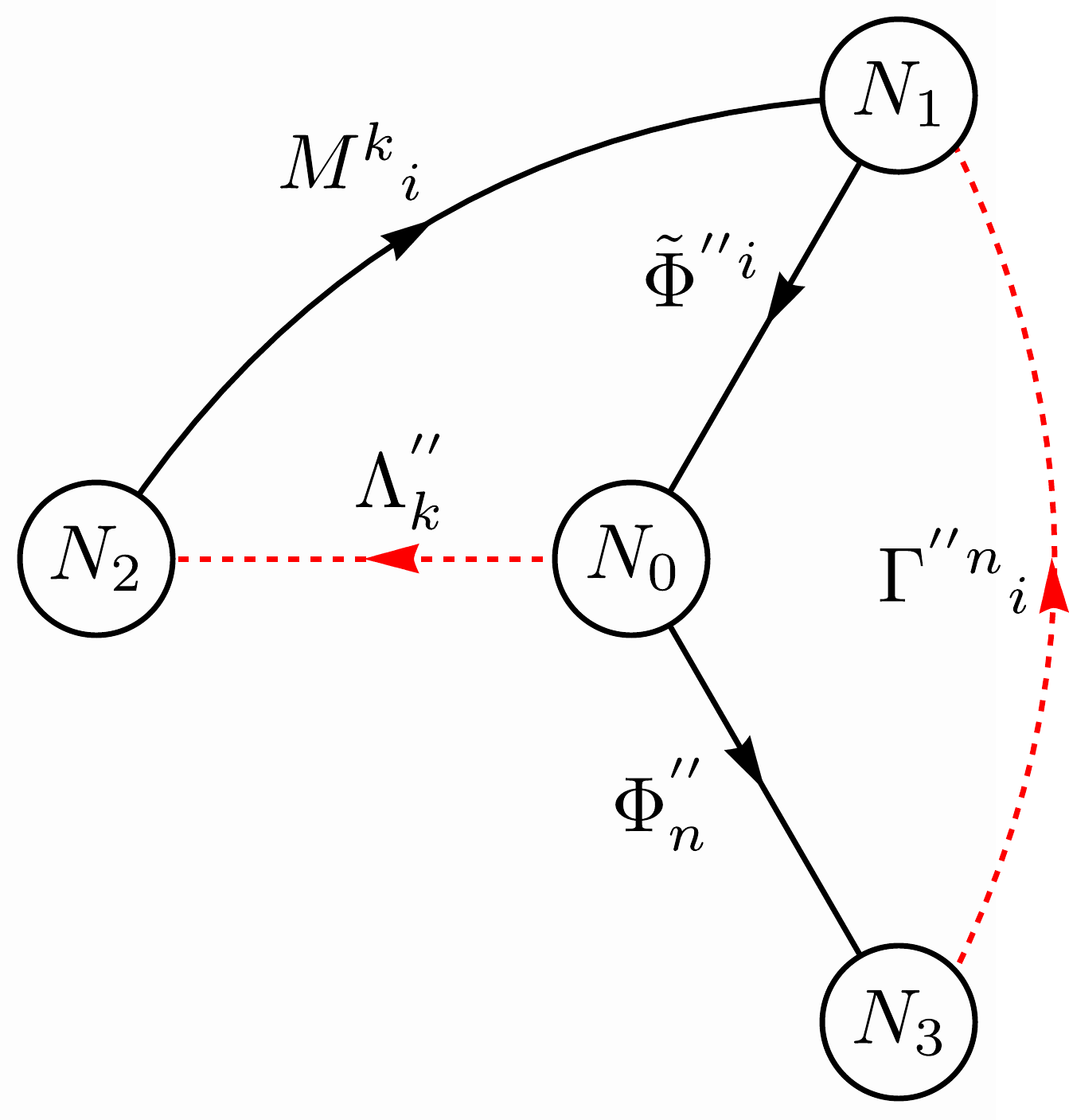}\label{fig:triality iii}}
\caption{Triality as a local operation at a quiver node. The transformation of the interaction terms under triality is discussed in the main text.}
\label{fig: triality review}
\end{center}
\end{figure}

Let us first review GGP triality and its action on quiver gauge theories \cite{Gadde:2013lxa}. The {\it triality transformation} can be formulated as a local operation at a single node $e_0$ of an $\CN=(0,2)$ supersymmetric quiver without adjoint matter fields, as depicted in Figure~\ref{fig: triality review}. 
The central node $e_0$ is a $U(N_0)$ gauge group, while the nodes $e_1, e_2, e_3$ realize a  ``flavor'' group $U(N_1)\times U(N_2) \times U(N_3)$ from the point of view of $U(N_0)$.~\footnote{For simplicity, we write down a single arrow $e_0 \rightarrow e_i$ ($i=1,2,3$) for the matter fields of the $U(N_0)$ gauge group at node $e_0$. In general, the ``effective flavor group'' $U(N_i)$ at the node $e_0$ corresponds to a combination of both quiver gauge groups and actual flavor symmetries, which may be broken explicitly by interaction terms. We choose the slightly schematic depiction of Figure~\protect\ref{fig: triality review} to avoid clutter.}
In the ``original'' theory, shown in Figure~\ref{fig:triality i}, we have chiral multiplets $\Phi^i$ in the fundamental representation of $U(N_0)$, chiral multiplets $\t\Phi_k$ in the antifundamental representation of $U(N_0)$, and fermi multiplets $\Lambda_n$ in the fundamental representation of $U(N_0)$. (The flavor indices $i, k, n$ run over $i=1, \cdots, N_1$;  $k=1, \cdots, N_2$;  $n= 1, \cdots, N_3$.) 
We must have:
\be
2 N_0= N_1 + N_2- N_3~, 
\ee
to cancel the non-abelian gauge anomaly.
The theory can also have non-trivial interaction terms. Let $\Xi$  and $X$ denote any additional fermi and chiral multiplets, respectively, distinct from $\Lambda$ and $\Phi, \t \Phi$, in any larger $\CN=(0,2)$ quiver in which Figure~\ref{fig:triality i} might be embedded. We have:
\be\label{E and J th i}
E_{\Xi}(X, \t\Phi \Phi)~, \qquad J_{\Xi}(X, \t\Phi\Phi)~, \qquad   E_{\Lambda}(X, \Phi, \t\Phi)~, \qquad 
  J_{\Lambda}(X, \Phi, \t\Phi)~,
\ee
which must be such that $\Tr(EJ)=\Tr( \sum_{\Xi} E_\Xi J_\Xi + \sum_\Lambda E_\Lambda J_\Lambda)=0$.

The ``triality move'' can be described as follows: Given  the above Theory $(i)$ with gauge group $U(N_0)$, we obtain Theory $(ii)$ as shown in Figure~\ref{fig:triality ii}. 
The dual gauge group is $U(N_0')$ with dual rank given by the number of antifundamental chiral multiplets minus $N_0$:
\be
N_0' \equiv N_2 - N_0~.
\ee
The dual charged matter fields in chiral and fermi multiplets, denoted by $\Phi'_k$, ${\t\Phi}^{'n}$ and $\Lambda'_i$, transform under the ``flavor'' group as indicated on the Figure. In addition, the new theory also contains some ``mesonic fields'' ${M^k}_i$ and ${\Gamma^{' k}}_n$. Those fields are identified with the following $U(N_0)$-invariant combinations of matter fields in Theory $(i)$:
\be
{M^k}_i = \t \Phi^k \Phi_i~, \qquad \qquad {\Gamma^{' k}}_n = \t \Phi^k \Lambda_n~.
\ee
To fully specify the new theory, we need to determine the new interaction terms. Given that the original theory has interaction terms \eqref{E and J th i}, the interaction terms for any ``spectator'' fermi multiplet $\Xi$ are obtained by substituting $\t\Phi\Phi= M$ inside $E_{\Xi}$ and $J_{\Xi}$:
\be\label{E and J th ii}
 E_{\Xi}(X, M)~,  \qquad J_{\Xi}(X, M)~.
\ee
In addition,  the interaction terms of the new fermi multiplets $\Lambda'$ and $\Gamma'$ are given by:~\footnote{These transformation rules were left implicit in most of the literature. They where recently studied explicitly in \cite{Franco:2017lpa}.}
\bea\label{E and J th ii Full}
&E_{\Lambda'_i}= \Phi'_k  {M^k}_i~, \qquad \qquad 
&&E_{{\Gamma^{'k}}_n }= - {M^k}_i  \left(\d E_{\Lambda_n}\ov \d \Phi_i\right)~,\cr
&J_{\Lambda'_i}= \left(\d E_{\Lambda_n}\ov \d \Phi_i\right) \, \t\Phi^{'n}~, \qquad \qquad
 && J_{{\Gamma^{'k}}_n }= \t\Phi^{'n} \Phi'_k- {\d J_{\Lambda_n}\ov \d \t\Phi_k}~,
\eea
as holomorphic functions of $X$,  $\Phi'$ and  $\t\Phi'$. One easily sees that the constraint $\Tr(EJ)=0$ is again satisfied (given that it is satisfied in the original theory).

Theory $(iii)$, shown in Figure~\ref{fig:triality iii}, is similarly obtained from Theory $(ii)$ by the {\it same} triality operation. The new gauge group is $U(N_0'')$ with:
\be
N_0''= N_3-N_0'~.
\ee
The new matter fields $\Lambda''_k$, $\Phi^{'' i}$ and $\t \Phi''_n$ are as indicated. We also have the new mesons $M'$ and $\Gamma''$, which are identified with the $U(N_0')$-invariant combinations:
\be
{M^{'n}}_k = \t \Phi^{'n}\Phi'_k~, \qquad { \Gamma^{'' n}}_i= \t \Phi^{'n}\Lambda'_i~,
\ee
in Theory $(ii)$. Applying the rules above for the interaction terms, one finds  that:
\be
J_{\Gamma^{'k}_n } = {M^{'n}}_k - {\d J_{\Lambda_n}\ov \d \t\Phi_k}~,
\ee
in particular.
This implies that $\Gamma'$ and $M'$ are both massive, and can be integrated out by imposing the linear relation $J_{\Gamma' } =0$. We are left with the mesons $M$ and $\Gamma''$ only, as shown in Figure~\ref{fig:triality iii}. 

Finally, one can check that another triality move, starting from Theory $(iii)$, gives a theory which is identical to Theory $(i)$ after integrating out all the massive fields. Thus, we confirm that the triality operation is indeed a ``duality'' of order three. More precisely, this is the case if we act repeatedly on a {\it single} node of a given $\CN=(0,2)$ quiver. If we act subsequently on different nodes, one uncovers very rich, infinite-dimensional ``triality trees''.

\subsection{Triality and the $\C^4/\Z_4$ quiver}\label{triality on C4Z4 quiver}
\begin{figure}[t]
\begin{center}
\subfigure[\small Theory (I).]{
\includegraphics[height=4.7cm]{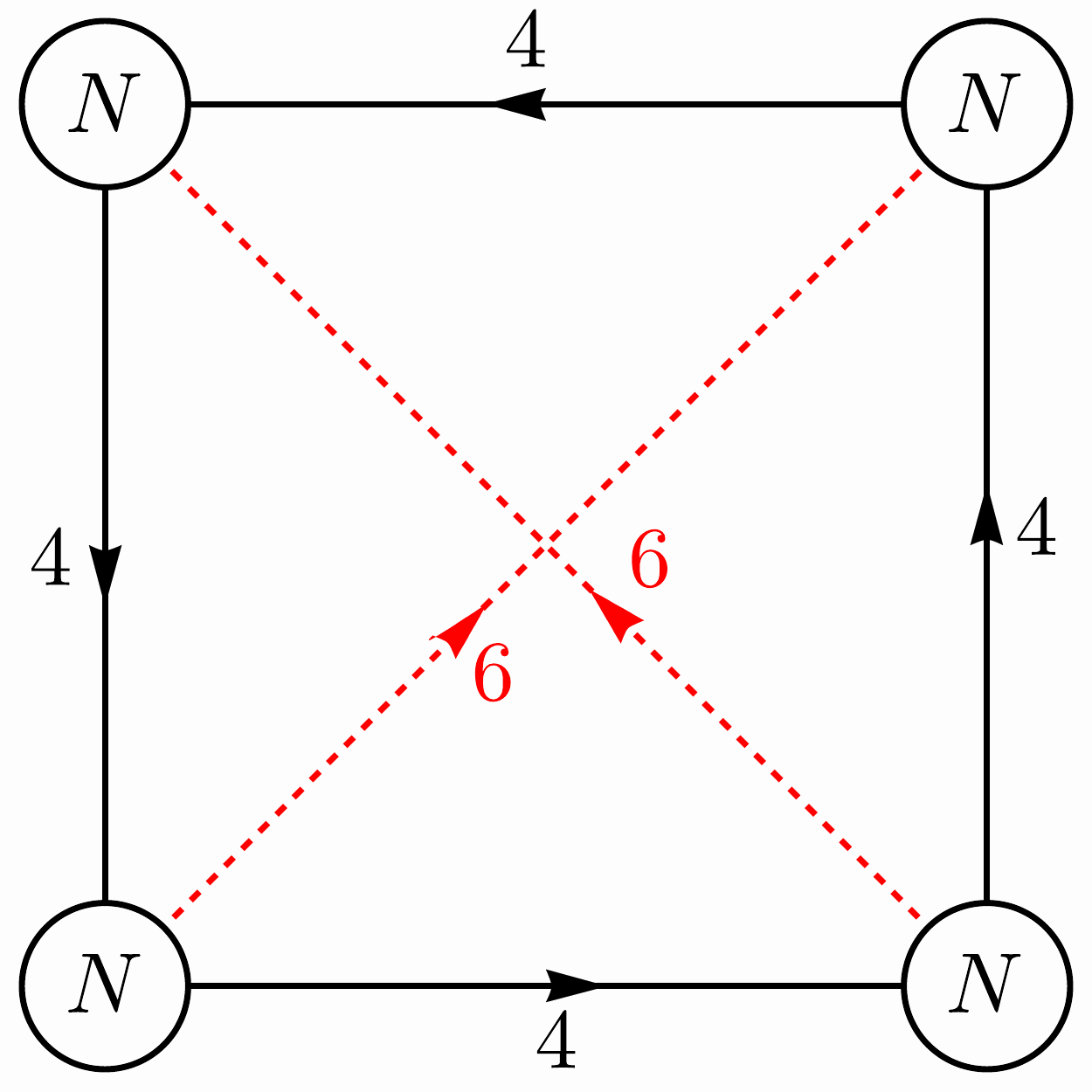}\label{fig:triality C4Z4 i}}\;\;
\subfigure[\small Theory (II).]{
\includegraphics[height=4.7cm]{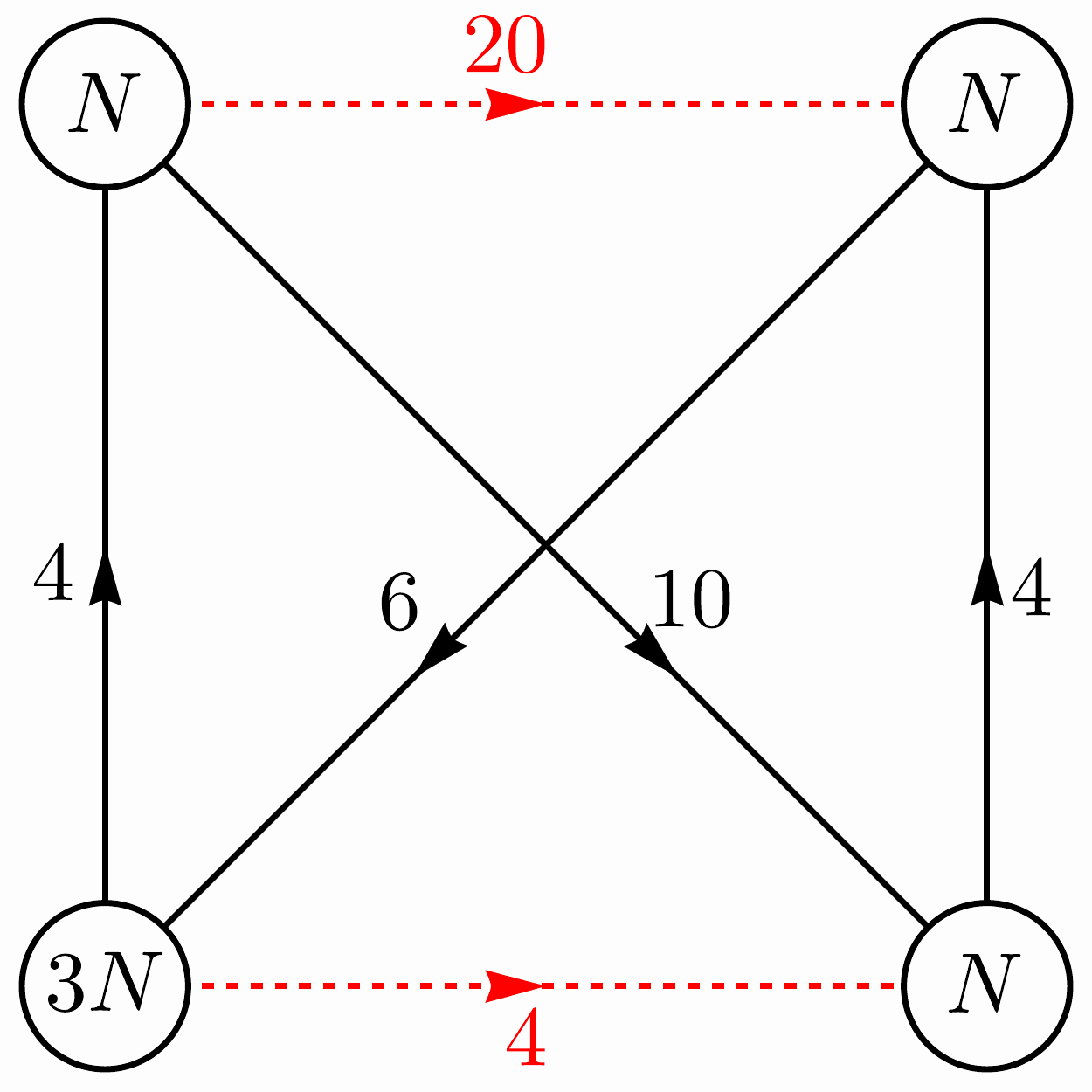}\label{fig:triality C4Z4 ii}}\;\;
\subfigure[\small Theory (III).]{
\includegraphics[height=4.7cm]{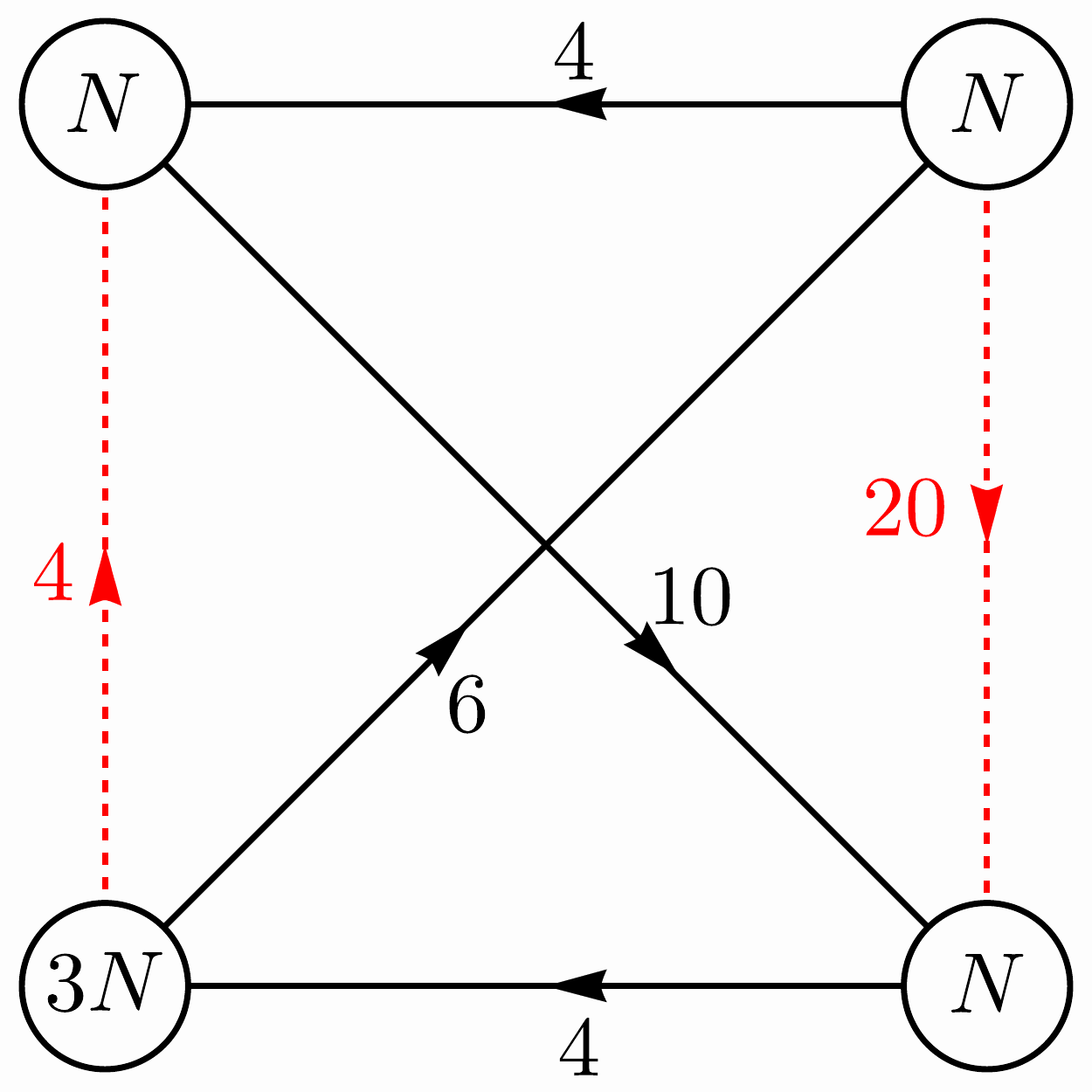}\label{fig:triality C4Z4 iii}}
\caption{Three $\CN=(0,2)$ quiver theories describing $N$ D1-branes at the resolved orbifold  $\widetilde{\C^4/\Z_4}(1,1,1,1)$. The numbers denote the multiplicities of the arrows.}
\label{fig: triality C4Z4}
\end{center}
\end{figure}
Let us now discuss an example of the triality operation on a full-fledged D1-brane quiver. Consider the $\C^4/\Z_4$ singularity with crepant resolution the local $\mathbb{P}^3$ geometry. Two distinct quiver gauge theories were derived in section \ref{sec local P3},  which we dubbed ``Theory (I)'' and ``Theory (III).'' They are reproduced in Figure~\ref{fig:triality C4Z4 i} and \ref{fig:triality C4Z4 iii}, respectively.

\begin{figure}[t]
\begin{center}
\subfigure[\small Theory (I).]{
\includegraphics[height=4.7cm]{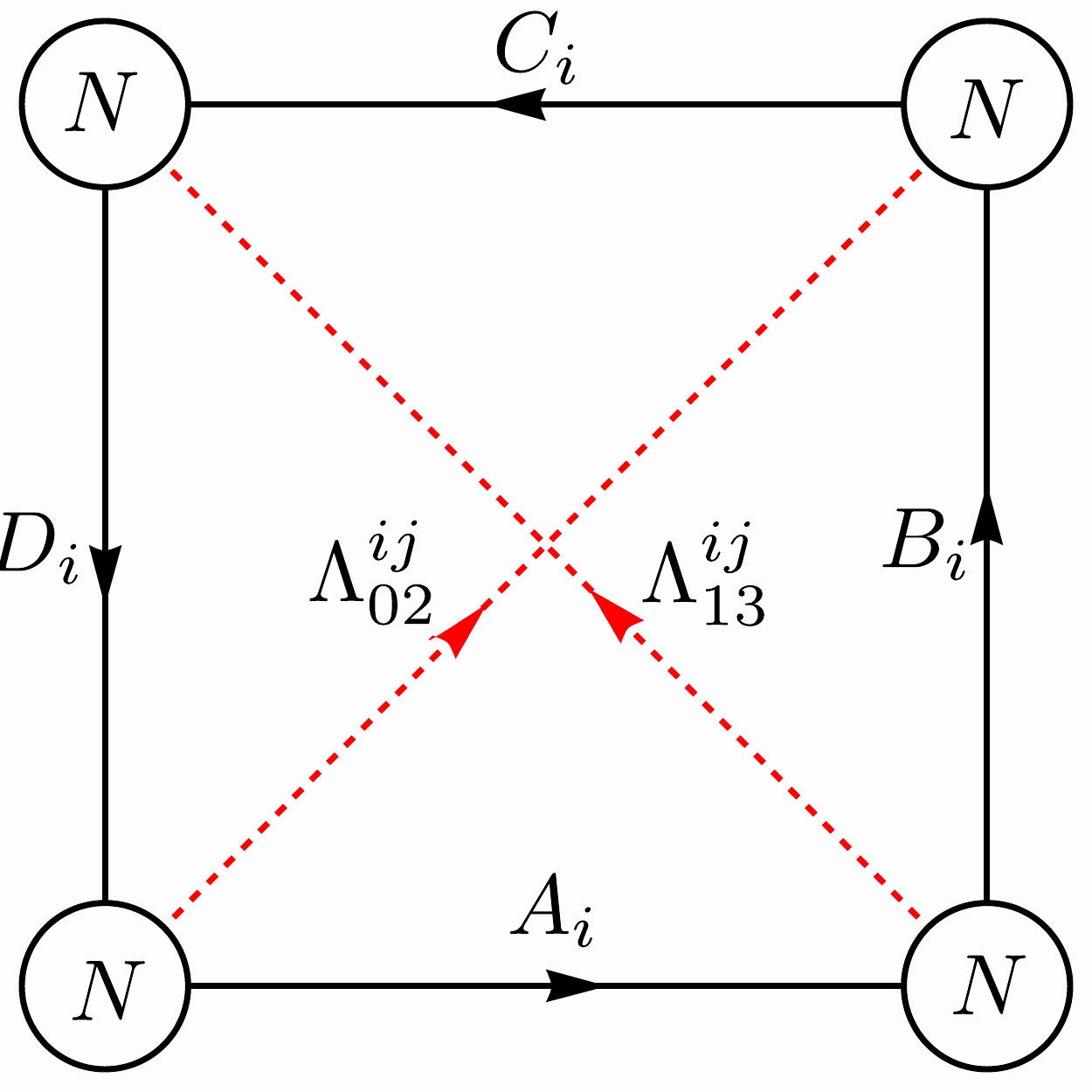}\label{fig: C4Z4 with rank I}}
\quad
\subfigure[\small Triality-transformed.]{
\includegraphics[height=4.7cm]{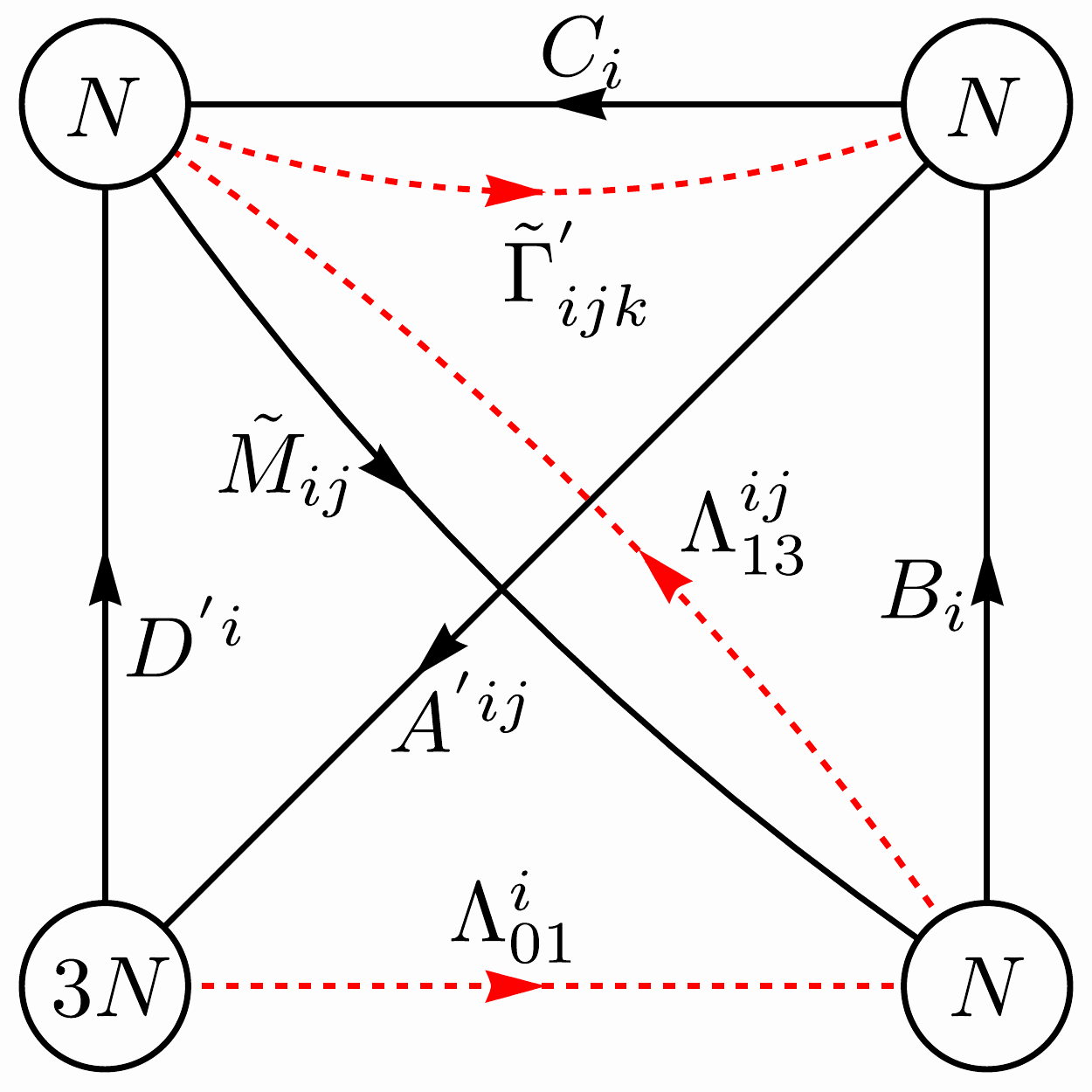}\label{fig: C4Z4 with rank II inter}}\quad
\subfigure[\small Theory (II).]{
\includegraphics[height=4.7cm]{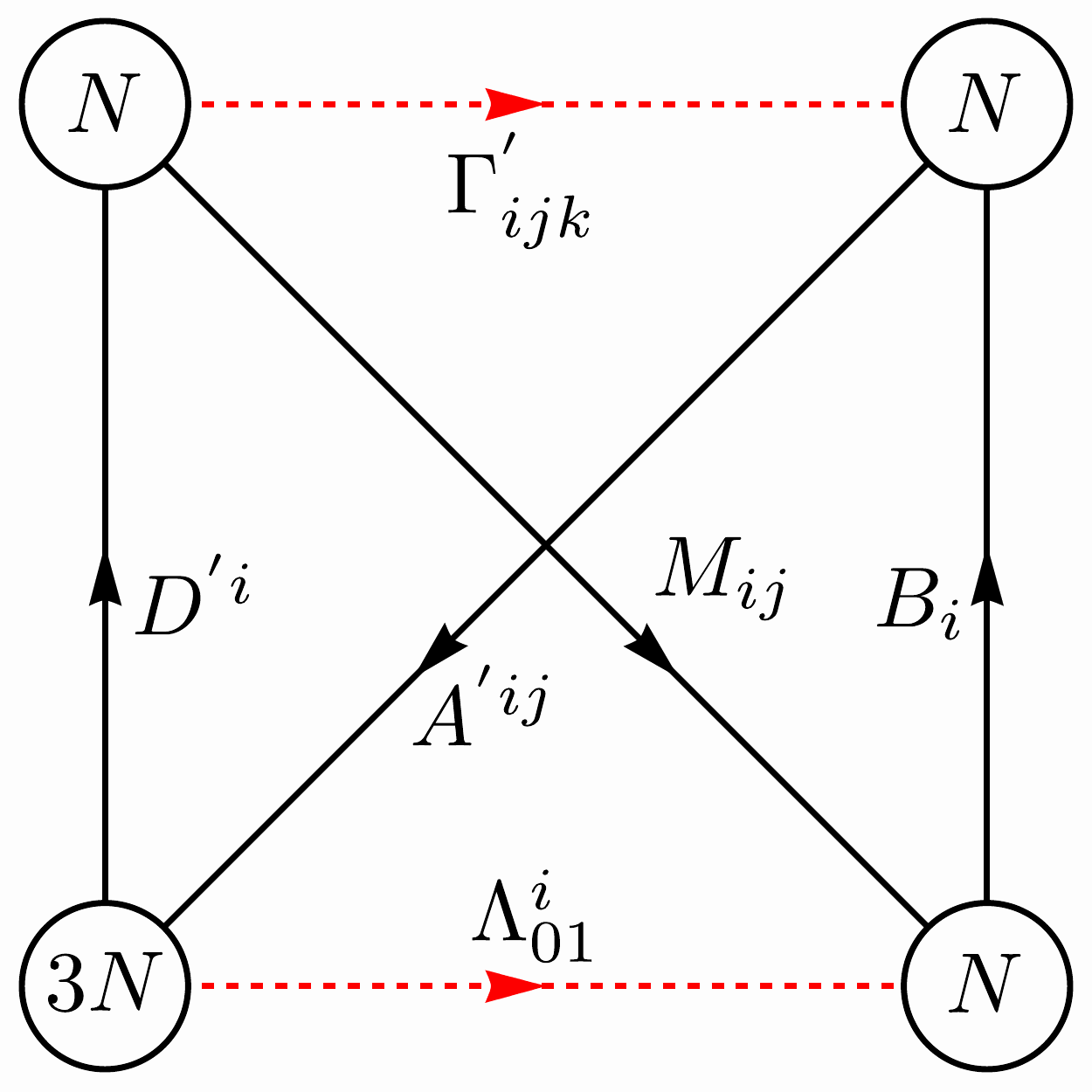}\label{fig: C4Z4 with rank II}}
\caption{A triality move on node $e_0$ (lower left) of Theory (I) gives the quiver shown in the middle. Integrating out the massive fields, which corresponds to cancelling fermi-chiral loops between pairs of nodes, we obtain Theory (II).} 
\label{fig: triality on Theory I}
\end{center}
\end{figure}

\paragraph{From Theory (I) to (II).}  It is straightforward to apply the ``triality operation'' of subsection~\ref{subsec: triality review} to these quiver gauge theories. For definiteness, conside ``Theory (I),'' whose field content is shown  in Figure~\ref{fig: C4Z4 with rank I}. The interaction terms read:
\bea\label{inter ThI}
&J_{\Lambda_{02}^{ij}} =\epsilon^{ijkl} C_k D_l~, \qquad \qquad 
&&J_{\Lambda_{13}^{ij}} =\epsilon^{ijkl} D_k A_l~,\cr
&E_{\Lambda_{02}^{ij}} = A_i B_j - A_j B_i~, \qquad \qquad 
&&E_{\Lambda_{13}^{ij}} = B_i C_j - B_j C_i~.
\eea
A triality operation on the node $e_0$ (lower left) leads to the quiver shown in Figure~\ref{fig: C4Z4 with rank II inter}, with the chiral and fermi multiplets as indicated. In particular, we have the mesonic fields $\t M$ and $\t \Gamma'$, which are given in terms of the elementary fields of Theory $(I)$ by:
\be
{\t M}_{ij}= D_i A_j~, \qquad \quad {\t \Gamma'}_{ijk}= D_i \Lambda_{02}^{jk}~.
\ee
By contruction, the mesons $\t M_{ij}$ sit in the ${\bf 4} \otimes  {\bf 4}$ of $SU(4)$, which decomposes into ${\bf 10} \oplus {\bf 6}$. Similarly, the fermionic fields $\t\Gamma'$ sit in the ${\bf 4}\otimes{\bf 6}\cong {\bf 20}\oplus {\bf 4}'$. From the matter content shown in Figure~\ref{fig: C4Z4 with rank II inter}, we see that the ${\bf 6}$ component of $\t M_{ij}$ and the ${\bf 4}'$ components of $\t \Gamma'_{ijk}$ can become massive by pairing with $\Lambda_{13}^{ij}$ and $C_i$, respectively. 
To see that this indeed happens, we simply look at the interaction terms, which are obtained by applying the triality rules \eqref{E and J th ii}-\eqref{E and J th ii Full}.
In particular, from \eqref{inter ThI} we find:
\be
J_{\Lambda_{13}^{ij}} =\epsilon^{ijkl} \t M_{kl}~,
\ee
which states that the antisymmetric part of $\t M_{ij}$ is massive, and can be set to zero in the low-energy theory. Let us denote by
$M_{ij}= \half (\t M_{ij}+ \t M_{ji} )$
the remaining light mesons, which span the ${\bf 10}$ of $SU(4)$. Similarly, it follows from \eqref{E and J th ii Full} and \eqref{inter ThI}  that the fields $C_i$ are massive. The corresponding constraint reads:
\be
J_{\t\Gamma'_{ijk}} = A^{' jk} D^{'i}  + \epsilon^{ijkl} C_l =0~.
\ee
This sets the ${\bf 4}'$ (fully antisymmetric) component of $\t\Gamma'_{ijk}$ to zero. Let us denote by:
\be
\Gamma'_{ijk}= \{\t\Gamma'_{ijk}  \}~,
\ee
the remaining fields, spanning the ${\bf 20}$ of $SU(4)$. Here and in the following, the notation $\{X_{ijk}\}$ denotes the projection of the three-tensor $X_{ijk}$ with two antisymmetrized indices onto the ${\bf 20}$ of $SU(4)$. 
We are thus left with the quiver shown in Figure~\ref{fig: C4Z4 with rank II}. The interaction terms are given explicitly by:
\bea\label{inter ThII}
&J_{\Lambda_{01}^{i}} = B_k A^{' ik}~, \qquad \qquad 
&&J_{\Gamma'_{ijk}} =   \{A^{'jk} D^{'i} \}~,\cr
&E_{\Lambda_{01}^{i}} =  D^{'j} M_{ji}~, \qquad \qquad 
&&E_{\Gamma'_{ijk}} =-M_{ij} B_k + M_{ik} B_j~.
\eea
One can again verify that $\Tr(EJ)=0$.

\begin{figure}[t]
\begin{center}
\subfigure[\small Triality transformation on Theory (II).]{
\includegraphics[height=4.8cm]{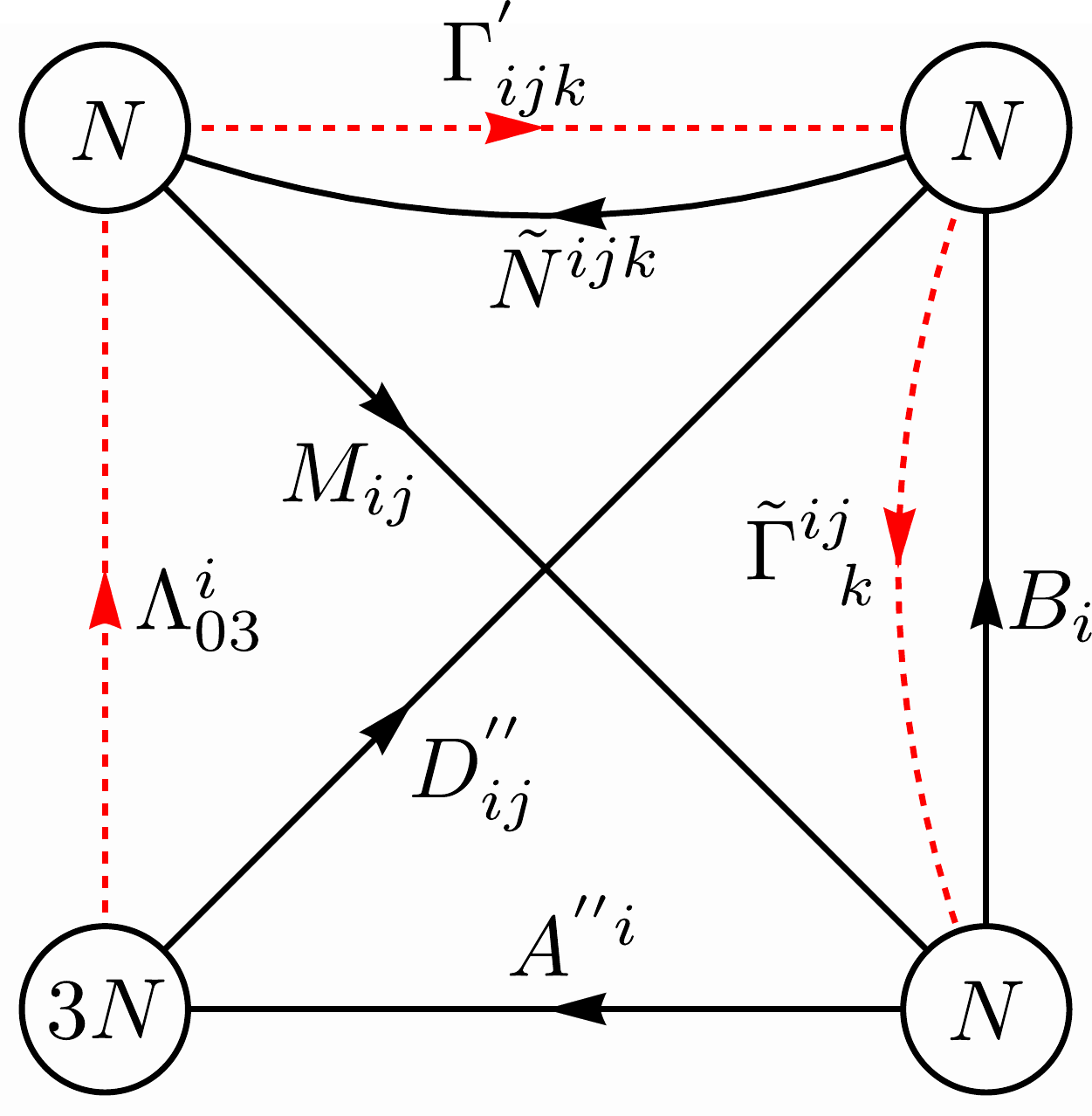}\label{fig: C4Z4 with rank III inter}}\quad
\subfigure[\small Theory (III).]{
\includegraphics[height=4.7cm]{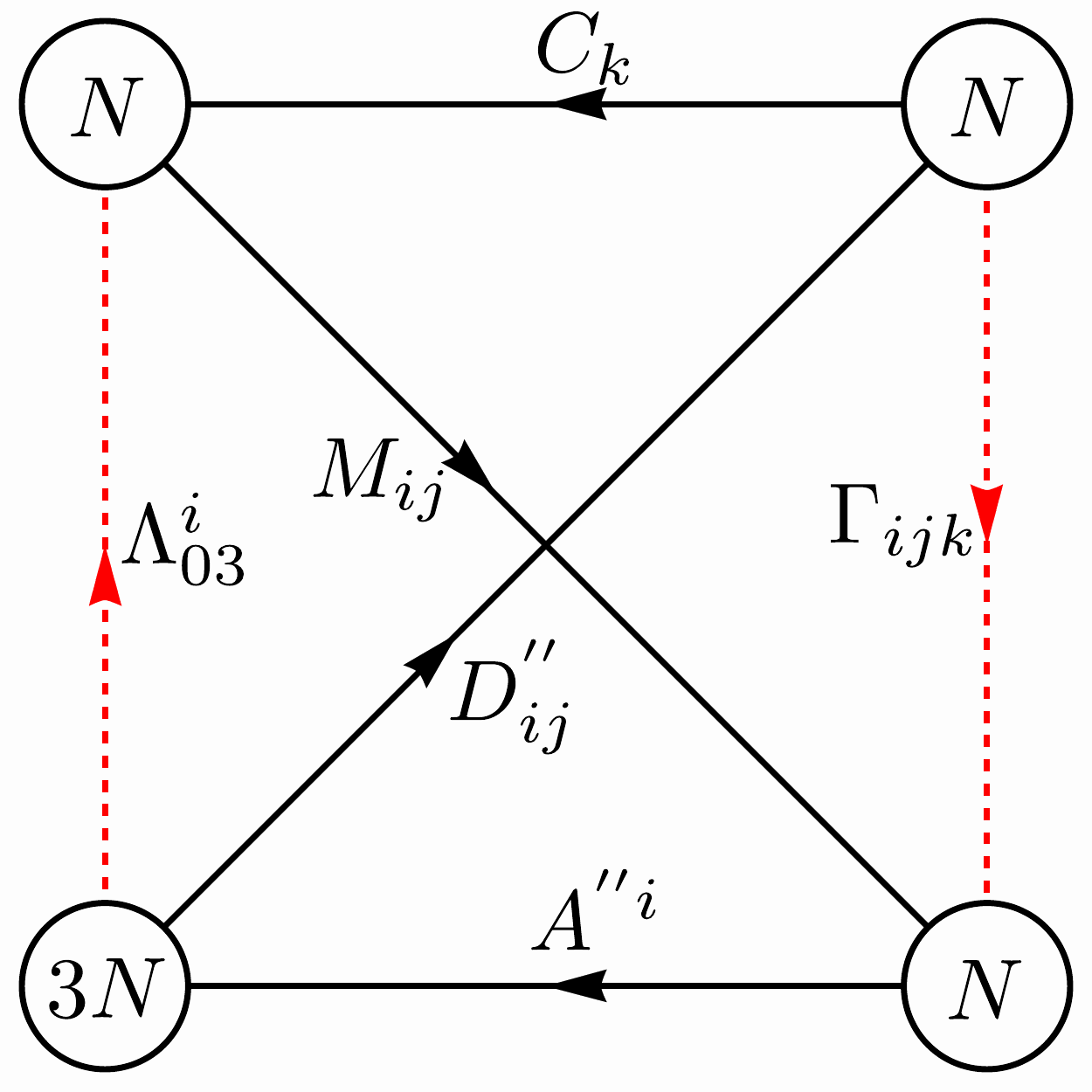}\label{fig: C4Z4 with rank III}}\;
\subfigure[\small Triality transformation on Theory (III).]{
\includegraphics[height=4.7cm]{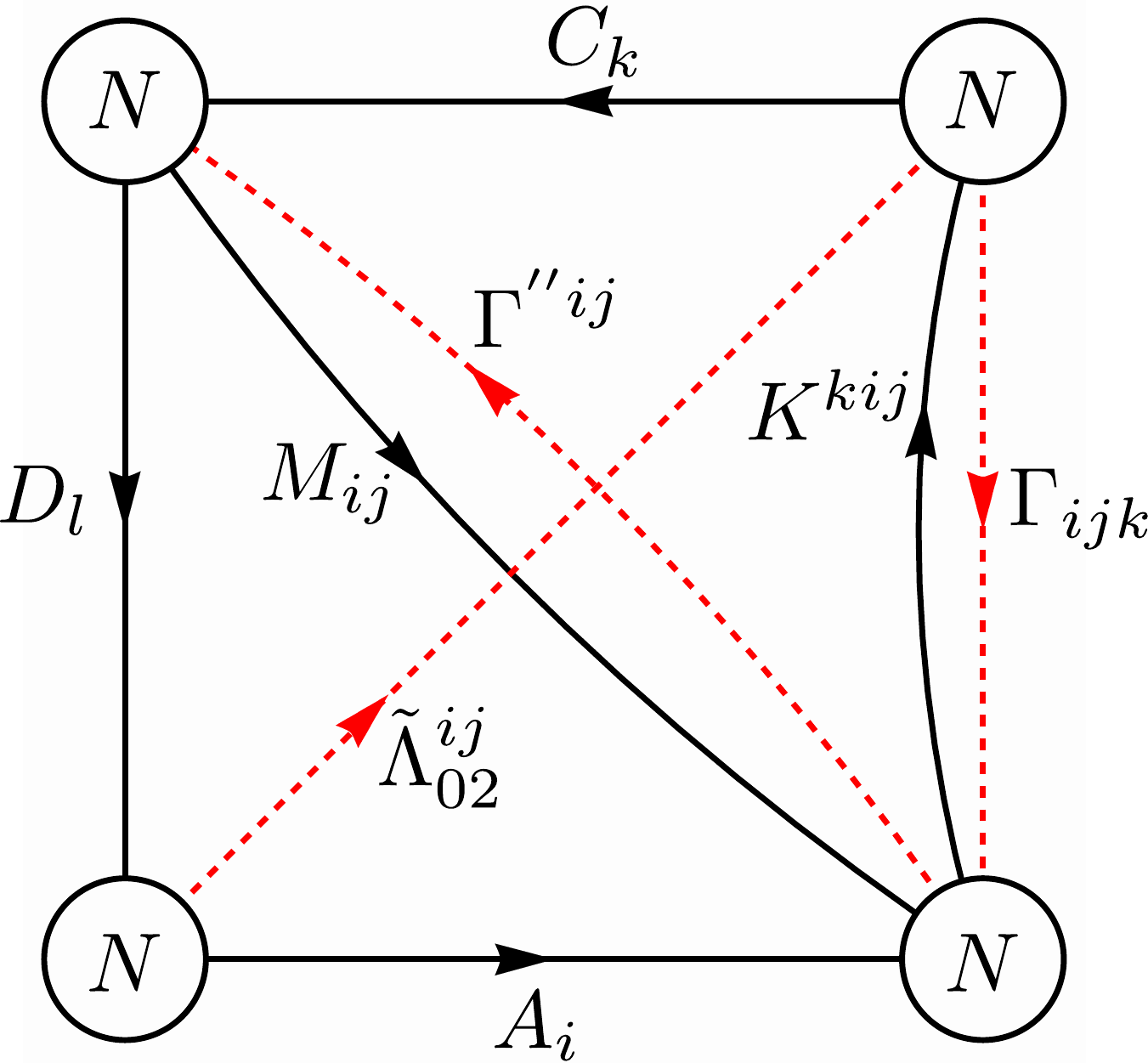}\label{fig: C4Z4 with rank I inter}}
\caption{A triality move on node $e_0$ of Theory (II) gives the quiver shown on the left. Integrating out the massive fields,  we obtain Theory (III), shown in the middle. Another triality move on Theory (III) gives the theory on the right, which is equivalent to Theory (I).}
\label{fig: triality on Theory II}
\end{center}
\end{figure}

\paragraph{From Theory (II) to (III).}   Starting from Theory (II) with the interaction terms \eqref{inter ThII}, we can again perform a triality operation on node $e_0$.
The process of integrating out massive fields is similar, as depicted in Figure~\ref{fig: triality on Theory II}. 
At the intermediate step (Figure~\ref{fig: C4Z4 with rank III inter}), we have the new mesons $\t N$ and $\t \Gamma$, which are identified with the fundamental fields of Theory~(II) according to:
\be
\t N^{ijk} = A^{' ij} D^{' k}~, \qquad {\t \Gamma^{ij}}_{\;\;k}= A^{' ij} \Lambda_{01}^k~.
\ee
We see from \eqref{inter ThII} that  the ${\bf 20}$ part of $\t N$ couple with $\Gamma'$ to form a mass term $J_{\Gamma'_{ijk}} = \{ \t N^{ijk}\}$. Setting $\{ \t N^{ijk}\}$ to zero, we are left with chiral fields in the ${\bf 4}$ of $SU(4)$, which we denote by $C_i$, defined such that:
\be
\t N^{ijk} = - \epsilon^{ijkl} C_l~.
\ee
Similarly, we have the following mass term for $B_i$:
\be
J_{{\t\Gamma^{ij}}_{\;\;k}} = A^{'' k} D^{''}_{ij} - \delta^k_i B_j~.
\ee
Integrating out $B_i$, we are left with the ${\bf 20}$ component of $ {\t \Gamma^{ij}}_{\;\;k}$. It is convenient to define the new fields:
\be 
\Gamma_{ijk}=-\half \epsilon_{ijln}\{ \t \Gamma^{ln}_{\;\;\; k} \}~, \qquad \t D^{ij}= \half \epsilon^{ijkl} D^{''}_{kl}~.
\ee
We then obtain Theory (III) shown in Figure~\ref{fig: C4Z4 with rank III}, with the interaction terms:
\bea\label{inter ThIII}
&J_{\Lambda_{03}^{i}} =  M_{ij} A^{''j}~, \qquad \qquad 
&&J_{\Gamma_{ijk}} =   - \{A^{'k} \t D^{ij} \}~,\cr
&E_{\Lambda_{03}^{i}} =   - \t D^{ij} C_j~, \qquad \qquad 
&&E_{\Gamma_{ijk}} =C_i M_{jk}-C_i M_{jk}~.
\eea
This is precisely the $\C^4/\Z_4$ quiver derived  in section \ref{subsec: quiver theory III derived}.

\paragraph{From Theory (III) to (I).}  Finally,  we can close this triality cycle by performing a triality operation on node $e_0$ of Theory (III). The intermediate step is shown in Figure~\ref{fig: C4Z4 with rank I inter}.
We have the new mesons $K$ and $\Gamma''$, which are identified with the fundamental fields of Theory (III) as:
\be
K^{k ij}= A^{'' k} \t D^{ij}~, \qquad \Gamma^{'' ij}= A^{'' i}\Lambda_{03}^j~.
\ee
It follows from $J_{\Gamma_{ijk}}= -\{K^{kij}\}$ that the ${\bf 20}$ component of $K^{kij}$ is massive. The remaining light fields, denoted by $B_i$, are defined by:
\be
K^{kij}= \epsilon^{ijkl} B_l~.
\ee
Similarly, we have the term:
\be
J_{\Gamma^{'' ij}} = D_j A_i - M_{ij}~,
\ee
which gives a mass to $M_{ij}$ and the symmetric part of $\Gamma^{'' ij}$. If we define the new fermi multiplets:
\be
\Lambda_{02}^{ij}= \half \epsilon^{ij kl}(\t \Lambda_{02})_{kl}~,\qquad
\Lambda_{13}^{ij}= - \epsilon^{ijkl} (\Gamma^{''})_{kl}~,
\ee
we precisely reproduce Theory (I) in Figure \ref{fig: C4Z4 with rank I}, with the interaction terms \eqref{inter ThI}.

\vskip0.3cm
These three $\CN=(0,2)$ quiver gauge theories are thus related by a triality cycle. Note that the quiver ranks of Theory (I) are $(N,N,N,N)$, while the quiver ranks of both Theory (II) and (III) are $(3N, N, N, N)$. In each case, this is the only rank assignment compatible with the non-abelian anomaly-free condition.  (Abelian anomalies are not cancelled; they are expected to be cancelled by the contribution of bulk modes in string theory.) Theories (II) and (III) are examples of ``non-toric'' quivers.

\subsection{Triality from mutation---a conjecture}\label{subsec: mutations}
We expect that the triality relations of $\CN=(0,2)$ quiver gauge theories are realized in string theory in the same way that all known Seiberg-like dualities are realized: by a change of ``brane basis''.  This intuition was realized in the type IIB mirror picture in \cite{Franco:2016qxh}, where triality was related to certain permutations of Lagrangian $4$-cycles. We would like to understand the analogous notion in the $B$-model.

\paragraph{Fractional branes from strongly exceptional collections.} Following previous work  \cite{Herzog:2003zc, Herzog:2004qw, Aspinwall:2004vm, Herzog:2006bu}, we consider the local Fano setup. We focus on $B_3= \mathbb{P}^3$, although we expect that most of the following is valid more generally.~\footnote{Including more general local geometries, such as the local $\mathbb{P}^1\times \mathbb{P}^1$ of section \protect\ref{sec local P1P1}.} Let us denote by $E_k$ the sheaves on $B_3$. A sheaf $E$ is called {\it exceptional} if $\Ext_{B_3}^i(E, E)= \delta_{i,0} \C$. 
A strongly exceptional collection:
\be\label{excpt coll}
{\bf E}= \{E_1, \cdots, E_n\}
\ee
 on $B_3$ is a collection of exceptional    objects such that~\footnote{An {\it exceptional collection} ${\bf E}$ is such that $\Ext_{B_3}^i(E_k, E_l)=0$ for $k>l$, $\forall i$. In this section, we consider the stronger condition of strong exceptionality, following  \cite{Herzog:2004qw}.}
 \be   
     \Ext^i_{B_3}(E_k, E_l) = 0 \;\;\; \forall \,i \neq 0~,\;  \forall k, l~, \qquad 
      \Ext^i_{B_3}(E_k, E_l) = 0 \;\; \; \forall \, i, \;  k > l~.
\ee
In particular, each sheaf in ${\bf E}$ is exceptional.  To describe fractional branes, we also need our collection to be ``maximal'' in some appropriate sense. Let $b_n= \dim H^n(B_3, \R)$ denote the Betti numbers of $B_3$. We call the strongly exceptional collection ${\bf E}$ {\it complete} if it contains $n=2+b_2+b_4$ sheaves---physically, this corresponds to the most general $D$-brane wrapping the $0$-, $2$-, $4$- and $6$-cycles \cite{Herzog:2004qw}. We have $n=4$ on $\mathbb{P}^3$.

Given a complete strongly exceptional collection \eqref{excpt coll} on $B_3$, we propose that there exists a good set of fractional branes on $\t {\bf X}_4 = {\rm Tot}(\CK \rightarrow B_3)$ given by:
\be\label{def frac brane from E}
\CE_I = i_\ast E_{n-I}[I]~, \qquad I= 0, \cdots, n-1~.
\ee
From the strong exceptionality condition on ${\bf E}$, it follows that the sheaves $\t E_I \equiv E_{n-I}[I]$ are ordered such that $\Ext^1(\t E_I, \t E_J)$ is non-vanishing only if $I =J+1$. Thus, we have:
\be\label{inter quiver}
E_1[n-1] \longrightarrow E_2[n-2]\longrightarrow \cdots \longrightarrow E_{n-1}[1] \longrightarrow E_n~,
\ee
where the arrows denote the $\Ext^1_{B_3}(\t E_I, \t E_J)$ elements. The pushforward to $\t{\bf X}_4$ will  ``close the quiver,'' by adding additional $\Ext$ groups due to the contribution of the embedding.

\vskip0.2cm \noindent
As an example, consider the following strongly exceptional collections on $\mathbb{P}^3$:
${\bf E}_I \equiv \left\{\Omega^3(3)~,\, \Omega^2(2)~,\, \Omega(1)~,\, \CO\right\}$.
The intermediate quiver \eqref{inter quiver} reads:
\be
\Omega^3(3)[3] \overset{4}{\longrightarrow}   
 \Omega^2(2)[2] \overset{4}{\longrightarrow}  \Omega(1)[1]
  \overset{4}{\longrightarrow} \CO~,
\ee
with the dimension of the $\Ext^1$ groups indicated over the arrows. The corresponding fractional branes were discussed in section \ref{sect:fourfold:p2:first}.

\paragraph{Triality and mutations.}  
A natural geometric operation on these fractional branes is provided by mutations of exceptional collections \cite{Rudakov:1619155}.
Consider the strongly exceptional collection \eqref{excpt coll}. A {\it mutation at position $k$}, with $k<n$, is a braiding operation on the exceptional collection:
\be\label{mutation}
(E_1, \cdots, E_k, E_{k+1},  \cdots E_n)     \quad  \rightsquigarrow\quad 
(E_1,\cdots, L_{E_k} E_{k+1},  E_{k}, \cdots \cdots E_n)~.
\ee
Here, the new sheaf $L_{E_k} E_{k+1}$ at position $k$ is given by a {\it left mutation}.
Note that a  left mutation of an exceptional pair of sheaves $(E, F)$ produces another exceptional pair $(L_E F, E)$. The precise definition of $L_E F$  can be found in  \cite{Rudakov:1619155}. For our purposes here, we just note the properties $L_{E[1]}F= L_E F$ and $L_{E}(F[1])=(L_{E} F)[1]$ under  the translation functor.

 The effect of \eqref{mutation} on the fractional branes may also be called a {\it mutation at node $e_I$}, with $I= n-k>0$. Given the ordered fractional branes \eqref{def frac brane from E}, a mutation at $e_I$ corresponds to:
\be
(\CE_1, \cdots, \CE_{I-1}, \CE_I, \cdots, \CE_{n-1})\quad  \rightsquigarrow\quad 
(\CE_1, \cdots, \CE_{I}[-1], L_{\CE_I}(\CE_{I-1})[1], \cdots, \CE_{n-1})
\ee
Here we defined the new fractional brane:
\be
L_{\CE_I}(\CE_{I-1}) = i_* \left(L_{E_k} E_{k+1}[n-k-1]\right)~, \qquad I= n-k~,
\ee
by abuse of notation. We conjecture that mutations of a strongly exceptional collection which preserve the strongly exceptional condition realize the field-theory triality operation of section \ref{subsec: triality review}.~\footnote{While a mutation of an exceptional collection gives another exceptional collection, it is not clear that is also preserves the strongly exceptional condition. This will be the case in our examples. See \cite{Herzog:2004qw} for further discussions.} 
 This proposal passes some obvious sanity checks. First of all, note that the pair $(\CE_{I-1}, \CE_I)$ involved in the mutation has:
\be\label{ext1 bm}
\Ext^{i+1}(\CE_I, \CE_{I-1})=  \delta_{i, 0}  \C^{n_a}~.
\ee
According to our general rules, $n_a$ is the number of incoming arrows at node $e_I$---in the language of section \ref{subsec: triality review}, the number of antifundamental chiral multiplets $\t\Phi$ under $U(N_0)$ is $n_a N_A$, and we have:
\be\label{incoming arrow conj}
\CE_{I} \xrightarrow{n_a} \CE_{I-1} \qquad \cong \qquad e_I \xleftarrow{n_a} e_{I-1}~.
\ee
The condition that the new collection is strongly exceptional leads to:
\be\label{ext1 am}
\Ext^{i+1}(L_{\CE_I}(\CE_{I-1})[1],  \CE_{I}[-1]) = \delta_{i, 0} \C^{n_a}~.
\ee
in the new quiver.~\footnote{More precisely, the dimension $n_a$ in \eqref{ext1 bm} and \eqref{ext1 am} might still differ. They must actually match for our conjecture to hold.}
This means that we now have {\it outgoing} arrow from $e_I$ to $e_{I-1}$ in the supersymmetric quiver: $e_I \xrightarrow{n_A} e_{I-1}$. This matches the fact that the antifundamental multiplets are dualized to  fundamental multiplets under triality  ($\t \Phi  \rightsquigarrow \Phi'$). We also see that:
\be
\Ext^{3}(\CE_I, \CE_J) = \Ext^2(\CE_I[-1], \CE_J)~, \qquad 
\Ext^{2}(\CE_I, \CE_J) = \Ext^1(\CE_I[-1], \CE_J)~.
\ee
These relations imply that the fundamental chiral multiplets of the original theory are dualized to fermi multiplets ($\Phi \rightsquigarrow \Lambda'$), and the fermi multiplets are dualized to antifundamental chiral multiplets ($\Lambda \rightsquigarrow \t\Phi'$). In this way, we elegantly reproduce the simplest aspects of the triality, as summarized in Figures~\ref{fig:triality i} and \ref{fig:triality ii}.

\vskip0.2cm
\paragraph{Examples:} 
Consider $B_3= \mathbb{P}^3$, as discussed above.
For definiteness, we start from the strongly exceptional collection:
\be\label{except coll III}
{\bf E}_{III}=  \left\{\CO(-1)~,\, \Omega^2(2)~,\, \CO~,\, \CO(1)\right\}~.
\ee
The corresponding fractional branes were discussed in section \ref{sect:fourfold:p2:second}---they were dubbed $\{\CE_1, \CE_0, \CE_2, \CE_3 \}$, where the ordering matters. The intermediate $\Ext^1$ quiver \eqref{inter quiver} reads:
\be
\CO(-1)[3] \overset{4}{\longrightarrow}   
 \Omega^2(2)[2] \overset{6}{\longrightarrow}  \CO(1)[1]
  \overset{4}{\longrightarrow} \CO(1)~,
\ee
and the full $\Ext^{1,2}$ quiver on the local CY$_4$ reads:
 \[
\xymatrix{ \mathcal{O}(-1)[3]\ar[rr]^4 \ar@{--}[dr]_{4} & &
 \Omega^2(2)[2]  \ar[dd]^6 \\
& \ar@{--}[dr] \ar@{--}[ur] & \\
\mathcal{O}(1) \ar[uu]^{10} \ar@{--}[ur]^{20} & &  \mathcal{O}[1] \ar[ll]_4
}
\]
This gives the supersymmetric quiver that we called ``Theory (III)'' above. Now, consider a mutation at $\CE_0$, which is a mutation at the third position in \eqref{except coll III}, at $E_3= \CO$. It is a well-known result that:
\be
L_{\mathcal{O}}\mathcal{O}(1)= \Omega^1(1)
\ee
on $\mathbb{P}^3$. Therefore, the new strongly exceptional collection is given:
\be\label{exp exc coll I}
{\bf E}_I = \left\{\CO(-1)~,\, \Omega^2(2)~,\, \Omega^1(1)~,\, \CO\right\}~.
\ee
 The corresponding $\Ext^{1,2}$ quiver on the CY fourfold reads:
 \[
\xymatrix{ \mathcal{O}(-1)[3] \ar[rr]^4 \ar@{--}[dr]_6 & &
\Omega^2(2)[2] \ar[dd]^4 \\
& \ar@{--}[dr] \ar@{--}[ur] & \\
\mathcal{O} \ar[uu]^4 \ar@{--}[ur]^6 & & \Omega^1(1)[1] \ar[ll]_4
}
\]
corresponding to ``Theory (I)'' above. (Recall that $\CO(-1)\cong \Omega^3(3)$ on $\mathbb{P}^3$.)
 This matches the field theory expectation: a mutation at $\CE_0= i_* \CO[1]$ should be a triality operation on node $e_0$ of Theory (III). This indeed gives Theory (I), as explained in section \ref{triality on C4Z4 quiver}.

 A triality operation at node $e_0$ of Theory (I) gives theory (II). Unfortunately, we cannot directly realize it by mutation, because $\CE_0$ corresponds to the last sheaf in  the exceptional collection \eqref{exp exc coll I}. However, remark that Theory (I) has a $\Z_4$ symmetry that rotates the four nodes of the quiver. Therefore, a triality at any node of Theory (I) gives Theory (II), up to a rotation of the nodes. We can then consider any other mutation of adjacent sheaves in \eqref{exp exc coll I} to obtain Theory (II). Consider a mutation at position $1$. One can show that:
 \be
 L_{\CO(-1)}  \Omega^2(2) = \CO(-2)~.
 \ee
We thus obtain the new strongly exceptional collection on $\mathbb{P}^3$:
\be\label{exp exc coll II}
{\bf E}_{II} = \left\{\CO(-2)~,\,\CO(-1)~,\, \Omega^1(1)~,\, \CO\right\}~.
\ee
The corresponding fractional branes are:
\be
\CE_0'= i_\ast \CO(-2)[3]~, \qquad \CE_1'= i_\ast \CO(-1)[2]~, \qquad
\CE_2'= i_\ast \Omega^1(1)[1]~, \qquad \CE_3'= i_\ast \CO~.
\ee
One can again compute the $\Ext^{1,2}$ quiver. It reads:
 \[
\xymatrix{ \mathcal{O}(-2)[3] \ar[rr]^4 \ar@{--}[dr]_{20} & &
\CO(-1)[2] \ar[dd]^6 \\
& \ar@{--}[dr] \ar@{--}[ur] & \\
\mathcal{O} \ar[uu]^{10} \ar@{--}[ur]^4 & & \Omega^1(1)[1] \ar[ll]_4
}
\]
Relabelling the fractional branes $(\CE_0', \CE_1', \CE_2', \CE_3')= (\CE_1, \CE_2, \CE_0, \CE_3)$, we precisely reproduce the ``Theory (II)'' quiver shown in Figure~\ref{fig:triality C4Z4 ii}.

\section{D-instanton quivers and gauged matrix models}\label{sec: GMM}
Zero-dimensional ``gauge theories''---gauged matrix models (GMM)---naturally arise as the low-energy description of D-instantons in type IIB string theory \cite{Ishibashi:1996xs}. In particular, gauged matrix models with $\CN=1$ supersymmetry can describe D-instantons at Calabi-Yau fivefold singularities \cite{Franco:2016tcm}. 

Since 0d $\CN=1$ superspace is spanned by a single Grassmanian coordinate $\theta$, any superfield is of the form $X= x + \theta y$, with $x$, $y$ some variables of opposite Grassmann parity.~\footnote{In this section, by an abuse  of language, we call the  Grassmann-even and Grassmann-odd  integration variables in the matrix integral the ``bosons'' and ``fermions,'' respectively.} The single supersymmetry is generated by $\CQ= \d_\theta$.
A generic $\CN=1$ GMM can be conveniently described using three elementary supermultiplets. The $\CN=1$ {\it chiral multiplet} $(\Phi, \b \Phi)$ consists of a complex boson $\phi, \b \phi$ and a fermion $\b\psi$. In superspace, we have:
\be
\Phi= \phi~, \qquad \b \Phi = \b \phi + \theta \b\psi~.
\ee
The chiral multiplet $\Phi$ has a single component, with $\CQ \phi=0$, while the anti-chiral multiplet $\b\Phi$ has two components, with $\CQ \b\phi = \b \psi$ and $\CQ \b\psi=0$. 
The bosons $\phi$ and $\b\phi$ should be considered as complex conjugate in the matrix integral, while there is a single fermion $\b\psi$. 
The second type of multiplet is the {fermi multiplet} $\Lambda$, with a single fermionic component $\lambda$, such that:
\be
\Lambda= \lambda + \theta F_\lambda~, \qquad \qquad  F_\lambda= F_\lambda(\phi)~.
\ee
Here, the $\CN=1$ superpotential $F_\lambda$ is an holomorphic function of the bosons $\phi$ in chiral multiplets. Given the chiral multiplets $\phi_i$ and fermi multiplets $\lambda_a$,  one can write the supersymmetric action:
\be\label{F term}
S_F = \int d\theta\, \b F^a(\b \Phi) \Lambda_a =  \b F^a(\b \phi) F_a(\phi) + \b \psi^i {\d \b F^a \ov \d \b\phi^i} \lambda_a~.
\ee
Another quadratic action in the fermions can be written in terms of an holomorphic potential $H^{ab}(\phi)=- H^{ba}(\phi)$:
\be\label{H term}
S_H = H^{ab}(\phi) \lambda_a \lambda_b~.
\ee
This is supersymmetric provided that $H^{ab}F_b=0$. The third type of sypersymmetry multiplet is the gaugino multiplet, which implements a gauge constraint on field space. The gaugino multplet $\CV$ consists of two components, the fermion $\chi$ and the real boson $D$, with:
\be
\CV= \chi + \theta D~.
\ee
Given a theory of chiral and fermi multiplets with some non-trivial Lie group symmetry, we can gauge a subgroup $\GG$ (with Lie algebra $\Fg$) of that symmetry by introducing an $\Fg$-valued gaugino multiplet, with the action:
\be
S_{\rm gauge}= \int d\theta\, \left(\half D \chi - i \xi \chi + i \b\phi \chi \phi\right)~,
\ee
with $\chi$ acting on $\phi$ in the appropriate representation, and an overall trace over the gauge group is implicit. Here $\xi$ is a 0d Fayet-Iliopoulos parameter.  Integrating out $D$, we obtain:
\be
S_{\rm gauge} \cong   \mu^2 - i \psi \chi \phi~.
\ee
where $\mu\equiv \b\phi \phi - \xi$ (schematically), which is the moment map (minus the ``level'' $\xi$) of the $\GG$ action on the bosonic field space.

\subsection{$\CN=1$ gauged matrix model from $B$-branes at a CY$_5$ singularity}
D-instantons at CY$_5$ singularities engineer precisely such gauged matrix models with gauge group $\prod_I U(N_I)$.
For each node $e_I$ in the 0d $\CN=1$ quiver, we have a $U(N_I)$ gaugino multiplet. The matter fields are either chiral or fermi multiplets, in adjoint or bifundamental representations. We have thus a quiver with two type of oriented arrows:  $e_I \rightarrow e_J$  for chiral multiplets $X_{IJ}$, and   $e_I \dashrightarrow  e_J$ for fermi multiplets $\Lambda_{IJ}$. Finally, we also have the $F$- and $H$-type interaction terms. To each fermi multiplet $\Lambda_{IJ}$, we associate the element $F_{IJ}$, a direct sum over oriented paths $p$ from $e_I$ to $e_J$, of length $k$:
\be\label{Fterms paths}
F_{IJ}(X) = \sum_{{\rm paths}\; p} c_{p}^{IJ}    X_{I K_1} X_{K_1K_2} \cdots X_{K_{k-1} J}~,
\ee
 similarly to \eqref{E and J superpot quiver}, with given coefficients $c_p^{IJ}$.
In addition, to every pair of fermi multiplets $\Lambda_{IJ}$ and $\Lambda_{KL}$, we associate the $H$-term action $S_{H^{IJ, KL}}$, which is a sum over closed loops $\t p$ from $e_I$ back to itself, which includes both $\Lambda_{IJ}$ and $\Lambda_{KL}$, in addition to chiral multiplets $X$:
\be\label{Hterms paths}
S_{H^{IJ, KL}}= \sum_{{\rm paths}\; \t p}  c_{\t p}^{IJ, KL}\Tr(  \Lambda_{IJ} \, X_{J M_1}\cdots X_{M_{k-1}K} \,\Lambda_{KL} \, X_{L N_1}\cdots X_{N_{k'-1}I})~.
\ee
Note that the closed path $\t p$ has length $k+k'+2$, including the two fermions.

This quiver structure naturally arises from open strings between fractional $D(-1)$-branes at a CY$_5$  singularity, where each node $e_I$ corresponds to a fractional brane $\CE_I$.
As before, we must have:
\be
\Ext^{0}_{{\bf X}_5}(\CE_I, \CE_J) = \C \delta_{IJ}~.
\ee
The non-vanishing $\Ext^0$ elements are identified with the gaugino multiplets. The degree-one $\Ext$ groups are identified with chiral multiplets:
\be
\Ext^1_{{\bf  X}_5}(\CE_J, \CE_I) \qquad \Leftrightarrow \qquad e_I \longrightarrow e_J   \qquad \Leftrightarrow \qquad X_{IJ}~,
\ee
in bifundamental (if $I\neq J$) or adjoint (if $I=J$) representations.
Similarly, the degree-two $\Ext$ groups are identified with the fermi multiplets:
\be
\Ext^2_{{\bf  X}_5}(\CE_J, \CE_I) \qquad \Leftrightarrow \qquad e_I \dashrightarrow e_J   \qquad \Leftrightarrow \qquad \Lambda_{IJ}~.
\ee
By Serre duality, we also have $\Ext^4_{{\bf  X}_5}(\CE_J, \CE_I) \cong \Ext^1_{{\bf  X}_5}(\CE_I, \CE_J)$ and $\Ext^3_{{\bf  X}_5}(\CE_J, \CE_I) \cong \Ext^2_{{\bf  X}_5}(\CE_I, \CE_J)$.

\vskip0.3cm 
\paragraph{Interaction terms.}
The $F$-terms \eqref{F term} and $H$-terms \eqref{H term} also arise naturally in the $B$-model. As discussed in section \ref{subsec: Ainf and EJ}, the $\Ext$-group generators satisfy an $A_\infty$ algebra with multi-products $m_k$. 
Consider a  fermi multiplet $\Lambda_{IJ}$ corresponding to $\alpha\in \Ext^2(\CE_J, \CE_I)$, and let us denote by $\b\alpha \in \Ext^3(\CE_I, \CE_J)$ the Serre dual generator.
For each path $p$ as in \eqref{Fterms paths}, we have the elements $x\in \Ext^1$ corresponding to the chiral multiplets $X$.  We propose that:
\be\label{cp def 0d}
c_p^{IJ} = \left\langle \b\alpha\, \, x_{I K_1} \cdots  \,  x_{K_{k-1}J} \right\rangle =  \gamma\big(m_2(\b\alpha, \, m_k(x_{I K_1} \cdots  \,  x_{K_{k-1}J})) \big)~,
\ee
 for the $F$-term coefficients in \eqref{Fterms paths}.
Similarly, consider  the fermi multiplets $\Lambda_{IJ}$ and $\Lambda_{KL}$ corresponding to $\alpha\in \Ext^2(\CE_J, \CE_I)$ and $\beta\in \Ext^2(\CE_L, \CE_K)$, respectively. We propose that the $H$-term coefficients in \eqref{Hterms paths} are given by:
\bea\label{cpt def 0d}
& c_{\t p}^{IJ, KL} = \left\langle \alpha  \,   x_{J M_1}  \cdots x_{M_{k-1}K} \,  \beta  \,  x_{L N_1} \cdots x_{N_{k'-1} I} \right\rangle \cr
& \qquad \quad \qquad \qquad= \gamma\big(m_2(\alpha, \, m_{\t k}(x_{J M_1},  \cdots, x_{M_{k-1}K},  \beta  ,  x_{L N_1}, \cdots, x_{N_{k'-1} I})) \big)~,
\eea
with $\t k=k+k'+1$.
We will check this prescription in some examples below. Note that this corresponds exactly to computing the formal 0d $\CN=1$ superpotential:
\be
\CW = \Tr\left(\b \Lambda^a F_a(X) + \Lambda_a \Lambda_b H^{ab}(X) \right)~,
\ee
which was recently introduced in \cite{Franco:2017lpa}.

\subsection{$D(-1)$-brane on $\C^5$}\label{exp: C5}
We can work out the very simplest case, a $D(-1)$ brane on ${\bf X}_5= \C^5$, exactly like in section \ref{exp: C4}. Consider the skyscraper sheaf $\CO_p$ at the origin of $\C^5$. We have:
\bea\label{ext0 C5}
&\Ext^0(\CO_p, \CO_p) \cong\Ext^5(\CO_p, \CO_p) \cong \C~,\cr
&\Ext^1(\CO_p, \CO_p) \cong\Ext^4(\CO_p, \CO_p) \cong \C^5~, \cr
&\Ext^2(\CO_p, \CO_p) \cong  \Ext^3(\CO_p, \CO_p)\cong  \C^{10}~.
\eea
Using the above dictionary to $\CN=1$ superfields, this reproduces  the expected field content of the maximally-supersymmetric $\CN=16$ matrix model, as we will review below.

\subsubsection{The $\Ext$ algebra of $\C^5$}

Proceeding as before, the Koszul resolution of $ \mathcal{O}_p $ on $\mathbb{C}^5$ reads:
\begin{equation}\label{koszul4a}
0 \xrightarrow[]{} \mathcal{O} \xrightarrow[]{E} \mathcal{O}^5
\xrightarrow[]{D} \mathcal{O}^{10} \xrightarrow[]{C} \mathcal{O}^{10}
\xrightarrow[]{B} \mathcal{O}^5 \xrightarrow[]{A} \mathcal{O} \xrightarrow[]{} \mathcal{O}_p
\xrightarrow[]{} 0,
\end{equation}
where:
{\small \[
A = \left(\begin{array}{ccccc} x & y & z & w & u
\end{array} \right)~,\quad
\qquad
B = \left(\begin{array}{cccccccccc}
y & z & w & u & 0 & 0 & 0 & 0 & 0 & 0 \\
-x & 0 & 0 & 0 & z & w & u & 0 & 0 & 0 \\
0 & -x & 0 & 0 & -y & 0 & 0 & w & u & 0 \\
0 & 0 & -x & 0 & 0 & -y & 0 & -z & 0 & u \\
0 & 0 & 0 & -x & 0 & 0 & -y & 0 & -z & -w
\end{array} \right)~,
\]
\[
C = \left(\begin{array}{cccccccccc}
z & w & u & 0 & 0 & 0 & 0 & 0 & 0 & 0 \\
-y & 0 & 0 & w & u & 0 & 0 & 0 & 0 & 0 \\
0 & -y & 0 & -z & 0 & u & 0 & 0 & 0 & 0 \\
0 & 0 & -y & 0 & -z & -w & 0 & 0 & 0 & 0\\
x & 0 & 0 & 0 & 0 & 0 & w & u & 0 & 0\\
0 & x & 0 & 0 & 0 & 0 & -z & 0 & u & 0 \\
0 & 0 & x & 0 & 0 & 0 & 0 & -z & -w & 0\\
0 & 0 & 0 & x & 0 & 0 & y & 0 & 0 & u\\
0 & 0 & 0 & 0 & x & 0 & 0 & y & 0 & -w\\
0 & 0 & 0 & 0 & 0 & x & 0 & 0 & y & z
\end{array} \right)~,\;
D = \left(\begin{array}{ccccc}
w & u & 0 & 0 & 0\\
-z & 0 & u & 0 & 0\\
0 & -z & -w & 0 & 0 \\
y & 0 & 0 & u & 0\\
0 & y & 0 & -w & 0\\
0 & 0 & y & z & 0\\
-x & 0 & 0 & 0 & u\\
0 & -x & 0 & 0 & -w\\
0 & 0 & -x & 0 & z\\
0 & 0 & 0 & -x & -y
\end{array} \right)~,\;
E = \left(\begin{array}{c}
u \\ -w \\ z \\ -y \\ x
\end{array} \right)
\]
}
\vskip-0cm\noindent
Similarly to section \ref{exp: C4}, 
we choose as bases of the Ext groups the commutative diagrams
whose leftmost nonzero vertical map has $1$ at an entry and $0$ elsewhere.
We denote them by $X^i_j$, following the same conventions.
The multiplication rule is again determined by composition.
The products $m_2(X_i^1, X_j^1)= X_i^1 \cdot X_j^1$ are given by:
\be
\begin{tabular}{c|ccccc}
& $X^1_1$ & $X^1_2$ & $X^1_3$ & $X^1_4$ & $X^1_5$ \\
\hline
$X^1_1$ & 0 & $-X^2_1$ & $-X^2_2$ & $-X^2_4$ & $-X^2_7$ \\
$X^1_2$ & $X^2_1$ & 0 & $-X^2_3$ & $-X^2_5$ & $-X^2_8$ \\
$X^1_3$ & $X^2_2$ & $X^2_3$ & 0 & $-X^2_6$ & $-X^2_9$ \\
$X^1_4$ & $X^2_4$ & $X^2_5$ & $X^2_6$ & 0 & $-X^2_{10}$\\
$X^1_5$ & $X^2_7$ & $X^2_8$ & $X^2_9$ & $X^2_{10}$ & 0
\end{tabular}
\ee
The elements in this table are the products of the
elements in the first column multiplied by elements in the first row. (For example, $X^1_1 \cdot X^1_2 = -X^2_1$.) Similarly, we have non-zero products $m_2(X^1_i, X^2_j)$  mapping $\Ext^1 \otimes \Ext^2$ to $\Ext^3$, according to:
\be
\begin{tabular}{c|cccccccccc}
& $X^2_1$ & $X^2_2$ & $X^2_3$ & $X^2_4$ & $X^2_5$ & $X^2_6$ & $X^2_7$ & $X^2_8$ & $X^2_9$ & $X^2_{10}$ \\
\hline
$X^1_1$ & 0 & 0 & $X^3_1$ & 0 & $X^3_2$ & $X^3_3$ & 0 & $X^3_5$ & $X^3_6$ & $X^3_8$ \\
$X^1_2$ & 0 & $-X^3_1$ & 0 & $-X^3_2$ & 0 & $X^3_4$ & $-X^3_5$ & 0 & $X^3_7$ & $X^3_9$ \\
$X^1_3$ & $X^3_1$ & 0 & 0 & $-X^3_3$ & $-X^3_4$ & 0 & $-X^3_6$ & $-X^3_7$ & 0 & $X^3_{10}$\\
$X^1_4$ & $X^3_2$ & $X^3_3$ & $X^3_4$ & 0 & 0 & 0 & $-X^3_8$ & $-X^3_9$ & $-X^3_{10}$ & 0\\
$X^1_5$ & $X^3_5$ & $X^3_6$ & $X^3_7$ & $X^3_8$ & $X^3_9$ & $X^3_{10}$ & 0 & 0 & 0 & 0
\end{tabular}
\ee
We also find the following Serre dual elements to  $X^2_i$:
\be
\begin{tabular}{c|cccccccccc}
& $X^2_1$ & $X^2_2$ & $X^2_3$ & $X^2_4$ & $X^2_5$ & $X^2_6$ & $X^2_7$ & $X^2_8$ & $X^2_9$ & $X^2_{10}$ \\
\hline
 {\rm dual:} & $X^3_{10}$  & $-X^3_{9}$ & $X^3_{8}$ & $X^3_{7}$ & $-X^3_{6}$ & $X^3_{5}$ & $-X^3_{4}$ & $X^3_{3}$  & $-X^3_{2}$  & $X^3_{1}$
 \end{tabular}
\ee
Using the multiplication rule (and the cyclic property of the open-string correlators), we see that
any nonzero correlation function can be computed in terms the following
$\langle X^3 X^1 X^1 \rangle $-type correlators:
\bea\label{311}
&\langle X^3_1 X^1_5 X^1_4 \rangle=1~,\quad ~ &&\langle X^3_2 X^1_5 X^1_3 \rangle=-1~,\quad,~&&\langle X^3_3 X^1_5 X^1_2 \rangle=1~,\\
&\langle X^3_4 X^1_5 X^1_1 \rangle=-1~,\quad~&&\langle X^3_5 X^1_4 X^1_3 \rangle=1~,\quad~&&\langle X^3_6 X^1_4 X^1_2 \rangle=-1~,\\
&\langle X^3_7 X^1_4 X^1_1 \rangle=1~,\quad~&&\langle X^3_8 X^1_3 X^1_2 \rangle=1~,\quad~&&\langle X^3_9 X^1_3 X^1_1 \rangle=-1~,\\
&\langle X^3_{10} X^1_2 X^1_1 \rangle=1~,
\eea
and the following $\langle X^2 X^2 X^1 \rangle $-type correlators:
\bea\label{221}
&\langle X^2_3 X^2_{10} X^1_1 \rangle=1~,\quad~&&\langle X^2_5 X^2_9 X^1_1 \rangle=-1~,\quad~&&\langle X^2_6 X^2_8 X^1_1 \rangle=1~,\\
&\langle X^2_2 X^2_{10} X^1_2 \rangle=-1~,\quad~&&\langle X^2_4 X^2_9 X^1_2 \rangle=1~,\quad~&&\langle X^2_7 X^2_6 X^1_2 \rangle=-1~,\\
&\langle X^2_1 X^2_{10} X^1_3 \rangle=1~,\quad~&&\langle X^2_4 X^2_8 X^1_3 \rangle=-1~,\quad~&&\langle X^2_5 X^2_7 X^1_3 \rangle=1~,\quad\\
&\langle X^2_1 X^2_9 X^1_4 \rangle=-1~,\quad~&&\langle X^2_2 X^2_8 X^1_4 \rangle=1~,\quad~&&\langle X^2_3 X^2_7 X^1_4 \rangle=-1~,\quad\\
&\langle X^2_1 X^2_6 X^1_5 \rangle=1~,\quad~&&\langle X^2_2 X^2_5 X^1_5 \rangle=-1~,\quad~&&\langle X^2_3 X^2_4 X^1_5 \rangle=1.\\
\eea

\subsubsection{The $\C^5$ quiver: $\CN=16$ SYM}
Consider the $\CN=16$ supersymmetric GMM with gauge group $U(N)$, corresponding to 
$N$ $D(-1)$-branes in flat space. Its field content can be deduced from dimensional reduction of 2d $\CN=(8,8)$ SYM in section \ref{88SYM}. In $\CN=1$ language, we have a single $U(N)$ gaugino multiplet, $5$ chiral multiplets in the adjoint representation, and $10$ fermi multiplets in the adjoint representation.  It is convenient to denote the chiral and fermi multiplets by $\Phi_n$ and $\Lambda_{mn}= - \Lambda_{nm}$, with $n=1, \cdots 5$, since $\Phi_n$ and $\Lambda_{nm}$ transform in the ${\bf 5}$ and ${\bf 10}$ of an  $SU(5)$ flavor symmetry. This spectrum is reproduced by the $\Ext$ groups above. We identify the fields with the $\Ext$ elements according to $X^1_n = \phi_n$, $n=1, \cdots, 5$, and:
\bea\label{def lambda SU5}
X_1^2= \lambda_{21}~, \quad
X_2^2= \lambda_{31}~, \quad
X_3^2= \lambda_{32}~, \quad
X_4^2= \lambda_{41}~, \quad
X_5^2= \lambda_{42}~, \cr
X_6^2= \lambda_{43}~, \quad
X_7^2= \lambda_{51}~, \quad
X_8^2= \lambda_{52}~, \quad
X_9^2= \lambda_{53}~, \quad
X_{10}^2= \lambda_{54}~.
\eea
The interaction terms are determined by the $F$- and $H$-terms \cite{Franco:2016tcm}:
\be
F_{mn}= \phi_m \phi_n - \phi_n \phi_m~, \qquad \quad H^{mn, pq}= \epsilon^{mnpq r}\phi_r~.
\ee
One can check that the open-string correlators \eqref{311}-\eqref{221} precisely reproduce these interactions. Note that, to check that the $H$-term:
\be
S_H = {1\ov 4} \epsilon^{mnpqr} \Tr(\lambda_{mn}\lambda_{pq} \phi_r)
\ee
is supersymmetric, we need to use the Jacobi identity for $U(N)$. This is equivalent to the non-trivial condition $H^{ab}F_b=0$ mentioned above, which must always be realized by the $B$-brane correlators.

\subsection{Orbifolds $\C^5/\Gamma$}
Given the above results for $\C^5$, we can easily study various $\CN=1$-preserving orbifolds $\C^5 / \Gamma$, where $\Gamma$ is any discrete subgroup of $SU(5)$.  

\subsubsection{$\C^5/\Z_5(1,1,1,1,1)$}

\begin{figure}[t]
\begin{center}
\subfigure[\small $\C^5/\Z_5(1,1,1,1,1)$  quiver.]{
\includegraphics[height=5cm]{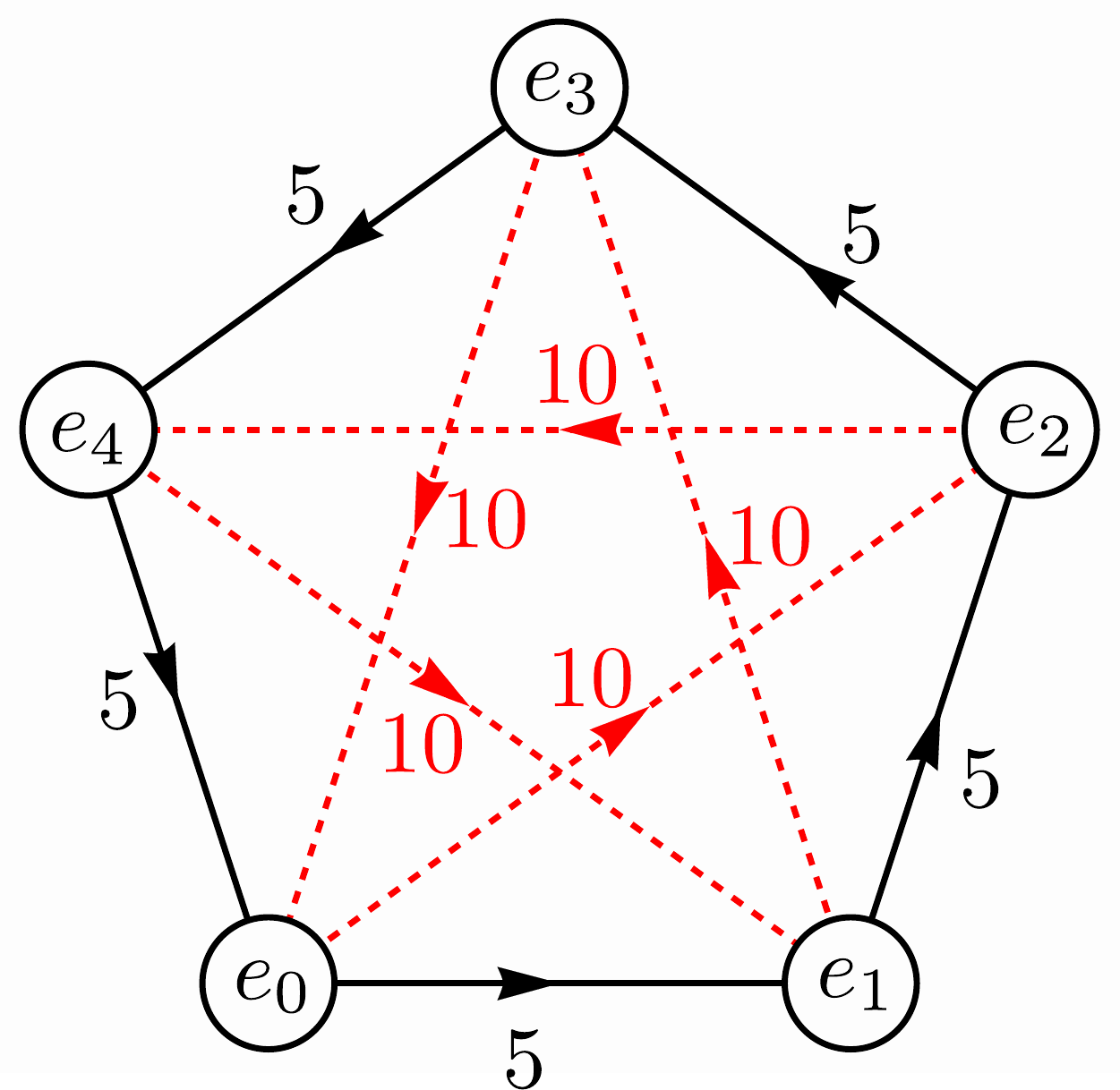}\label{fig:C5Z5}}\quad\qquad
\subfigure[\small $\C^5/\Z_3(1,1,1,1,2)$  quiver.]{
\includegraphics[height=5cm]{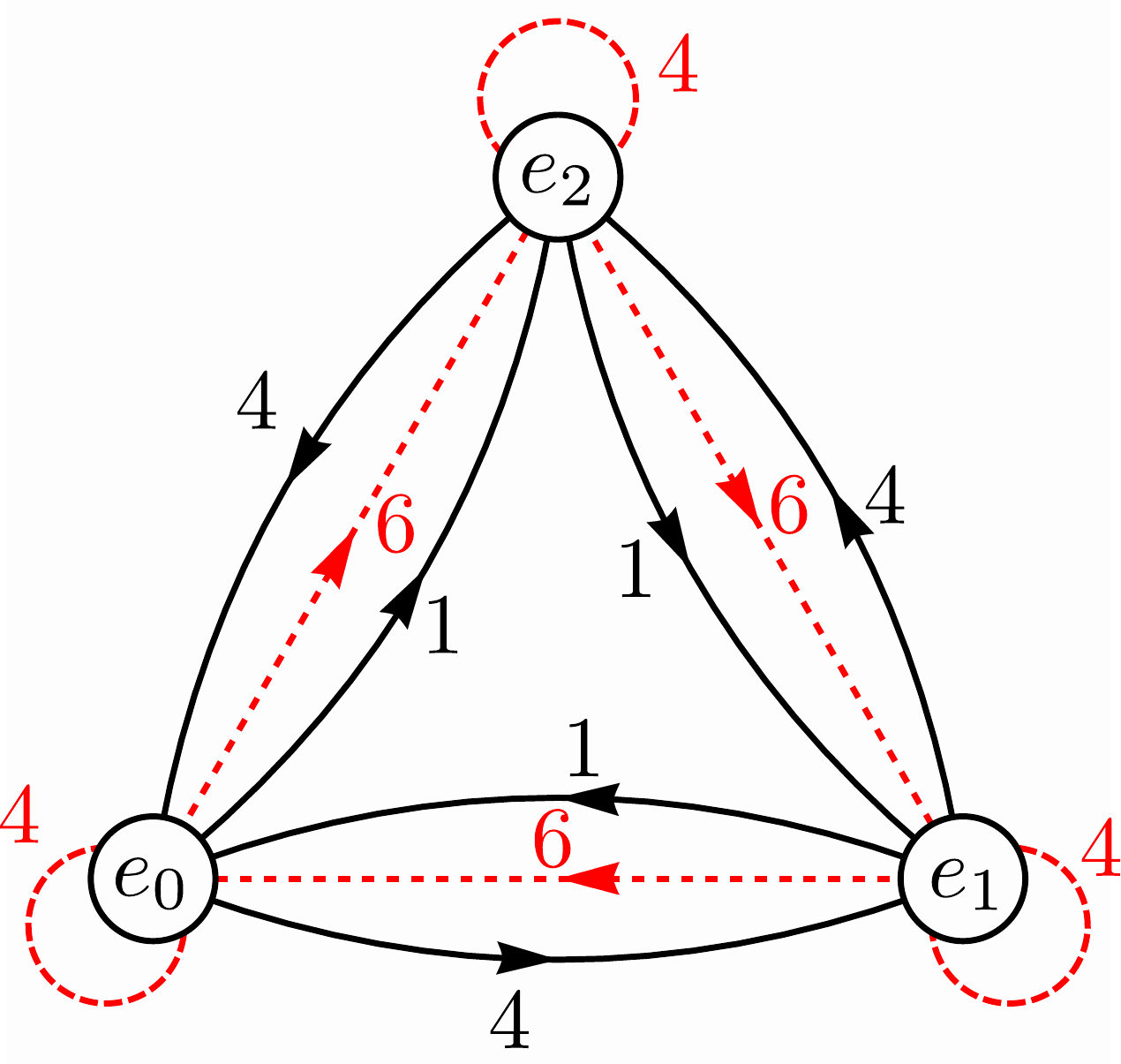}\label{fig:C5Z3}}\quad\;
\caption{Examples of 0d $\CN=1$ quivers for orbifold singularities $\C^5/\Gamma$. The numbers on the chiral and fermi multiplet arrows indicate their  multiplicities.}
\label{fig: C5orb}
\end{center}
\end{figure}

Consider for instance $\C^5/\Z_5$, where $\Z_5$ acts as:
\be
(x,y,z,w,t) \mapsto (\omega x,\omega y,\omega z,\omega w,\omega t)~, \qquad \omega= e^{2 \pi i \ov 5}
\ee
on the $\C^5$ coordinates. We have five fractional branes denoted by $\CE_i$, $i=0,\cdots, 4$. The weights for the  sheaves in the Koszul resolution of $\mathcal{E}_i$ are given by:
{\small\[
\left( \begin{array} {c}
i \end{array} \right) \xrightarrow[]{E} \left( \begin{array} {c}
i+1 \\ i+1 \\ i+1 \\ i+1 \\ i+1 \end{array} \right)
\xrightarrow[]{D} \left( \begin{array} {c}
i+2 \\ i+2 \\ i+2 \\ i+2 \\ i+2 \\ i+2 \\ i+2 \\ i+2 \\ i+2 \\ i+2 \end{array} \right)
\xrightarrow[]{C} \left( \begin{array} {c}
i+3 \\ i+3 \\ i+3 \\ i+3 \\ i+3 \\ i+3 \\ i+3 \\ i+3 \\ i+3 \\ i+3 \end{array} \right)
\xrightarrow[]{B} \left( \begin{array} {c}
i+4 \\ i+4 \\ i+4 \\ i+4 \\ i+4 \end{array} \right)
\xrightarrow[]{A} \left( \begin{array} {c}
i \end{array} \right)
\]}
We then find the spectrum:
\bea\nn
&\mathrm{Ext}^0_{[\mathbb{C}^5/\mathbb{Z}_5]}(\mathcal{E}_i, \mathcal{E}_j)=\left\{
\begin{array}{ll}
\mbox{Span}_{\mathbb{C}}\{X^0_1\} \quad&{\rm if}\; j \equiv i ~\mbox{mod} \, 5~,\\
0 & \mbox{otherwise,}
\end{array}
\right.\cr
&\mathrm{Ext}^1_{[\mathbb{C}^5/\mathbb{Z}_5]}(\mathcal{E}_i, \mathcal{E}_j)=\left\{
\begin{array}{ll}
\mbox{Span}_{\mathbb{C}}\{X^1_1,X^1_2,X^1_3,X^1_4,X^1_5\} \quad&{\rm if}\;  j+1 \equiv i ~\mbox{mod} \, 5~,\\
0 & \mbox{otherwise,}
\end{array}
\right.\cr
&\mathrm{Ext}^2_{[\mathbb{C}^5/\mathbb{Z}_5]}(\mathcal{E}_i, \mathcal{E}_j)=\left\{
\begin{array}{ll}
\mbox{Span}_{\mathbb{C}}\{X^2_1,X^2_2,X^2_3,X^2_4,X^2_5,X^2_6,X^2_7,X^2_8,X^2_9,X^2_{10}\} \;&{\rm if}\;  j+2 \equiv i ~\mbox{mod} \,5~,\\
0 & \mbox{otherwise.}
\end{array}
\right.\cr
\eea
The higher $\Ext$ groups are obtained by Serre duality.
The correlation functions can be read off from \eqref{311}-\eqref{221}. 
Let us introduce the chiral multiplets:
\be
\Phi^n_I : e_I \longrightarrow e_{I+1}~, \qquad \Lambda_I^{mn} : e_I\dashrightarrow e_{I+2}~,
\ee
with $I$ an integer mod $5$, $m, n= 1, \cdots, 5$, and $\Lambda_I^{mn}= - \Lambda_I^{nm}$. The gauged matrix model quiver is shown in Figure~\ref{fig:C5Z5}. The interaction terms are:
\be
F_{\Lambda_{I}^{mn}} = \Phi_I^m \Phi_{I+1}^n-  \Phi_I^n \Phi_{I+1}^m~, \qquad
H^{\Lambda_{I}^{mn}, \Lambda_{I+2}^{pq}}= \epsilon_{mnpqr} \Phi_{I-1}^r~.
\ee
Note the obvious $SU(5)$ flavor symmetry.
This quiver was discussed in \cite{Diaconescu:2000ec, Douglas:2002fr, Franco:2016tcm}.

\subsubsection{$\mathbb{C}^5/\mathbb{Z}_3$(11112)}
 As a last example, consider the $\C^5/\Z_3$ orbifold:
\be
(x,y,z,w,t) \mapsto (\omega x, \omega y, \omega z, \omega w, \omega^2 t)~, \qquad \omega= e^{2\pi i \ov 3}~.
\ee
We have three fractional branes $\CE_i$, $i=0,1,2$.
The weights for the sheaves in the Koszul resolution of $\mathcal{E}_i$ are:
{\small\[
\left( \begin{array} {c}
i \end{array} \right) \xrightarrow[]{E} \left( \begin{array} {c}
i+2 \\ i+1 \\ i+1 \\ i+1 \\ i+1 \end{array} \right)
\xrightarrow[]{D} \left( \begin{array} {c}
i \\ i \\ i+2 \\ i \\ i+2 \\ i+2 \\ i \\ i+2 \\ i+2 \\ i+2 \end{array} \right)
\xrightarrow[]{C} \left( \begin{array} {c}
i+1 \\ i+1 \\ i+1 \\ i \\ i+1 \\ i+1 \\ i \\ i+1 \\ i \\ i \end{array} \right)
\xrightarrow[]{B} \left( \begin{array} {c}
i+2 \\ i+2 \\ i+2 \\ i+2 \\ i+1 \end{array} \right)
\xrightarrow[]{A} \left( \begin{array} {c}
i \end{array} \right)
\]}
The spectrum consists of:
\bea\nn
& \mathrm{Ext}^0_{[\mathbb{C}^5/\mathbb{Z}_3]}(\mathcal{E}_i, \mathcal{E}_j)=\left\{
\begin{array}{ll}
\mbox{Span}_{\mathbb{C}}\{X^0_1\} \quad &{\rm if}\; j \equiv i ~\mbox{mod}~3~,\\
0 & \mbox{otherwise,}
\end{array}\right. 
\eea
\bea
& \mathrm{Ext}^1_{[\mathbb{C}^5/\mathbb{Z}_3]}(\mathcal{E}_i, \mathcal{E}_j)=\left\{
\begin{array}{ll}
0  \quad &{\rm if}\; j \equiv i ~\mbox{mod}~3~,\\
\mbox{Span}_{\mathbb{C}}\{X^1_2,X^1_3,X^1_4,X^1_5\}  \quad &{\rm if}\; j+1 \equiv i ~\mbox{mod}~3~,\\
\mbox{Span}_{\mathbb{C}}\{X^1_1\} \quad &{\rm if}\;  j+2 \equiv i ~\mbox{mod}~3~,
\end{array}\right.
\cr
&\mathrm{Ext}^2_{[\mathbb{C}^5/\mathbb{Z}_3]}(\mathcal{E}_i, \mathcal{E}_j)=\left\{
\begin{array}{ll}
\mbox{Span}_{\mathbb{C}}\{X^2_1,X^2_2,X^2_4,X^2_7\}\quad &{\rm if}\; j \equiv i ~\mbox{mod}~3~,\\
\quad &{\rm if}\; j+1 \equiv i ~\mbox{mod}~3~,\\
\mbox{Span}_{\mathbb{C}}\{X^2_3,X^2_5,X^2_6,X^2_8,X^2_9,X^2_{10}\}\quad &{\rm if}\; j+2 \equiv i ~\mbox{mod}~3~.
\end{array}\right.
\eea
The corresponding 0d $\CN=1$ quiver is shown in Figure \ref{fig:C5Z3}. 
The correlation functions can be read off from \eqref{311}-\eqref{221}. 
Taking advantage of the residual $SU(4)$ flavor symmetry, 
let us introduce the chiral multiplets:
\be
A_I \, : e_I \overset{X_1^1}{\longrightarrow} e_{I-1}~, \qquad
B_I^a\, :  e_I \overset{X_1^{a+1}}{\longrightarrow} e_{I+1}~, 
\ee
with $a=1,\cdots 4$, and $I$ an integer mod $3$. Similarly, we define the fermi multiplets:
\be
\Lambda_{I}^{ab} \, : e_I \overset{\lambda_{a+1, b+1}}{\dashrightarrow} e_{I-1}~, \qquad
\Xi_{I}^{a} \, : e_I \overset{\lambda_{a+1, 1}}{\dashrightarrow} e_{I}~, 
\ee
where the $\Ext^1$ elements $\lambda_{mn}= - \lambda_{nm}$ are defined as in \eqref{def lambda SU5}.
In this notation, the interaction terms read:
\bea
&F_{\Lambda^{ab}_I}= B_I^a B_{I+1}^b-B_I^b B_{I+1}^a~, \qquad 
&& F_{\Xi_I^a} = - A_I B_{I-1}^a~, \cr
&H_{\Xi_I^a, \Lambda^{bc}_{I}}= - \epsilon_{abcd} B^d_{I-1}~, \qquad
&&  H_{\Lambda^{ab}_{I}, \Xi_{I-1}^c}=  \epsilon_{abcd} B^d_{I-1} \cr
&H_{\Lambda^{ab}_I, \Lambda^{cd}_{I-1}}=  \epsilon_{abcd} A_{I+1}~.
\eea
Many more $\CN=1$ matrix models can be worked out in this way. It would also be  instructive to study fractional branes on local Fano fourfold varieties, such as the resolution of the $\C^5/\Z_5(1,1,1,1,1)$ to ${\rm Tot}(\CO(-5) \rightarrow \mathbb{P}^4)$. We leave this and many other related questions for future work.

\section*{Acknowledgements}
We would like to thank P.~Aspinwall, S.~Franco, D.~Ghim,  C.~Herzog, S.~Katz, W.~Lerche, I.~Melnikov, T.~Pantev, and R.~K.~Seong for useful conversations and comments.
E.S. was partially supported by NSF grants PHY-1417410 and PHY-1720321.

\appendix

\section{Dimensional reductions}\label{app:dim red}

\paragraph{Fourfolds versus threefolds.}

Let $X$ be a Calabi-Yau orbifold $[X_c/G]$ of complex dimension $3$,
with a set of fractional branes $\{ {\cal E}_i \}$ supported at a point
$p \in X_c$, a fixed point of the $G$-action.
Let $N_3$ denote the normal bundle $N_{p/X}$.

Let us build another Calabi-Yau orbifold $Y = {\mathbb C} \times X$,
which again has an isomorphic set of fractional branes $\{ {\cal E}_i \}$,
supported at $x \equiv \{0\} \times p \in {\mathbb C} \times X$,
of codimension four.
Let $N$ denote the normal bundle to $x$ in $Y$, and
$\rho_0$ the structure sheaf with trivial $G$-equivariant structure.
Then,
\begin{eqnarray*}
N & = & \rho_0 \oplus N_3~, \\
\wedge^2 N & = & \rho_0 \otimes N_3 \oplus \wedge^2 N_3 \: = \:
N_3 \oplus \wedge^2 N_3~, \\
\wedge^3 N & = & \rho_0 \otimes \wedge^2 N_3 \oplus \wedge^3 N_3
\: = \: \wedge^2 N_3 \oplus \wedge^3 N_3~, \\
\wedge^4 N & = & \rho_0 \otimes \wedge^3 N_3 \: = \: \wedge^3 N_3~.
\end{eqnarray*}
We then have:
\begin{eqnarray*}
{\rm Ext}^0_Y( {\cal E}_i, {\cal E}_j ) & = &
H^0(x, \rho_i^* \otimes \rho_j )^G \\
& = & {\rm Ext}^0_X( {\cal E}_i, {\cal E}_j)~, 
\\
{\rm Ext}^1_Y({\cal E}_i, {\cal E}_j) & = &
H^0(x, \rho_i^* \otimes \rho_j \otimes N )^G \: = \:
H^0(x, \rho_i^* \otimes \rho_j \otimes (\rho_0 \oplus N_3) )^G \\
& = & {\rm Ext}^0_X({\cal E}_i, {\cal E}_j) \: \oplus \:
{\rm Ext}^1_X({\cal E}_i, {\cal E}_j)~, 
\\
{\rm Ext}^2_Y( {\cal E}_i, {\cal E}_j) & = &
H^0(x, \rho_i^* \otimes \rho_j \otimes \wedge^2 N )^G
\: = \: H^0(x, \rho_i^* \otimes \rho_j \otimes (N_3 \oplus \wedge^2 N_3) )^G\\
& = & {\rm Ext}^1_X({\cal E}_i, {\cal E}_j) \: \oplus \:
{\rm Ext}^2_X( {\cal E}_i, {\cal E}_j)~, 
\\
{\rm Ext}^3_Y({\cal E}_i, {\cal E}_j) & = &
H^0(x, \rho_i^* \otimes \rho_j \otimes \wedge^3 N)^G \: = \:
H^0(x, \rho_i^* \otimes \rho_j \otimes (\wedge^2 N_3 \oplus \wedge^3 N_3) )^G\\
& = & {\rm Ext}^2_X({\cal E}_i,{\cal E}_j) \: \oplus \:
{\rm Ext}^3_X({\cal E}_i,{\cal E}_j)~, 
\\
{\rm Ext}^4_Y({\cal E}_i, {\cal E}_j) & = &
H^0(x,\rho_i^* \otimes \rho_j \otimes \wedge^4 N)^G \: = \:
H^0(x, \rho_i^* \otimes \rho_j \otimes \wedge^3 N_3)^G \\
& = & {\rm Ext}^3_X({\cal E}_i,{\cal E}_j)~.
\end{eqnarray*}
This directly confirms \eqref{3 to 4 ext} in the case of an orbifold singularity. We conjecture that it holds more generally.

\vskip0.5cm
\paragraph{Fourfolds versus twofolds.}

Similarly, we may consider $X$ a Calabi-Yau orbifold $[X_c/G]$ of complex dimension $2$,
with a set of fractional branes $\{ {\cal E}_i \}$ supported at a point
$p \in X_c$, a fixed point of the $G$-action.
Let $N_X$ denote the normal bundle $N_{p/X}$.
Let us build another Calabi-Yau orbifold $Y = {\mathbb C}^2 \times X$,
which again has an isomorphic set of fractional branes $\{ {\cal E}_i \}$,
supported at $x \equiv \{(0,0)\} \times p \in {\mathbb C}^2 \times X$,
of codimension four.
Let $N$ denote the normal bundle to $x$ in $Y$, and
$\rho_0$ the structure sheaf with trivial $G$-equivariant structure.
Then,
\begin{eqnarray*}
N & = & \rho_0^2 \oplus N_X~, \\
\wedge^2 N & = & \rho_0\otimes\rho_0 \oplus (\rho_0 \otimes N_X)^{\oplus 2} 
\oplus \wedge^2 N_X \: = \: \rho_0 \oplus N_X^{\oplus 2} \oplus \wedge^2 N_X, \\
\wedge^3 N & = & \rho_0\otimes \rho_0 \otimes N_X \oplus ( \rho_0 \otimes
\wedge^2 N_X)^{\oplus 2} \: = \:  N_X \oplus (\wedge^2 N_X)^{\oplus 2}~, \\
\wedge^4 N & = & \rho_0 \otimes \rho_0 \otimes \wedge^2 N_X \: = \:
\wedge^2 N_X~.
\end{eqnarray*}
We then find:
\begin{eqnarray*}
{\rm Ext}^0_Y( {\cal E}_i, {\cal E}_j) & = &
H^0(x, \rho_i^* \otimes \rho_j)^G \\
& = & {\rm Ext}^0_X( {\cal E}_i, {\cal E}_j)~, 
\\
{\rm Ext}^1_Y( {\cal E}_i, {\cal E}_j ) & = &
H^0(x, \rho_i^* \otimes \rho_j \otimes N)^G \: = \:
H^0(x, \rho_i^* \otimes \rho_j \otimes ( \rho_0^2 \oplus N_X) )^G \\
& = & 
{\rm Ext}^0_X( {\cal E}_i, {\cal E}_j) \oplus
{\rm Ext}^0_X( {\cal E}_i, {\cal E}_j) \oplus
{\rm Ext}^1_X( {\cal E}_i, {\cal E}_j)~, 
\\
{\rm Ext}^2_Y( {\cal E}_i, {\cal E}_j) & = &
H^0(x, \rho_i^* \otimes \rho_j \otimes \wedge^2 N)^G \: = \:
H^0(x, \rho_i^* \otimes \rho_j \otimes ( \rho_0 \oplus (N_X)^{\oplus 2} \oplus
\wedge^2 N_X) )^G \\
& = &
{\rm Ext}^0_X({\cal E}_i,{\cal E}_j) \oplus
{\rm Ext}^1_X({\cal E}_i,{\cal E}_j) \oplus
{\rm Ext}^1_X({\cal E}_i,{\cal E}_j) \oplus
{\rm Ext}^2_X({\cal E}_i, {\cal E}_j)~, 
\\
{\rm Ext}^3_Y({\cal E}_i,{\cal E}_j) & = &
H^0(x, \rho_i^* \otimes \rho_j \otimes \wedge^3 N)^G \: = \:
H^0(x, \rho_i^* \otimes \rho_j \otimes (N_X \oplus (\wedge^2 N_X)^{\oplus 2})
)^G \\
& = &
{\rm Ext}^1_X({\cal E}_i, {\cal E}_j) \oplus
{\rm Ext}^2_X({\cal E}_i, {\cal E}_j) \oplus
{\rm Ext}^2_X({\cal E}_i, {\cal E}_j)~,
\\
{\rm Ext}^4_Y({\cal E}_i,{\cal E}_j) & = &
H^0(x, \rho_i^* \otimes \rho_j \otimes \wedge^4 N)^G \: = \:
H^0(x, \rho_i^* \otimes \rho_j \otimes \wedge^2 N_X)^G \\
& = &
{\rm Ext}^2_X({\cal E}_i, {\cal E}_j)~.
\end{eqnarray*}
This decomposition corresponds to the dimensional reduction of a 6d $\CN=1$ quiver theory (or, equivalently, of a 4d $\CN=2$ theory) to 2d, giving rise to an $\CN=(4,4)$ quiver theory. Each $\CN=(4,4)$ vector  multiplet splits into one $\CN=(2,2)$ vector multiplet, two chiral multiplets and one fermi multiplet. Each $\CN=(4,4)$ hypermultiplet splits into two chiral and two fermi multiplets. This is precisely the decomposition seen here.

\section{Fractional D3-branes on a local ${\mathbb P}^2$}
\label{sect:warmup:local-p2}
Consider the well-known case of fractional D3-branes on the Calabi-Yau threefold:
\be\label{X3 expl}
\t{\bf X}_3 = {\rm Tot}(\CO(-3) \rightarrow \mathbb{P}^2)~,
\ee
which is a crepant resolution of the orbifold singularity ${\bf X}_3= \C^3/\Z_3$. The corresponding 4d $\CN=1$ quiver gauge theory is very well studied---see {\it e.g.} \cite{Douglas:1997de, Kachru:1998ys, Douglas:2000qw, Cachazo:2001sg}. In this Appendix, we review this 4d $\CN=1$ quiver using the $B$-brane language. This will help to  illustrate,   in a more familiar context, the tools that we similarly use to study D1-brane quivers.

\subsection{Fractional branes and supersymmetric quivers}
Let us discuss two particular sets of fractional branes. Below, we will see how they are related by mutation of exceptional collections, providing a geometric realization of Seiberg duality \cite{Cachazo:2001sg}.
\subsubsection{A first set of fractional branes: Theory (I)}
Fractional branes on the resolution \eqref{X3 expl} can be constructed from the data of a strongly exceptional collection on $\mathbb{P}^2$, as in section \ref{subsec: mutations}. Let us first consider the exceptional collection:
\be\label{exp coll I P2}
{\bf E}_I = \{\Omega^2(2)~, \; \Omega^1(1)~, \; \CO\}~.
\ee
The corresponding three fractional branes on $\t{\bf X}_3$ are:
\be
\CE_0 = i_* {\cal O}~, \qquad  \CE_1=  i_* \Omega^1(1)[1]~, \qquad \CE_2= i_* \Omega^2(2)[2]~,
\ee
where $i$ is the inclusion from ${\mathbb P}^2$ into $\t{\bf X}_3$.

Let $z_0, z_1, z_2$ be the homogeneous coordinates of $\mathbb{P}^2$
and $U_i$ be the open set in which $z_i \neq 0$. Denote the local
coordinates in $U_i$ by $(x_i, y_i)$ and the coordinate of the fiber
of $\mathcal{O}(-3)$ in $U_i$ by $w_i$. We have $w_1 = x_0^3 w_0,
w_2 = y_0^3 w_0 = y_1^3 w_1$. In the following we will take Koszul
resolutions:
\[
\begin{split}
&0 \xrightarrow[]{} \mathcal{O}(k+3) \xrightarrow[]{w_0}
\mathcal{O}(k) \xrightarrow[]{} i_* \mathcal{O}_{\mathbb{P}^2}(k)
\xrightarrow[]{} 0\\
&0 \xrightarrow[]{} \Omega(k+3) \xrightarrow[]{w_0} \Omega(k)
\xrightarrow[]{} i_* \Omega_{\mathbb{P}^2}(k) \xrightarrow[]{} 0
\end{split}
\]
\vskip-0.1cm\noindent
It is straightforward to compute the Ext groups themselves. The $\Ext^1$ quiver reads:
\[
\xymatrix{
 & i_* \mathcal{O}(-1)[2] \ar[dr]^c_3 &  \\
 i_* \mathcal{O} \ar[ur]_3^a & &
 i_* \Omega(1)[1] \ar[ll]^b_3
}
\]
\vskip0.0cm\noindent
A basis of the Ext groups can be chosen as follows:

\vskip0.2cm\noindent
$\Ext^1(i_*  \CO, i_* \mathcal{O}(-1)[2])$ is generated by 
$a_i \in \check{C}^2(\mathit{Hom}^{-1}(i_* \mathcal{O}, i_*\mathcal{O}(-1)[2]))$:
\[
\begin{CD}
@. \mathcal{O}(3) @>{}>> \mathcal{O} \\
@. @Va_iVV @.\\
\mathcal{O}(2) @>{}>> \mathcal{O}(-1) @.
\end{CD}
\]
\[
a_1 = \frac{1}{x_0 y_0}~,\qquad  a_2 = \frac{1}{x_0^2 y_0}~,\qquad
 a_3 = \frac{1}{x_0 y_0^2}~.
\]
\vskip0.2cm\noindent
$\Ext^1(i_* \Omega(1)[1], i_*  \CO)$ is generated by 
$b_i \in \check{C}^0(\mathit{Hom}^{1}(i_* \Omega(1)[1], i_*\mathcal{O}))$:
\[
b_1:~~~~~
\begin{CD}
\Omega(4) @>{}>> \Omega(1) \\
@V{(x_0,y_0)}VV @V{(-x_0,-y_0)}VV \\
\mathcal{O}(3) @>{}>> \mathcal{O}
\end{CD}~~~
b_2:~~~~~
\begin{CD}
\Omega(4) @>{}>> \Omega(1) \\
@V{(-1,0)}VV @V{(1,0)}VV \\
\mathcal{O}(3) @>{}>> \mathcal{O}
\end{CD}~~~
b_3:~~~~~
\begin{CD}
\Omega(4) @>{}>> \Omega(1) \\
@V{(0,-1)}VV @V{(0,1)}VV \\
\mathcal{O}(3) @>{}>> \mathcal{O}
\end{CD}~~~
\]
\vskip0.2cm\noindent
$\Ext^1( i_*  \CO(-1)[2], i_* \Omega(1)[1])$ is generated by 
$c_i \in \check{C}^0(\mathit{Hom}^{1}(i_* \mathcal{O}(-1)[2], i_*
\Omega(1)[1]))$:
\[
c_1:~~~~~~
\begin{CD}
\mathcal{O}(2) @>{}>> \mathcal{O}(-1) \\
@V{\tiny \left(\begin{array}{c} y_0 \\ -x_0 \end{array}\right)}VV
@V{\tiny \left(\begin{array}{c} -y_0 \\ x_0 \end{array}\right)}VV \\
\Omega(4) @>{}>> \Omega(1)
\end{CD}~~~
c_2:~~~~~
\begin{CD}
\mathcal{O}(2) @>{}>> \mathcal{O}(-1) \\
@V{\tiny \left(\begin{array}{c} 0 \\ 1 \end{array}\right)}VV
@V{\tiny \left(\begin{array}{c} 0 \\ -1 \end{array}\right)}VV \\
\Omega(4) @>{}>> \Omega(1)
\end{CD}~~~
c_3:~~~~~
\begin{CD}
\mathcal{O}(2) @>{}>> \mathcal{O}(-1) \\
@V{\tiny \left(\begin{array}{c} -1 \\ 0 \end{array}\right)}VV
@V{\tiny \left(\begin{array}{c} 1 \\ 0 \end{array}\right)}VV \\
\Omega(4) @>{}>> \Omega(1)
\end{CD}~~~
\]
\vskip0.2cm\noindent
The generator of $\Ext^3(i_* \mathcal{O}, i_* \mathcal{O})$ can be
chosen to be $t \in \check{C}^2(X, \mathit{Hom}^{1}(i_* \mathcal{O},
i_* \mathcal{O}))$ with
\[
t|_{U_0} = \frac{1}{x_0 y_0}~.
\]
One can then compute:
\be\label{products C3Z3}
\gamma(m_2(a_i, m_2(b_j, c_k))) = \epsilon_{ijk}~.
\ee
Note that there is a $GL(3)$ symmetry inherited from $\mathbb{P}^2$, and a corresponding $SU(3)$ flavor symmetry in the $\CN=1$  gauge theory.

\begin{figure}[t]
\begin{center}
\subfigure[\small $\C^3/\Z_3$  quiver.]{
\includegraphics[height=4.5cm]{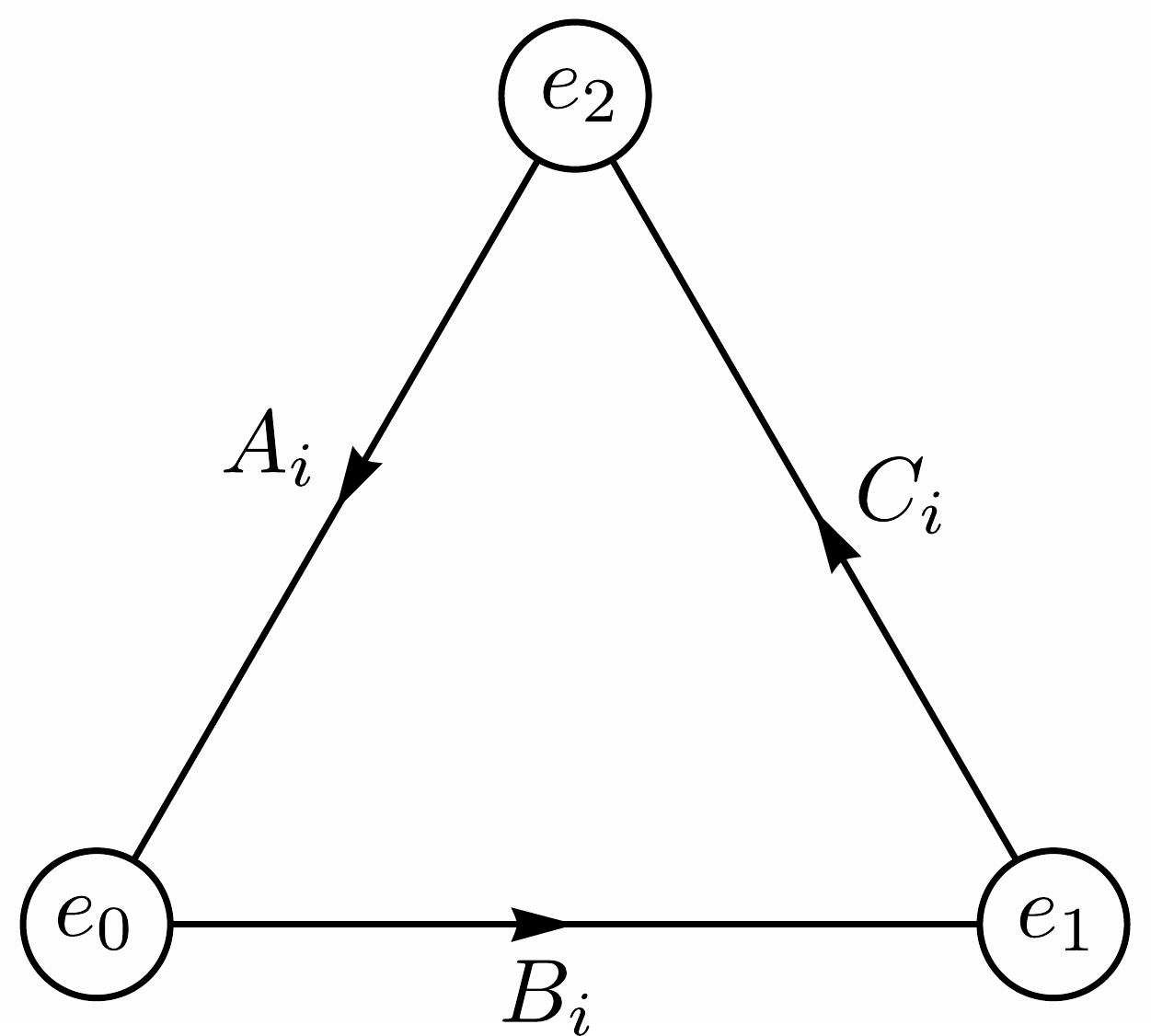}\label{fig:C3Z3 I}}\quad\qquad
\subfigure[\small A Seiberg dual quiver.]{
\includegraphics[height=4.5cm]{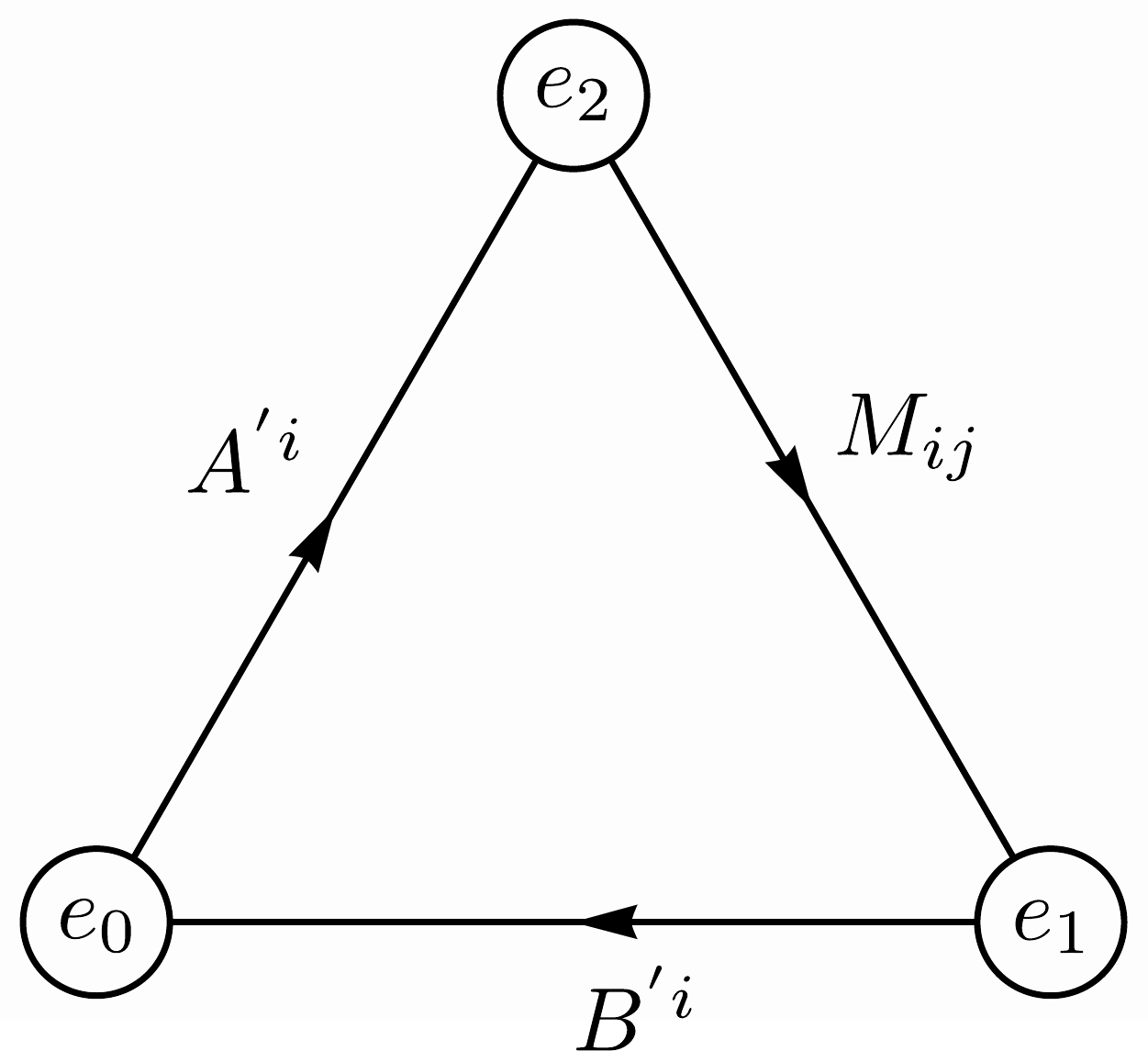}\label{fig:C3Z3 II}}\quad\;
\caption{The $\C^3/\Z_3$ quiver and a Seiberg dual. The arrows $A, B, C$ and $A'$, $B'$ each have multiplicity $3$, while the arrow $M$ has multiplicity $6$.}
\label{fig: C3Z3}
\end{center}
\end{figure}

The $\CN=1$ quiver gauge theory is the one shown in Figure~\ref{fig:C3Z3 I}, with a gauge group 
$U(N)\times U(N) \times U(N)$. 
The bifundamental chiral multiplets $A_i, B_i, C_i$ correspond to the $\Ext^1$ elements $a_i, b_i, c_i$, and the product structure \eqref{products C3Z3} leads to the $\CN=1$ superpotential:
\be\label{W th I}
W = \mathrm{Tr} (\epsilon^{ijk} A_i B_j C_k)~.
\ee
This quiver can also be obtained by  orbifold projection from 4d $\CN=4$ theory \cite{Douglas:1997de, Kachru:1998ys}.

\subsubsection{A second set of fractional branes: Theory (II)}
Consider another strongly exceptional collection on $\mathbb{P}^2$:
\be\label{exp coll II P2}
{\bf E}_{II} = \{ \CO(-1)~, \; \CO~, \CO(1) \}~.
\ee
The corresponding fractional branes are:
\be
\CE_0 = i_* {\cal O}[1]~, \qquad \CE_1= i_* {\cal O}(1)~, \qquad \CE_2= i_* {\cal O}(-1)[2]~.
\ee
We repeat the same analysis as before. The $\Ext^1$ quiver reads:
\[
\xymatrix{
 &i_* \mathcal{O}(-1)[2]   \ar[dl]^3_{a'} &  \\
 i_* \mathcal{O}[1]   \ar[rr]^3_{b'} & &
 i_* \mathcal{O}(1)  \ar[ul]^6_{d'}
}
\]
\vskip0.0cm\noindent
The corresponding $\CN=1$ quiver is shown in Figure~\ref{fig:C3Z3 II}.

\vskip0.2cm\noindent
 $\Ext^1(i_* \mathcal{O}(-1)[2], i_*  \CO[1])$ is generated by 
$a'_i  \in \check{C}^0(\mathit{Hom}^{1}(i_* \mathcal{O}(-1)[2], i_*\mathcal{O}[1]))$:
\[
a'_1:~~~~~~
\begin{CD}
\mathcal{O}(2) @>{}>> \mathcal{O}(-1) \\
@V-1VV
@V1VV \\
\mathcal{O}(3) @>{}>> \mathcal{O}
\end{CD}~~~
a'_2:~~~~~
\begin{CD}
\mathcal{O}(2) @>{}>> \mathcal{O}(-1) \\
@V-x_0VV
@Vx_0VV \\
\mathcal{O}(3) @>{}>> \mathcal{O}
\end{CD}~~~
a'_3:~~~~~
\begin{CD}
\mathcal{O}(2) @>{}>> \mathcal{O}(-1) \\
@V-y_0VV
@Vy_0VV \\
\mathcal{O}(3) @>{}>> \mathcal{O}
\end{CD}~~~
\]
\vskip0.2cm\noindent
$\Ext^1(i_* \mathcal{O}[1], i_*  \CO(1))$ is generated by 
$b'_i \in \check{C}^0(\mathit{Hom}^{1}(i_* \mathcal{O}[1], i_*
\mathcal{O}(1)))$:
\[
b'_1:~~~~~~
\begin{CD}
\mathcal{O}(3) @>{}>> \mathcal{O} \\
@V-1VV
@V1VV \\
\mathcal{O}(4) @>{}>> \mathcal{O}(1)
\end{CD}~~~
b'_2:~~~~~
\begin{CD}
\mathcal{O}(3) @>{}>> \mathcal{O} \\
@V-x_0VV
@Vx_0VV \\
\mathcal{O}(4) @>{}>> \mathcal{O}(1)
\end{CD}~~~
b'_3:~~~~~
\begin{CD}
\mathcal{O}(3) @>{}>> \mathcal{O} \\
@V-y_0VV
@Vy_0VV \\
\mathcal{O}(4) @>{}>> \mathcal{O}(1)
\end{CD}~~~
\]
\vskip0.2cm\noindent
$\Ext^1( i_*  \CO(1), i_* \mathcal{O}(-1)[2])$ is generated by 
$d'_n \in \check{C}^2(\mathit{Hom}^{-1}(i_* \mathcal{O}(1), i_*
\mathcal{O}(-1)[2]))$:
\[
\begin{CD}
@. \mathcal{O}(4) @>{}>> \mathcal{O}(1) \\
@. @Vd'_iVV @.\\
\mathcal{O}(2) @>{}>> \mathcal{O}(-1) @.
\end{CD}
\]
\[
d'_1 = \frac{1}{x_0 y_0}~,\qquad  d'_2 = \frac{1}{x_0 y_0^2}~,\qquad
 d'_3 =\frac{1}{x_0^2 y_0}~,
\]
\[
d'_4 = \frac{1}{x_0^2 y_0^2}~,\qquad  d'_5 = \frac{1}{x_0^3 y_0}~,\qquad
 d'_6 =\frac{1}{x_0 y_0^3}~.
\]

\vskip0.2cm\noindent The field theory is shown in Figure~\ref{fig:C3Z3 II}. The fields $A^{'i}$, $B^{'i}$ are both in the ${\bf 3}$ of the $SU(3)$ flavor symmetry, while the fields $M_{ij}= M_{ji}$ span the ${\bf 6}$ of $SU(3)$. They are identified with the $\Ext^1$ elements according to:
\be
A^{'i} = a_i~, \qquad B^{'i}= b_i~, \qquad
M_{ij}= \mat{d'_1 & d'_3 & d'_2 \\ d'_3 & d'_5 & d_4' \\ d'_2 & d'_4 & d'_6}~.
\ee
One can then derive the superpotential:
\be\label{W th II}
W= A^{' i} M_{ij} A^{'j}~.
\ee
Moreover, due to the non-abelian anomaly-cancellation condition, the gauge group must be $U(2N)\times U(N)\times U(N)$. This is also what is obtained from the usual rules of Seiberg duality.


\subsection{Seiberg duality as mutation}
The two $\CN=1$ quiver theories of Figure~\ref{fig: C3Z3} are related by a Seiberg duality on node $e_0$. Consider for instance the ``Theory (I)''. A Seiberg duality at node $e_0$ reverses the arrows $A_i$ and $B_j$ while generating the new mesons $\t M_{ij}$, with the identification $\t M_{ij}= A_i B_j$ across the duality. The superpotential dual to \eqref{W th I} reads:
\be
W= \epsilon^{ijk}\t M_{ij} C_k + A^{' i} \t M_{ij} B^{'j}~.
 \ee
 This contains a mass term for $C_i$ and the antisymmetric part of $\t M_{ij}$. Integrating those fields out, we are left with ``Theory (II)'', including the superpotential  \eqref{W th II}.  
 
 Similarly, if we start from Theory (II) and perform a Seiberg duality at node $e_0$, we flip the arrows $A^{'i}$, $B^{' j}$, and generate the dual mesons $N^{ij}= A^{'i}B^{'j} $, with the superpotential:
 \be
 W= M_{ij} N^{ji} + B_i N^{ij} A_j~.
 \ee
Integrating out the massive fields---$M_{ij}$ and the symmetric part of $N^{ij}$---we recover Theory (I) and \eqref{W th I}, with the identification $N^{ij}= - \epsilon^{ijk} C_k$.

\paragraph{Mutation of exceptional collection.} 
It was proposed in \cite{Cachazo:2001sg} that Seiberg duality could be realized
as mutation on exceptional collections of sheaves. 
Start with Theory (II) and the corresponding exceptional collection ${\bf E}_{II}$ \eqref{exp coll II P2}. Using the left mutation:
\be
L_\CO \CO(1) = \Omega^1(1)
\ee
on $\mathbb{P}^2$, we see that a left mutation of the collection ${\bf E}_{II}$ at the second sheaf precisely gives the collection ${\bf E}_{I}$ in \eqref{exp coll I P2}:
\be
\{ \CO(-1)~, \; \CO~, \CO(1) \} \quad  \rightsquigarrow\quad 
 \{ \CO(-1)~, \; \Omega^1(1)~, \; \CO\}~.
\ee
Therefore, the Seiberg duality at node $e_0$ of Theory (I) is indeed realized by a mutation of the underlying sheaves. This observation has been generalized to a number of other cases \cite{Herzog:2004qw}.

\section{$A_\infty$ structure and $\CN=(0,2)$ quiver}\label{app: Ainfinity}
In this Appendix, we discuss the $A_\infty$ structure of the $\Ext^\bullet$ algebra, and how it is related to the structure of the $\CN=(0,2)$ quiver. This discussion is a straightforward generalization of a similar discussion for 4d $\CN=1$ quivers by Aspinwall and Katz \cite{Aspinwall:2004bs}. See also \cite{1999math.....10179K, Brunner:1999jq, Douglas:2002fr}.

\subsection{An algebraic preliminary}\label{subsec: A and V}
Let $V$ be a graded vector space, and let $T(V)$ be the associated graded tensor algebra:
\be
T(V)= \bigoplus_{n=1}^\infty V^{\otimes n}~.
\ee
Let ${\bf d}$ be an derivative operator of degree $1$ acting on $T(V)$, satisfying the graded Leibniz rule:
\be
{\bf d}(A\otimes B) = {\bf d}A\otimes B + (-1)^{|A|} A\otimes {\bf d}B~,
\ee
with $A, B \in T(V)$, and $|A|$ denoting the degree of $A$. We also require that:
\be\label{d2zero}
{\bf d}^2=0~.
\ee
Using the Leibniz rule, the action of ${\bf d}$ on $T(V)$ is determined by its action on $V$ itself. Let us decompose ${\bf d}$ as:
\be\label{dk def}
{\bf d}\big|_V = d_1 + d_2+ \cdots~, \qquad\qquad {\rm with}\qquad d_k: V\rightarrow V^{\otimes k}~.
\ee
Let $V[1]$ denote the vector space $V$ with all degrees decreased by one, and let $s: V\rightarrow V[1]$ denote the corresponding map of degree $-1$. Given this data, we can define an $A_\infty$ algebra $A$ as being the {\it dual} of $V[1]$:
\be
A= \left(V[1]\right)^\ast~,
\ee
together with the multi-products:
\be
m_k \; : \; A^{\otimes k} \rightarrow A
\ee
given by the dual of the map $s^{\otimes k} \cdot d_k \cdot s^{-1} : V[1] \rightarrow V[1]^{\otimes k}$. The nilpotency condition \eqref{d2zero} is equivalent to the following $A_\infty$ relation on the multi-products:
\be\label{Ainfinty rel}
\sum_{r+s+t=n} (-1)^{r + st} m_{n+1-s}({\bf 1}^{\otimes r}\otimes m_s \otimes {\bf 1}^{\otimes t})=0~, \qquad \forall n>0~,
\ee
where the sum is over all $r, t\geq 0$, $s>0$, such that $r+s+t=n$ \cite{1999math.....10179K}.

\subsection{Ext algebra and $\CN=(0,2)$ quiver}
In our physical setup, the vector space $A$ is spanned by the various $\Ext^i$ groups ($i=0, \cdots, 4$) among the fractional branes on a CY$_4$ singularity. Schematically:
\be\label{A explicit}
A\cong \Ext^0 \oplus \Ext^1 \oplus \Ext^2 \oplus \Ext^3 \oplus \Ext^4~.
\ee
 The grading of $A$ is given by the degree $i$ of $\Ext^i$. Any $z\in A$ of degree $q$ is associated to a local vortex operator in the $B$-model, with the degree identified with the ghost number. Given $z\in A$, let $z^{(1)}$ denote the corresponding one-form descendant. The one-form operators can be used to deform the $B$-model according to \cite{Brunner:1999jq, Aspinwall:2004bs}: 
\be
S \rightarrow S + \sum_i Z_i \, z^{(1)}_i~.
\ee
The coupling $Z_i$ is identified with a ``quiver field'' in the space-time (D1-brane) theory. Note that $Z_i$ has degree $1-q_i$ if $z_i$ has degree $q_i$. The quiver fields are elements of the vector space $V$, in the notation of subsection \ref{subsec: A and V}. 

Let us denote by $\b z \in A$ the Serre dual of $z\in A$, with the $\Ext$ algebra $A$ given by  \eqref{A explicit}.  Let us then choose a basis of $A$ according to:
\be\label{basis of zi}
\{z_i \} = \{ e_0~,\, x_\alpha~,\, \alpha_I~, \, \b\alpha_I~,\,  \b x_\alpha~,\, \b e_0\}~,
\ee
with:
\be
e_0\in \Ext^0~, \quad x_\alpha \in \Ext^1~, \quad \alpha_I, \b \alpha_I \in \Ext^2~, \quad \b x_\alpha\in \Ext^3~, \quad \b e_0 \in \Ext^4~.
\ee
As discussed in the main text, the choice of basis for $\Ext^2$ is arbitrary. Any given choice introduces a distinction between the elements $\alpha$ and the Serre dual elements $\b\alpha$, which is a matter of convention.

The dual vector space $V$ spans the ``quiver fields''. We choose a basis of $V$:
\be\label{basis of Zis}
\{Z_i \} = \{ e~, \, X_\alpha~, \, \Lambda_I~, \, \b \Lambda^I~, \b X^\alpha~, \b e\}~,
\ee
dual to \eqref{basis of zi}. The element $e$ correspond to the vector multiplets, while $X_\alpha$  and $\Lambda_I$ correspond to the chiral and fermi multiplets, respectively. Note the degrees:
\be\nn
\begin{tabular}{c|cccccc}
& $e$ & $X_\alpha$ & $\Lambda_I$ & $ \b \Lambda^I~$ & $ \b X^\alpha$ & $\b e$  \\
\hline
 {\rm degree:} & $1$ & $0$ & $-1$ & $-1$ & $-2$& $-3$
 \end{tabular}
 \ee
 In particular, the chiral multiplets have degree $0$.
Given this explicit basis of $V$, we define a derivative ${\bf d}$ as follows. First, let us introduce the $\CN=(0,2)$ ``superpotential'':
\be\label{CW as trace}
\CW = \Tr\big( \Lambda_I \otimes J^I(X) + \b\Lambda^I \otimes  E_I(X)\big)~,
\ee
with $J_I(X)$ and $E_I(X)$ some arbitrary functions of the chiral multiplets $X_\alpha$. This $\CW$ is an arbitrary gauge-invariant function of degree $-1$ that is independent of $e$, except that we need to impose the constraint:
\be\label{EJ zero algebraic}
\Tr(E_I \otimes J^I)=0~.
\ee
Let us also define the derivatives:
\be
\d_\alpha \CW \equiv {\d \CW \ov \d X_\alpha}
\ee
by left derivation on $\CW$---that is, we use the cyclic property of the trace to write \eqref{CW as trace}
 with $X_\alpha$ on the left, and the derivative with respect to $X_\alpha$ is defined as the sum of all possible forms of $\CW$ with $X_\alpha$ in front, with $X_\alpha$ removed.
 Given the superpotential, we define the degree-one derivative ${\bf d}$  on $V$ as:~\footnote{This is the analogue of equations (30) and (39) of \cite{Aspinwall:2004bs}.}
\bea\label{def d explicit}
 &{\bf d} e\, &=&\;\;  -e \otimes e~, \cr
 &{\bf d} X_\alpha\, &=&\; \; X_\alpha \otimes e- e\otimes X_\alpha~,\cr
 &{\bf d} \Lambda_I\, &=&\;\;  E_I(X) - \Lambda_I \otimes e - e\otimes \Lambda_I~,\cr
  &{\bf d} \b\Lambda^I\, &=&\;\;  J^I(X) - \b\Lambda^I \otimes e - e\otimes \b\Lambda^I~,\cr
  &{\bf d} \b X^\alpha \, &=&\; \; \d_\alpha \CW - e\otimes \b X^\alpha+ \b X^\alpha \otimes e~, \cr
    &{\bf d} \b e \, &=&\; \;  \b X^\alpha \otimes X_\alpha - X_\alpha \otimes \b X^\alpha + \b \Lambda^I \otimes \Lambda_I + \Lambda_I \otimes \b \Lambda^I - \b e \otimes e - e\otimes \b e~.
\eea
(See also \cite{Eager:2018oww}[section 10.3].)
By direct computation, one can check that ${\bf d}^2=0$. The relations:
\be
{\bf d}^2e=0~, \qquad {\bf d}^2 X_\alpha=0~,\qquad  {\bf d}^2 \Lambda_I=0~, \qquad {\bf d}^2 \b\Lambda^I=0
\ee
are obvious.~\footnote{To check the last two relations, one uses that:
\be\nn
{\bf d}F(X) = F(X) \otimes e - e \otimes F(X)~,
\ee
for any degree-zero holomorphic function $F(X)$, which follows from the second line in \eqref{def d explicit}.}
The key relation is:
\be
{\bf d}^2 \b X^\alpha=0~,
\ee
which holds true if and only if the non-trivial constraint \eqref{EJ zero algebraic} is satisfied. This is nothing but the requirement that the $\CN=(0,2)$ superpotential be properly supersymmetric.
Since we explicitly displayed a nilpotent derivative ${\bf d}$ on the vector space $V$ spanned by the quiver fields, it follows from the general discussion above that the multi-products $m_k$ acting on the $\Ext$ vector space $A$ satisfy the $A_\infty$ relations \eqref{Ainfinty rel}. In this way, we see clearly that the $A_\infty$ relations on a CY$_4$ are intimately related to the supersymmetry constraint \eqref{EJ zero algebraic}.

We should also note that the differential ${\bf d}$ defined in \eqref{def d explicit} has:
\be\label{d1zero}
d_1=0~,
\ee
where $d_k$ is defined as in \eqref{dk def}, if and only if the potentials $E_I$ and $J^I$ do not contain any linear terms in $X_\alpha$. In such a case, we have $m_1=0$ in the dual $\Ext$ algebra, which gives us a minimal $A_\infty$ structure. Linear terms in $E_I$ or $J^I$ are mass terms, and the corresponding fields can always be integrated out, as discussed in examples in section~\ref{sec: duality}. Therefore, \eqref{d1zero} always holds for the low-energy quiver.

Similarly, we see from \eqref{dk def} that there exists non-zero higher products $m_k$ for $k= 2, \cdots, k_{\rm max}$, with $k_{\rm max}$ the highest order in the fields $X_\alpha$ that appear in the potentials $E_I$, $J^I$. In the simplest case when $E_I$, $J^I$ are all quadratic in the chiral multiplets, we have $m_k=0$ for $k \geq 3$, and the $A_\infty$ algebra reduces to an associative algebra with a product given by $m_2$.

\subsection{General procedure to compute the higher products}\label{app: mk algo}
Let us discuss in more detail the procedure to compute the higher products of the $\Ext^\bullet$ $A_\infty$ algebra \cite{Aspinwall:2004bs}, which we outlined in section \ref{subsec: Ainf and EJ}. Consider an $A_\infty$ algebra $\t A$ and the $A_\infty$ map:
\be
f: H^{\bullet}(\t A) \rightarrow \t A~.
\ee 
Let the first map:
\be
f_1= i: H^{\bullet}(\t A) \rightarrow \t A~,
\ee
be the inclusion map defined by picking representatives of cohomology classes, and let $d=\t m_1 : \t A \rightarrow \t A$ denote the differential on $\t A$. 
The first $A_{\infty}$ constraint on the maps $f_k$ reads:
\be\nn
i \circ m_2(\alpha, \beta) = i(\alpha) \circ i(\beta) + d f_2(\alpha, \beta)~.
\ee
We can compute $i(\alpha) \circ i(\beta)$, and use the result
to define $m_2(\alpha,\beta)$ and $f_2(\alpha,\beta)$.
The next $A_{\infty}$ constraint  is of the form:
\begin{eqnarray*}
\lefteqn{
i \circ m_3(\alpha, \beta, \gamma) =
} \\
& & 
f_2(\alpha, m_2(\beta, \gamma)) - f_2(m_2(\alpha, \beta),\gamma) +
i(\alpha) \circ f_2(\beta,\gamma) - f_2(\alpha,\beta) \circ i(\gamma) +
df_3(\alpha,\beta, \gamma).
\end{eqnarray*}
Using the previously-computed $m_2$ and $f_2$, this expression  allows us  to compute $m_3$ and $f_3$.  Proceeding inductively in this fashion,
one can construct $m_k$ and $f_k$ to any order $k$.



\section{Higher products on a local $\mathbb{P}^1\times \mathbb{P}^1$}\label{app: higher prod P1P1}
In this Appendix, we spell out the computation of the higher products on the local $\mathbb{P}^1\times \mathbb{P}^1$ geometry of section \ref{sec local P1P1}, using the procedure summarized in Appendix~\ref{app: mk algo}.

Consider the $\Ext$ element representations discussed in section \ref{subsec: Ext P1P1}.
From the composition of the  chain maps, one
finds the products:
\begin{equation}\label{m2 P1P1}
\begin{split}
&m_2(b_1, d_1) = \beta'_1~,\quad m_2(b_1, d_2) = \beta'_2~,\quad m_2(b_2, d_1) =
\beta'_3~,\quad m_2(b_2, d_2) = \beta'_4~, \\
&m_2(a_1, e_1) = \beta'_1~,\quad m_2(a_2, e_1) = \beta'_2~,\quad m_2(a_1, e_2) =
\beta'_3~,\quad m_2(a_2, e_2) = \beta'_4~.
\end{split}
\end{equation}
It follows that $f_2(b,d)=f_2(a,e)=0$.
Define the 1-cochains $\lambda$ and $\delta$ as follows:
\[
\begin{split}
&(\lambda)_{01}=(\lambda)_{02}=(\lambda)_{03}=(\lambda)_{12}=
(\lambda)_{13}=0~,\quad (\lambda)_{23}=x^{-1}u^{-1}~, \\
&(\delta)_{03}=(\delta)_{12}=(\delta)_{23}=0~,\quad
(\delta)_{01}=(\delta)_{02}=-(\delta)_{03}=u^{-1}.
\end{split}
\]
One can compute
\[
\begin{split}
&d_1 \cdot c_1 = d\lambda_1~,\qquad d_1 \cdot c_2 = d\lambda_2~,\\
&d_2 \cdot c_2 = d\delta_1~,\qquad d_2 \cdot c_2 = d\delta_2~,
\end{split}
\]
where the chain maps are defined by
\bea\nn
&\lambda_1 = \left(\begin{array}{c} -\lambda  \\ 0
\end{array}\right) (0,-\lambda)~,\quad 
&\lambda_2 = \left(\begin{array}{c} 0 \\ -\lambda
\end{array}\right) (\lambda,0)~,\quad \\
&\delta_1 = \left(\begin{array}{c} -\delta  \\ 0
\end{array}\right) (0,-\delta)~,\quad 
&\delta_2 = \left(\begin{array}{c} 0 \\ -\delta
\end{array}\right) (\delta,0)~,\quad \\
\eea
between the corresponding complexes. This implies $m_2(d,c)=0$ and:
\begin{equation}\label{dc}
\begin{split}
&f_2(d_1,c_1)=-\lambda_1~,\quad f_2(d_1,c_2)=-\lambda_2~, \\
&f_2(d_2,c_1)=-\delta_1~,\quad f_2(d_2,c_2)=-\delta_2~.
\end{split}
\end{equation}
Similarly, one can show:
\[
\begin{split}
&c_1 \cdot a_1 = d\lambda_1~,\quad   c_2 \cdot a_1 = d\lambda_2~,\\
&c_1 \cdot a_2 = d\delta_1~,\quad  c_2 \cdot a_2 = d\delta_2~.
\end{split}
\]
Thus, $m_2(c,a)=0$ and:
\[
\begin{split}
&f_2(c_1,a_1)=-\lambda_1~,\quad  f_2(c_2,a_1)=-\lambda_2~, \\
&f_2(c_1,a_2)=-\delta_1~,\quad   f_2(c_2,a_2)=-\delta_2~.
\end{split}
\]
Plugging these results into the $A_\infty$ map constraint:
\[
i m_3(d,c,a)=f_2(d,m_2(c,a))-f_2(m_2(d,c),a)+d \cdot
f_2(c,a)-f_2(d,c)\cdot a + df_3(d,c,a)~,
\]
we get:
\bea\label{m3_1}
&m_3(d_1,c_1,a_1) = 0~, \qquad && m_3(d_1,c_1,a_2) = \alpha_1~,\cr
&m_3(d_1,c_2,a_1) = 0~, \qquad &&  m_3(d_1,c_2,a_2) = \alpha_2~,\cr
&m_3(d_2,c_1,a_1) = -\alpha_1~, \qquad && m_3(d_2,c_1,a_2) = 0~,\cr
&m_3(d_2,c_2,a_1) = -\alpha_2~, \qquad && m_3(d_2,c_2,a_2) = 0~.
\eea
Similarly, if we define $\tau$ and $\theta$ by:
\[
\begin{split}
&(\tau)_{01}=(\tau)_{03}=(\tau)_{13}=(\tau)_{23}=0~,\quad
(\tau)_{02}=(\tau)_{12}=x^{-1}u^{-1}~, \\
&(\theta)_{01}=(\theta)_{02}=(\theta)_{12}=(\theta)_{23}=0~,\quad
(\theta)_{03}=(\theta)_{13}=-x^{-1}~,
\end{split}
\]
we get $m_2(e,c)=0, m_2(c,b)=0$ and
\begin{equation}\label{tau_theta}
\begin{split}
&f_2(e_1,c_1)=\tau_1~, \quad f_2(e_1,c_2)=\tau_2~, \\
&f_2(e_2,c_1)=\theta_1~, \quad f_2(e_2,c_2)=\theta_2~, \\
&f_2(c_1,b_1)=-\tau_1~, \quad f_2(c_2,b_1)=-\tau_2~, \\
&f_2(c_1,b_2)=-\theta_1~, \quad f_2(c_2,b_2)=-\theta_2~,
\end{split}
\end{equation}
where:
\bea\nn
&\tau_1 = \left(\begin{array}{c} \tau  \\ 0
\end{array}\right) (0,\tau)~, \qquad
&\tau_2 = \left(\begin{array}{c} 0 \\ \tau
\end{array}\right) (-\tau, 0)~, \\
&\theta_1 = \left(\begin{array}{c} \theta  \\ 0
\end{array}\right) (0,\theta)~, \qquad
&\theta_2 = \left(\begin{array}{c} 0 \\ \theta
\end{array}\right) (-\theta,0)~. \\
\eea
Plugging these results into:
\[
i m_3(e,c,b)=f_2(e,m_2(c,b))-f_2(m_2(e,c),b)+e \cdot
f_2(c,b)-f_2(e,c)\cdot b + df_3(e,c,b)~,
\]
we get:
\bea\label{m3_2}
&m_3(e_1,c_1,b_1) = 0~,\qquad && m_3(e_1,c_1,b_2) = -\alpha'_1~,\\
&m_3(e_1,c_2,b_1) = 0~,\qquad && m_3(e_1,c_2,b_2) = -\alpha'_2~,\\
&m_3(e_2,c_1,b_1) = \alpha'_1~,\qquad && m_3(e_2,c_1,b_2) = 0~,\\
&m_3(e_2,c_2,b_1) = \alpha'_2~,\qquad && m_3(e_2,c_2,b_2) = 0~.
\eea
This completes the computation of the three-product $m_3$. As a consistency check, one can verify that the $m_2$ and $m_3$ just computed satisfy the relevant $A_\infty$ relations. 

By the same procedure, we could also find the $m_4$ product, while the higher products vanish. We can use various short-cuts to the correct answer, however. For instance, we can impose $\Tr(EJ)=0$ in the gauge theory quiver, which is equivalent to imposing the $A_\infty$ relations. This leads to the result \eqref{m4 P1P1} for the $m_4$ products amongst the $\Ext^1$ elements.

\bibliographystyle{utphys}
\bibliography{bibBbranes}{}

\providecommand{\href}[2]{#2}\begingroup\raggedright\begin{thebibliography}{10}

\bibitem{Douglas:1996sw}
M.~R. Douglas and G.~W. Moore, ``{D-branes, quivers, and ALE instantons},''
\href{http://arxiv.org/abs/hep-th/9603167}{{\ttfamily arXiv:hep-th/9603167
  [hep-th]}}.

\bibitem{Douglas:1997de}
M.~R. Douglas, B.~R. Greene, and D.~R. Morrison, ``{Orbifold resolution by
  D-branes},'' \href{http://dx.doi.org/10.1016/S0550-3213(97)00517-8}{{\em
  Nucl. Phys.} {\bfseries B506} (1997) 84--106},
\href{http://arxiv.org/abs/hep-th/9704151}{{\ttfamily arXiv:hep-th/9704151
  [hep-th]}}.

\bibitem{Kachru:1998ys}
S.~Kachru and E.~Silverstein, ``{4-D conformal theories and strings on
  orbifolds},'' \href{http://dx.doi.org/10.1103/PhysRevLett.80.4855}{{\em Phys.
  Rev. Lett.} {\bfseries 80} (1998) 4855--4858},
\href{http://arxiv.org/abs/hep-th/9802183}{{\ttfamily arXiv:hep-th/9802183
  [hep-th]}}.

\bibitem{Klebanov:1998hh}
I.~R. Klebanov and E.~Witten, ``{Superconformal field theory on three-branes at
  a Calabi-Yau singularity},''
  \href{http://dx.doi.org/10.1016/S0550-3213(98)00654-3}{{\em Nucl. Phys.}
  {\bfseries B536} (1998) 199--218},
\href{http://arxiv.org/abs/hep-th/9807080}{{\ttfamily arXiv:hep-th/9807080
  [hep-th]}}.

\bibitem{Feng:2000mi}
B.~Feng, A.~Hanany, and Y.-H. He, ``{D-brane gauge theories from toric
  singularities and toric duality},''
  \href{http://dx.doi.org/10.1016/S0550-3213(00)00699-4}{{\em Nucl. Phys.}
  {\bfseries B595} (2001) 165--200},
\href{http://arxiv.org/abs/hep-th/0003085}{{\ttfamily arXiv:hep-th/0003085
  [hep-th]}}.

\bibitem{Wijnholt:2002qz}
M.~Wijnholt, ``{Large volume perspective on branes at singularities},''
  \href{http://dx.doi.org/10.4310/ATMP.2003.v7.n6.a6}{{\em Adv. Theor. Math.
  Phys.} {\bfseries 7} no.~6, (2003) 1117--1153},
\href{http://arxiv.org/abs/hep-th/0212021}{{\ttfamily arXiv:hep-th/0212021
  [hep-th]}}.

\bibitem{Herzog:2003zc}
C.~P. Herzog, ``{Exceptional collections and del Pezzo gauge theories},''
  \href{http://dx.doi.org/10.1088/1126-6708/2004/04/069}{{\em JHEP} {\bfseries
  04} (2004) 069},
\href{http://arxiv.org/abs/hep-th/0310262}{{\ttfamily arXiv:hep-th/0310262
  [hep-th]}}.

\bibitem{Herzog:2004qw}
C.~P. Herzog, ``{Seiberg duality is an exceptional mutation},''
  \href{http://dx.doi.org/10.1088/1126-6708/2004/08/064}{{\em JHEP} {\bfseries
  08} (2004) 064},
\href{http://arxiv.org/abs/hep-th/0405118}{{\ttfamily arXiv:hep-th/0405118
  [hep-th]}}.

\bibitem{Aspinwall:2004vm}
P.~S. Aspinwall and I.~V. Melnikov, ``{D-branes on vanishing del Pezzo
  surfaces},'' \href{http://dx.doi.org/10.1088/1126-6708/2004/12/042}{{\em
  JHEP} {\bfseries 12} (2004) 042},
\href{http://arxiv.org/abs/hep-th/0405134}{{\ttfamily arXiv:hep-th/0405134
  [hep-th]}}.

\bibitem{Aspinwall:2004bs}
P.~S. Aspinwall and S.~H. Katz, ``{Computation of superpotentials for
  D-branes},'' \href{http://dx.doi.org/10.1007/s00220-006-1527-6}{{\em Commun.
  Math. Phys.} {\bfseries 264} (2006) 227--253},
\href{http://arxiv.org/abs/hep-th/0412209}{{\ttfamily arXiv:hep-th/0412209
  [hep-th]}}.

\bibitem{Hanany:2005ve}
A.~Hanany and K.~D. Kennaway, ``{Dimer models and toric diagrams},''
\href{http://arxiv.org/abs/hep-th/0503149}{{\ttfamily arXiv:hep-th/0503149
  [hep-th]}}.

\bibitem{Franco:2005rj}
S.~Franco, A.~Hanany, K.~D. Kennaway, D.~Vegh, and B.~Wecht, ``{Brane dimers
  and quiver gauge theories},''
  \href{http://dx.doi.org/10.1088/1126-6708/2006/01/096}{{\em JHEP} {\bfseries
  01} (2006) 096},
\href{http://arxiv.org/abs/hep-th/0504110}{{\ttfamily arXiv:hep-th/0504110
  [hep-th]}}.

\bibitem{Franco:2005sm}
S.~Franco, A.~Hanany, D.~Martelli, J.~Sparks, D.~Vegh, and B.~Wecht, ``{Gauge
  theories from toric geometry and brane tilings},''
  \href{http://dx.doi.org/10.1088/1126-6708/2006/01/128}{{\em JHEP} {\bfseries
  01} (2006) 128},
\href{http://arxiv.org/abs/hep-th/0505211}{{\ttfamily arXiv:hep-th/0505211
  [hep-th]}}.

\bibitem{Feng:2005gw}
B.~Feng, Y.-H. He, K.~D. Kennaway, and C.~Vafa, ``{Dimer models from mirror
  symmetry and quivering amoebae},''
  \href{http://dx.doi.org/10.4310/ATMP.2008.v12.n3.a2}{{\em Adv. Theor. Math.
  Phys.} {\bfseries 12} no.~3, (2008) 489--545},
\href{http://arxiv.org/abs/hep-th/0511287}{{\ttfamily arXiv:hep-th/0511287
  [hep-th]}}.

\bibitem{Herzog:2006bu}
C.~P. Herzog and R.~L. Karp, ``{On the geometry of quiver gauge theories
  (Stacking exceptional collections)},'' {\em Adv. Theor. Math. Phys.}
  {\bfseries 13} no.~3, (2009) 599--636,
\href{http://arxiv.org/abs/hep-th/0605177}{{\ttfamily arXiv:hep-th/0605177
  [hep-th]}}.

\bibitem{Morrison:1998cs}
D.~R. Morrison and M.~R. Plesser, ``{Nonspherical horizons. 1.},'' {\em Adv.
  Theor. Math. Phys.} {\bfseries 3} (1999) 1--81,
\href{http://arxiv.org/abs/hep-th/9810201}{{\ttfamily arXiv:hep-th/9810201
  [hep-th]}}.

\bibitem{Mohri:1997ef}
K.~Mohri, ``{D-branes and quotient singularities of Calabi-Yau fourfolds},''
  \href{http://dx.doi.org/10.1016/S0550-3213(98)00085-6}{{\em Nucl. Phys.}
  {\bfseries B521} (1998) 161--182},
\href{http://arxiv.org/abs/hep-th/9707012}{{\ttfamily arXiv:hep-th/9707012
  [hep-th]}}.

\bibitem{GarciaCompean:1998kh}
H.~Garcia-Compean and A.~M. Uranga, ``{Brane box realization of chiral gauge
  theories in two-dimensions},''
  \href{http://dx.doi.org/10.1016/S0550-3213(98)00725-1}{{\em Nucl. Phys.}
  {\bfseries B539} (1999) 329--366},
\href{http://arxiv.org/abs/hep-th/9806177}{{\ttfamily arXiv:hep-th/9806177
  [hep-th]}}.

\bibitem{Franco:2015tna}
S.~Franco, D.~Ghim, S.~Lee, R.-K. Seong, and D.~Yokoyama, ``{2d (0,2) quiver
  gauge theories and D-Branes},''
  \href{http://dx.doi.org/10.1007/JHEP09(2015)072}{{\em JHEP} {\bfseries 09}
  (2015) 072},
\href{http://arxiv.org/abs/1506.03818}{{\ttfamily arXiv:1506.03818 [hep-th]}}.

\bibitem{Franco:2015tya}
S.~Franco, S.~Lee, and R.-K. Seong, ``{Brane brick models, toric Calabi-Yau
  4-folds and 2d (0,2) quivers},''
  \href{http://dx.doi.org/10.1007/JHEP02(2016)047}{{\em JHEP} {\bfseries 02}
  (2016) 047},
\href{http://arxiv.org/abs/1510.01744}{{\ttfamily arXiv:1510.01744 [hep-th]}}.

\bibitem{Franco:2016qxh}
S.~Franco, S.~Lee, R.-K. Seong, and C.~Vafa, ``{Brane brick models in the
  mirror},''
\href{http://arxiv.org/abs/1609.01723}{{\ttfamily arXiv:1609.01723 [hep-th]}}.

\bibitem{Franco:2017lpa}
S.~Franco and G.~Musiker, ``{Higher cluster categories and QFT dualities},''
\href{http://arxiv.org/abs/1711.01270}{{\ttfamily arXiv:1711.01270 [hep-th]}}.

\bibitem{Gadde:2013lxa}
A.~Gadde, S.~Gukov, and P.~Putrov, ``{(0, 2) trialities},''
  \href{http://dx.doi.org/10.1007/JHEP03(2014)076}{{\em JHEP} {\bfseries 03}
  (2014) 076},
\href{http://arxiv.org/abs/1310.0818}{{\ttfamily arXiv:1310.0818 [hep-th]}}.

\bibitem{Gadde:2014ppa}
A.~Gadde, S.~Gukov, and P.~Putrov, ``{Exact solutions of 2d supersymmetric
  gauge theories},''
\href{http://arxiv.org/abs/1404.5314}{{\ttfamily arXiv:1404.5314 [hep-th]}}.

\bibitem{Gadde:2015kda}
A.~Gadde, ``{Holomorphy, triality, and nonperturbative beta function in 2D
  supersymmetric QCD},''
  \href{http://dx.doi.org/10.1103/PhysRevD.94.025024}{{\em Phys. Rev.}
  {\bfseries D94} no.~2, (2016) 025024},
\href{http://arxiv.org/abs/1506.07307}{{\ttfamily arXiv:1506.07307 [hep-th]}}.

\bibitem{Franco:2016nwv}
S.~Franco, S.~Lee, and R.-K. Seong, ``{Brane brick models and 2d (0, 2)
  triality},'' \href{http://dx.doi.org/10.1007/JHEP05(2016)020}{{\em JHEP}
  {\bfseries 05} (2016) 020},
\href{http://arxiv.org/abs/1602.01834}{{\ttfamily arXiv:1602.01834 [hep-th]}}.

\bibitem{Schafer-Nameki:2016cfr}
S.~Schafer-Nameki and T.~Weigand, ``{F-theory and 2d $(0, 2)$ theories},''
  \href{http://dx.doi.org/10.1007/JHEP05(2016)059}{{\em JHEP} {\bfseries 05}
  (2016) 059},
\href{http://arxiv.org/abs/1601.02015}{{\ttfamily arXiv:1601.02015 [hep-th]}}.

\bibitem{Apruzzi:2016iac}
F.~Apruzzi, F.~Hassler, J.~J. Heckman, and I.~V. Melnikov, ``{UV Completions
  for Non-Critical Strings},''
  \href{http://dx.doi.org/10.1007/JHEP07(2016)045}{{\em JHEP} {\bfseries 07}
  (2016) 045},
\href{http://arxiv.org/abs/1602.04221}{{\ttfamily arXiv:1602.04221 [hep-th]}}.

\bibitem{Apruzzi:2016nfr}
F.~Apruzzi, F.~Hassler, J.~J. Heckman, and I.~V. Melnikov, ``{From 6D SCFTs to
  dynamic GLSMs},'' \href{http://dx.doi.org/10.1103/PhysRevD.96.066015}{{\em
  Phys. Rev.} {\bfseries D96} no.~6, (2017) 066015},
\href{http://arxiv.org/abs/1610.00718}{{\ttfamily arXiv:1610.00718 [hep-th]}}.

\bibitem{Lawrie:2016axq}
C.~Lawrie, S.~Schafer-Nameki, and T.~Weigand, ``{Chiral 2d theories from N = 4
  SYM with varying coupling},''
  \href{http://dx.doi.org/10.1007/JHEP04(2017)111}{{\em JHEP} {\bfseries 04}
  (2017) 111},
\href{http://arxiv.org/abs/1612.05640}{{\ttfamily arXiv:1612.05640 [hep-th]}}.

\bibitem{Lawrie:2016rqe}
C.~Lawrie, S.~Schafer-Nameki, and T.~Weigand, ``{The gravitational sector of 2d
  (0, 2) F-theory vacua},''
  \href{http://dx.doi.org/10.1007/JHEP05(2017)103}{{\em JHEP} {\bfseries 05}
  (2017) 103},
\href{http://arxiv.org/abs/1612.06393}{{\ttfamily arXiv:1612.06393 [hep-th]}}.

\bibitem{Katz:2004nn}
S.~H. Katz and E.~Sharpe, ``{Notes on certain (0,2) correlation functions},''
  \href{http://dx.doi.org/10.1007/s00220-005-1443-1}{{\em Commun. Math. Phys.}
  {\bfseries 262} (2006) 611--644},
\href{http://arxiv.org/abs/hep-th/0406226}{{\ttfamily arXiv:hep-th/0406226
  [hep-th]}}.

\bibitem{Witten:2005px}
E.~Witten, ``{Two-dimensional models with (0,2) supersymmetry: Perturbative
  aspects},'' {\em Adv. Theor. Math. Phys.} {\bfseries 11} no.~1, (2007) 1--63,
\href{http://arxiv.org/abs/hep-th/0504078}{{\ttfamily arXiv:hep-th/0504078
  [hep-th]}}.

\bibitem{Adams:2005tc}
A.~Adams, J.~Distler, and M.~Ernebjerg, ``{Topological heterotic rings},''
  \href{http://dx.doi.org/10.4310/ATMP.2006.v10.n5.a2}{{\em Adv. Theor. Math.
  Phys.} {\bfseries 10} no.~5, (2006) 657--682},
\href{http://arxiv.org/abs/hep-th/0506263}{{\ttfamily arXiv:hep-th/0506263
  [hep-th]}}.

\bibitem{Tan:2006qt}
M.-C. Tan, ``{Two-dimensional twisted sigma models and the theory of chiral
  differential operators},''
  \href{http://dx.doi.org/10.4310/ATMP.2006.v10.n6.a1}{{\em Adv. Theor. Math.
  Phys.} {\bfseries 10} no.~6, (2006) 759--851},
\href{http://arxiv.org/abs/hep-th/0604179}{{\ttfamily arXiv:hep-th/0604179
  [hep-th]}}.

\bibitem{McOrist:2007kp}
J.~McOrist and I.~V. Melnikov, ``{Half-twisted correlators from the Coulomb
  branch},'' \href{http://dx.doi.org/10.1088/1126-6708/2008/04/071}{{\em JHEP}
  {\bfseries 04} (2008) 071},
\href{http://arxiv.org/abs/0712.3272}{{\ttfamily arXiv:0712.3272 [hep-th]}}.

\bibitem{Tan:2008mi}
M.-C. Tan and J.~Yagi, ``{Chiral algebras of (0,2) sigma models: beyond
  perturbation theory},''
  \href{http://dx.doi.org/10.1007/s11005-008-0249-4}{{\em Lett. Math. Phys.}
  {\bfseries 84} (2008) 257--273},
\href{http://arxiv.org/abs/0801.4782}{{\ttfamily arXiv:0801.4782 [hep-th]}}.

\bibitem{Donagi:2011uz}
R.~Donagi, J.~Guffin, S.~Katz, and E.~Sharpe, ``{A mathematical theory of
  quantum sheaf cohomology},''
  \href{http://dx.doi.org/10.4310/AJM.2014.v18.n3.a1}{{\em Asian J. Math.}
  {\bfseries 18} (2014) 387--418},
\href{http://arxiv.org/abs/1110.3751}{{\ttfamily arXiv:1110.3751 [math.AG]}}.

\bibitem{Donagi:2011va}
R.~Donagi, J.~Guffin, S.~Katz, and E.~Sharpe, ``{Physical aspects of quantum
  sheaf cohomology for deformations of tangent bundles of toric varieties},''
  \href{http://dx.doi.org/10.4310/ATMP.2013.v17.n6.a2}{{\em Adv. Theor. Math.
  Phys.} {\bfseries 17} no.~6, (2013) 1255--1301},
\href{http://arxiv.org/abs/1110.3752}{{\ttfamily arXiv:1110.3752 [hep-th]}}.

\bibitem{Dedushenko:2015opz}
M.~Dedushenko, ``{Chiral algebras in Landau-Ginzburg models},''
\href{http://arxiv.org/abs/1511.04372}{{\ttfamily arXiv:1511.04372 [hep-th]}}.

\bibitem{Closset:2015ohf}
C.~Closset, W.~Gu, B.~Jia, and E.~Sharpe, ``{Localization of twisted $
  \mathcal{N}=\left(0,\;2\right) $ gauged linear sigma models in two
  dimensions},'' \href{http://dx.doi.org/10.1007/JHEP03(2016)070}{{\em JHEP}
  {\bfseries 03} (2016) 070},
\href{http://arxiv.org/abs/1512.08058}{{\ttfamily arXiv:1512.08058 [hep-th]}}.

\bibitem{Guo:2015caf}
J.~Guo, Z.~Lu, and E.~Sharpe, ``{Quantum sheaf cohomology on Grassmannians},''
  \href{http://dx.doi.org/10.1007/s00220-016-2763-z}{{\em Commun. Math. Phys.}
  {\bfseries 352} no.~1, (2017) 135--184},
\href{http://arxiv.org/abs/1512.08586}{{\ttfamily arXiv:1512.08586 [hep-th]}}.

\bibitem{Gu:2017nye}
W.~Gu and E.~Sharpe, ``{A proposal for (0,2) mirrors of toric varieties},''
\href{http://arxiv.org/abs/1707.05274}{{\ttfamily arXiv:1707.05274 [hep-th]}}.

\bibitem{Sharpe:1999qz}
E.~R. Sharpe, ``{D-branes, derived categories, and Grothendieck groups},''
  \href{http://dx.doi.org/10.1016/S0550-3213(99)00535-0}{{\em Nucl. Phys.}
  {\bfseries B561} (1999) 433--450},
\href{http://arxiv.org/abs/hep-th/9902116}{{\ttfamily arXiv:hep-th/9902116
  [hep-th]}}.

\bibitem{Douglas:2000gi}
M.~R. Douglas, ``{D-branes, categories and N=1 supersymmetry},''
  \href{http://dx.doi.org/10.1063/1.1374448}{{\em J. Math. Phys.} {\bfseries
  42} (2001) 2818--2843},
\href{http://arxiv.org/abs/hep-th/0011017}{{\ttfamily arXiv:hep-th/0011017
  [hep-th]}}.

\bibitem{Katz:2002gh}
S.~H. Katz and E.~Sharpe, ``{D-branes, open string vertex operators, and Ext
  groups},'' \href{http://dx.doi.org/10.4310/ATMP.2002.v6.n6.a1}{{\em Adv.
  Theor. Math. Phys.} {\bfseries 6} (2003) 979--1030},
\href{http://arxiv.org/abs/hep-th/0208104}{{\ttfamily arXiv:hep-th/0208104
  [hep-th]}}.

\bibitem{Katz:2002jh}
S.~H. Katz, T.~Pantev, and E.~Sharpe, ``{D branes, orbifolds, and Ext
  groups},'' \href{http://dx.doi.org/10.1016/j.nuclphysb.2003.09.022}{{\em
  Nucl. Phys.} {\bfseries B673} (2003) 263--300},
\href{http://arxiv.org/abs/hep-th/0212218}{{\ttfamily arXiv:hep-th/0212218
  [hep-th]}}.

\bibitem{Sharpe:2003dr}
E.~Sharpe, ``{Lectures on D-branes and sheaves},''
\newblock 2003.
\newblock
\href{http://arxiv.org/abs/hep-th/0307245}{{\ttfamily arXiv:hep-th/0307245
  [hep-th]}}.
\newblock

\bibitem{Aspinwall:2004jr}
P.~S. Aspinwall,
  \href{http://dx.doi.org/10.1142/9789812775108_0001}{``{D-branes on Calabi-Yau
  manifolds},''} in {\em {Progress in string theory. Proceedings, Summer
  School, TASI 2003, Boulder, USA, June 2-27, 2003}}, pp.~1--152.
\newblock 2004.
\newblock
\href{http://arxiv.org/abs/hep-th/0403166}{{\ttfamily arXiv:hep-th/0403166
  [hep-th]}}.
\newblock

\bibitem{Franco:2016tcm}
S.~Franco, S.~Lee, R.-K. Seong, and C.~Vafa, ``{Quadrality for supersymmetric
  matrix models},''
\href{http://arxiv.org/abs/1612.06859}{{\ttfamily arXiv:1612.06859 [hep-th]}}.

\bibitem{Witten:1993yc}
E.~Witten, ``{Phases of N=2 theories in two-dimensions},''
  \href{http://dx.doi.org/10.1016/0550-3213(93)90033-L}{{\em Nucl. Phys.}
  {\bfseries B403} (1993) 159--222},
  \href{http://arxiv.org/abs/hep-th/9301042}{{\ttfamily arXiv:hep-th/9301042
  [hep-th]}}.
[AMS/IP Stud. Adv. Math.1,143(1996)].

\bibitem{1999math.....10179K}
B.~{Keller}, ``{Introduction to A-infinity algebras and modules},'' {\em ArXiv
  Mathematics e-prints} (Nov., 1999) ,
  \href{http://arxiv.org/abs/math/9910179}{{\ttfamily math/9910179}}.

\bibitem{Aspinwall:2005ur}
P.~S. Aspinwall and L.~M. Fidkowski, ``{Superpotentials for quiver gauge
  theories},'' \href{http://dx.doi.org/10.1088/1126-6708/2006/10/047}{{\em
  JHEP} {\bfseries 10} (2006) 047},
\href{http://arxiv.org/abs/hep-th/0506041}{{\ttfamily arXiv:hep-th/0506041
  [hep-th]}}.

\bibitem{eagertalk} R.~Eager, ``{Two dimensional $N=(0,2)$ theories
and Calabi-Yau 4-algebras},'' talk in Bern, Switzerland, available at
{\tt http://susy-gauge.itp.unibe.ch/2016/talks/01-Monday/01-07-Eager.pdf}.

\bibitem{Sharpe:2006qd}
E.~Sharpe, ``{Notes on certain other (0,2) correlation functions},''
  \href{http://dx.doi.org/10.4310/ATMP.2009.v13.n1.a2}{{\em Adv. Theor. Math.
  Phys.} {\bfseries 13} no.~1, (2009) 33--70},
\href{http://arxiv.org/abs/hep-th/0605005}{{\ttfamily arXiv:hep-th/0605005
  [hep-th]}}.

\bibitem{Berenstein:2002fi}
D.~Berenstein and M.~R. Douglas, ``{Seiberg duality for quiver gauge
  theories},''
\href{http://arxiv.org/abs/hep-th/0207027}{{\ttfamily arXiv:hep-th/0207027
  [hep-th]}}.

\bibitem{Ibanez:1998qp}
L.~E. Ibanez, R.~Rabadan, and A.~M. Uranga, ``{Anomalous U(1)'s in type I and
  type IIB D = 4, N=1 string vacua},''
  \href{http://dx.doi.org/10.1016/S0550-3213(98)00791-3}{{\em Nucl. Phys.}
  {\bfseries B542} (1999) 112--138},
\href{http://arxiv.org/abs/hep-th/9808139}{{\ttfamily arXiv:hep-th/9808139
  [hep-th]}}.

\bibitem{Herbst:2004jp}
M.~Herbst, C.-I. Lazaroiu, and W.~Lerche, ``{Superpotentials, A(infinity)
  relations and WDVV equations for open topological strings},''
  \href{http://dx.doi.org/10.1088/1126-6708/2005/02/071}{{\em JHEP} {\bfseries
  02} (2005) 071},
\href{http://arxiv.org/abs/hep-th/0402110}{{\ttfamily arXiv:hep-th/0402110
  [hep-th]}}.

\bibitem{kad}
T.~V. Kadeishvili, ``The algebraic structure in the homology of an
  {$A_{\infty}$} algebra,'' {\em Sobshch. Akad. Nauk. Gruzin. SSR} {\bfseries
  108} (1982) 249--252.

\bibitem{Cachazo:2001sg}
F.~Cachazo, B.~Fiol, K.~A. Intriligator, S.~Katz, and C.~Vafa, ``{A geometric
  unification of dualities},''
  \href{http://dx.doi.org/10.1016/S0550-3213(02)00078-0}{{\em Nucl. Phys.}
  {\bfseries B628} (2002) 3--78},
\href{http://arxiv.org/abs/hep-th/0110028}{{\ttfamily arXiv:hep-th/0110028
  [hep-th]}}.

\bibitem{Hanany:2006nm}
A.~Hanany, C.~P. Herzog, and D.~Vegh, ``{Brane tilings and exceptional
  collections},'' \href{http://dx.doi.org/10.1088/1126-6708/2006/07/001}{{\em
  JHEP} {\bfseries 07} (2006) 001},
\href{http://arxiv.org/abs/hep-th/0602041}{{\ttfamily arXiv:hep-th/0602041
  [hep-th]}}.

\bibitem{10.2307/1969996}
R.~Bott, ``Homogeneous vector bundles,'' {\em Annals of Mathematics} {\bfseries
  66} no.~2, (1957) 203--248.

\bibitem{10.2307/1970237}
B.~Kostant, ``Lie algebra cohomology and the generalized {Borel-Weil}
  theorem,'' {\em Annals of Mathematics} {\bfseries 74} no.~2, (1961) 329--387.

\bibitem{Guo:2015gha}
J.~Guo, B.~Jia, and E.~Sharpe, ``{Chiral operators in two-dimensional (0,2)
  theories and a test of triality},''
  \href{http://dx.doi.org/10.1007/JHEP06(2015)201}{{\em JHEP} {\bfseries 06}
  (2015) 201},
\href{http://arxiv.org/abs/1501.00987}{{\ttfamily arXiv:1501.00987 [hep-th]}}.

\bibitem{Hori:2011pd}
K.~Hori, ``{Duality in two-dimensional (2,2) supersymmetric non-abelian gauge
  theories},'' \href{http://dx.doi.org/10.1007/JHEP10(2013)121}{{\em JHEP}
  {\bfseries 10} (2013) 121},
\href{http://arxiv.org/abs/1104.2853}{{\ttfamily arXiv:1104.2853 [hep-th]}}.

\bibitem{Jia:2014ffa}
B.~Jia, E.~Sharpe, and R.~Wu, ``{Notes on nonabelian (0,2) theories and
  dualities},'' \href{http://dx.doi.org/10.1007/JHEP08(2014)017}{{\em JHEP}
  {\bfseries 08} (2014) 017},
\href{http://arxiv.org/abs/1401.1511}{{\ttfamily arXiv:1401.1511 [hep-th]}}.

\bibitem{Rudakov:1619155}
A.~N. Rudakov, {\em {Helices and vector bundles}}.
\newblock London Mathematical Society Lecture Note Series. Cambridge University
  Press, Cambridge, 1990.

\bibitem{Ishibashi:1996xs}
N.~Ishibashi, H.~Kawai, Y.~Kitazawa, and A.~Tsuchiya, ``{A large N reduced
  model as superstring},''
  \href{http://dx.doi.org/10.1016/S0550-3213(97)00290-3}{{\em Nucl. Phys.}
  {\bfseries B498} (1997) 467--491},
\href{http://arxiv.org/abs/hep-th/9612115}{{\ttfamily arXiv:hep-th/9612115
  [hep-th]}}.

\bibitem{Diaconescu:2000ec}
D.-E. Diaconescu and M.~R. Douglas, ``{D-branes on stringy Calabi-Yau
  manifolds},''
\href{http://arxiv.org/abs/hep-th/0006224}{{\ttfamily arXiv:hep-th/0006224
  [hep-th]}}.

\bibitem{Douglas:2002fr}
M.~R. Douglas, S.~Govindarajan, T.~Jayaraman, and A.~Tomasiello, ``{D branes on
  Calabi-Yau manifolds and superpotentials},''
  \href{http://dx.doi.org/10.1007/s00220-004-1091-x}{{\em Commun. Math. Phys.}
  {\bfseries 248} (2004) 85--118},
\href{http://arxiv.org/abs/hep-th/0203173}{{\ttfamily arXiv:hep-th/0203173
  [hep-th]}}.

\bibitem{Douglas:2000qw}
M.~R. Douglas, B.~Fiol, and C.~Romelsberger, ``{The spectrum of BPS branes on a
  noncompact Calabi-Yau},''
  \href{http://dx.doi.org/10.1088/1126-6708/2005/09/057}{{\em JHEP} {\bfseries
  09} (2005) 057},
\href{http://arxiv.org/abs/hep-th/0003263}{{\ttfamily arXiv:hep-th/0003263
  [hep-th]}}.

\bibitem{Brunner:1999jq}
I.~Brunner, M.~R. Douglas, A.~E. Lawrence, and C.~Romelsberger, ``{D-branes on
  the quintic},'' \href{http://dx.doi.org/10.1088/1126-6708/2000/08/015}{{\em
  JHEP} {\bfseries 08} (2000) 015},
\href{http://arxiv.org/abs/hep-th/9906200}{{\ttfamily arXiv:hep-th/9906200
  [hep-th]}}.

\bibitem{Eager:2018oww}
  R.~Eager and I.~Saberi,
  ``Holomorphic field theories and Calabi--Yau algebras,''
  {\ttfamily arXiv:1805.02084 [hep-th]}.


\end{thebibliography}\endgroup

\end{document}